\documentclass[]{jfm}
\usepackage{xpatch}
\usepackage{graphicx}
\usepackage{newtxtext}
\usepackage{newtxmath}
\usepackage[numbers]{natbib}
\usepackage{booktabs}
\usepackage{siunitx}
\usepackage{array}
\usepackage{hyperref}
\usepackage{subcaption}
\usepackage{comment}
\usepackage[mathlines]{lineno}  % mathlines numbers equations too
\hypersetup{
    colorlinks = true,
    urlcolor   = blue,
    citecolor  = blue,
    linkcolor = blue,
}

\newcommand{\RomanNumeralCaps}[1]
    {\MakeUppercase{\romannumeral #1}}
\newcommand{\mainref}[2]{\hyperref[#1]{#2}}
\newcommand{\PDfirst}[2]{\frac{\partial {#1}}{\partial #2}}

\newcommand{\Dfirst}[2]{\frac{\mathrm{d} #1}{\mathrm{d} {#2}} }

\usepackage[normalem]{ulem} % for strikeout (\sout)

\makeatletter

\xpatchcmd\NAT@citex
 {%
  \@citea\NAT@hyper@{%
    \NAT@nmfmt{\NAT@nm}%
    \hyper@natlinkbreak{\NAT@aysep\NAT@spacechar}{\@citeb\@extra@b@citeb}%
    \NAT@date
  }%
 }
 {%
  \@citea
  \NAT@nmfmt{\NAT@nm}%
  \NAT@aysep\NAT@spacechar
  \NAT@hyper@{\NAT@date}%
 }
 {}{}
\xpatchcmd\NAT@citex
 {%
  \@citea\NAT@hyper@{%
    \NAT@nmfmt{\NAT@nm}%
    \hyper@natlinkbreak{\NAT@spacechar\NAT@@open\if*#1*\else#1\NAT@spacechar\fi}%
    {\@citeb\@extra@b@citeb}%
    \NAT@date
  }%
 }
 {
  \@citea
    \NAT@nmfmt{\NAT@nm}%
    \NAT@spacechar\NAT@@open\if*#1*\else#1\NAT@spacechar\fi
    \NAT@hyper@{\NAT@date}%
 }
 {}{}

\makeatother

\title{Dynamics of a compressible gas {injected} into a confined porous layer}

\author{Peter Castellucci\aff{1}
  \corresp{\email{peter.castellucci@manchester.ac.uk}},
  Radha Boya\aff{2},
  Lin Ma\aff{3}, Igor L. Chernyavsky\aff{1} \and Oliver E. Jensen\aff{1}}

\affiliation{\aff{1}Department of Mathematics, University of Manchester, Manchester M13 9PL, UK
\aff{2}Department of Physics and Astronomy, University of Manchester, Manchester M13 9PL, UK
\aff{3}Department of Chemical Engineering, University of Manchester, Manchester M13 9PL, UK}

\begin{document}
%\linenumbers
\maketitle

\begin{abstract}   % 250 words
{Underground} gas storage is a critical technology in global efforts to mitigate climate change. In particular, hydrogen storage offers a promising solution for integrating renewable energy into the power grid. When injected into the subsurface, hydrogen’s low viscosity compared to the resident brine causes {a bubble of hydrogen trapped beneath caprock} to spread rapidly {into an aquifer through release of} a thin gas layer above the brine, complicating recovery. In long aquifers, the large viscous pressure drop between source and outlet induces significant pressure variations, potentially leading to substantial density changes in the injected gas. To examine the role of gas compressibility in the spreading dynamics, we use long-wave theory to derive coupled nonlinear evolution equations for the gas pressure and {gas/liquid} interface height, focusing on the limit of long domains, weak gas compressibility and low {gas/liquid viscosity ratio}. Simulations are supplemented with a comprehensive asymptotic analysis of parameter regimes.  Unlike the near-incompressible limit, in which gas spreading rates are dictated by the source strength and {viscosity ratio}, and compressive effects are transient, we show how compression of the main gas bubble can generate dynamic pressure changes that are coupled to those in the thin gas layer that spreads over the liquid, with compressive effects having a sustained influence along the layer.  This {coupling} allows compressibility to reduce spreading rates and gas pressures.  We characterise this behaviour via a set of low-order models that reveal dominant scalings, highlighting the role of compressibility in mediating the evolution of the gas layer.
\end{abstract}

\section{Introduction}
\label{sec:headings}

{Underground} gas storage plays a crucial role in addressing anthropogenic climate change, both by sequestering greenhouse gas emissions prior to atmospheric release and by enabling the transition to cleaner energy sources. The first approach is well-established through carbon capture and storage, where large quantities of CO$_2$ are injected  {underground} for long-term containment. More recently, underground hydrogen storage has gained significant attention for its potential to enable large-scale integration of renewable energy into the power grid (\citealt{heinemann_enabling_2021}; \citealt{zivar_underground_2021}; \citealt{muhammed_review_2022}; \citealt{lord_geologic_2014}; \citealt{tarkowski_underground_2019}). The concept is to use surplus renewable energy generated during the summer months to produce hydrogen via electrolysis, store it in underground formations, and subsequently recover it during the winter when demand exceeds supply. Hydrogen is a particularly attractive candidate for this large-scale energy storage owing to its high energy density and the fact that its combustion or use in a fuel cell produces minimal harmful emissions \citep{zivar_underground_2021}. Proposed underground storage sites for gas include depleted reservoirs, aquifers, and salt caverns, all of which are typically saturated with brine. Unlike CO$_2$ storage, where the gas is intended for permanent sequestration, hydrogen storage requires recoverability, making accurate predictions of gas plume evolution crucial. Due to its exceptionally low viscosity relative to brine, injected hydrogen may spread rapidly along the top of the brine within an aquifer, forming a thin gas layer that extends far from the injection site \citep{hagemann_mathematical_2015}. This behaviour not only reduces storage efficiency but also increases the risk of hydrogen loss.

The injected layer of hydrogen, through its displacement of heavier brine, is a form of viscous gravity current \citep{zheng2022}.  Much of the research on viscous gravity currents has focused on incompressible flows, where the interaction between buoyancy and viscosity governs the rate and extent of spreading. In the context of hydrogen storage, the low viscosity of the injected gas, coupled with buoyancy effects, plays a critical role in determining how far and how quickly the hydrogen spreads {underground}. Early studies primarily addressed gravity-driven spreading, where a dense fluid advances along a rigid boundary within an infinitely deep porous medium \citep{barenblatt_unsteady_1952, huppert_gravity-driven_1995, lyle_axisymmetric_2005}. However, when a fluid is actively injected into a confined porous layer, the resulting pressure gradient, caused by the pressure drop between the injection site and the far field, significantly alters the flow dynamics.

\citet{huppert_gravity-driven_1995} examined the exchange of two confined incompressible fluids with differing densities and pressures, but equal viscosities, between two aquifers. By deriving a similarity solution, they showed that buoyancy transports the denser fluid along the lower boundary and the lighter fluid along the upper boundary, while an imposed pressure gradient drives a net flow. Subsequent research expanded on this framework to incorporate fluids with different viscosities, highlighting how the viscosity contrast influences spreading behaviour (\citealt{nordbotten_similarity_2006}; \citealt{pegler_fluid_2014}; \citealt{zheng_flow_2015}). \citet{nordbotten_similarity_2006} generalised Huppert \& Woods' (\citeyear{huppert_gravity-driven_1995}) study to investigate the effect of viscosity contrasts in an axisymmetric geometry. \citet{pegler_fluid_2014} conducted a comprehensive study of flow behaviour in a two-dimensional channel. Their findings revealed that, at early times, spreading is primarily buoyancy-driven and can be approximated by the classical porous-medium equation of \citet{barenblatt_unsteady_1952}. At later times, low-viscosity injected fluid 
forms a thin layer along the upper boundary, with buoyancy becoming less dominant compared to viscous effects. By deriving a large-time similarity solution, they demonstrated that the movement of the upper contact line is controlled by the strength of the source. In this case, since the flow is incompressible, the length of the channel determines the pressure needed to drive the motion but does not affect the spreading dynamics. These theoretical predictions were validated through laboratory experiments in which freshwater was injected into a saltwater-saturated porous medium, confirming that the similarity solution accurately describes the interface evolution.

When the injected gas is compressible, however, the dynamics change: the source can in principle increase the gas pressure without advancing the gas-liquid interface, and the subsequent propagation of the upper contact line may be influenced by the length of the channel. In long channels, the viscous pressure drop between source and outlet may lead to spatial variations in pressure, which could in turn lead to noticeable variations in gas density along the channel. These variations could affect the mass flow rate and, consequently, influence the rate of gas spreading. Despite the potential for compressibility to significantly alter the flow dynamics, its effects in this context remain under-explored. While some studies have addressed compressibility, most have focused on fluid-structure interactions, particularly the impact of pressure buildup on the rock matrix integrity \citep{mathias_approximate_2009} and fluid migration in neighbouring reservoirs \citep{jenkins_impact_2019}. A common approach to coupling the mechanics of the pore structure with pressure build-up during injection is to introduce rock compressibility, defined as $c_r = ({1}/{\phi}) {\mathrm{d} \phi}/{\mathrm{d} p^*}$, where $\phi$ is the porosity and $p^*$ is the pressure. This parameter quantifies how the pore space deforms in response to pressure changes and is valid under conditions of constant vertical stress and negligible lateral strain \citep{jenkins_impact_2019}.

\cite{mathias_approximate_2009} examined pressure build-up during gas injection into an axisymmetric channel of infinite extent. Using matched asymptotics, they derived similarity solutions for pressure evolution in the case where the compressibilities of the gas and resident liquid are comparable (relative to rock and water), and where the interface propagates much more slowly than the diffusive pressure front. This work was later extended by \cite{mathias_pressure_2011} to finite-length channels, offering further insight into pressure evolution under more realistic conditions. \cite{jenkins_impact_2019} investigated pressure dissipation in a layered aquifer system, where multiple confined channels are stacked vertically. They analyzed how water leakage from the injection aquifer to the surrounding aquifers influences the pressure field. In their model, CO$_2$ displaces resident water in the injection channel, and weak vertical flow allows liquid to escape through the confining boundaries, relieving pressure within the system. By incorporating the compressibility of both gas and liquid through a linearized equation of state, they found that pressure reduction due to water escape slows the gas plume’s spreading rate. In contrast to these studies, \cite{cuttle_compression-driven_2023} examined how compressibility {interacts with surface tension and viscous forces in the liquid, to influence viscous fingering instabilities, when} a compressible gas is injected into a Hele-Shaw cell filled with water. By neglecting pressure variations in the gas due to viscous effects, they used Boyle’s law to couple the uniform pressure in the gas, which drives displacement, to the viscous effects in the liquid and the capillary forces at the interface. Their results showed that gas compressibility delays the onset of viscous fingering and reduces its severity, with the severity quantified by the isoperimetric ratio (which compares the length of the interface to the area it occupies). Additionally, compressibility was found to increase the breakthrough time, \hbox{i.e.} the moment when the interface reaches the outlet.

\begin{figure}
\centering
  \centerline{\includegraphics[scale = 0.8]{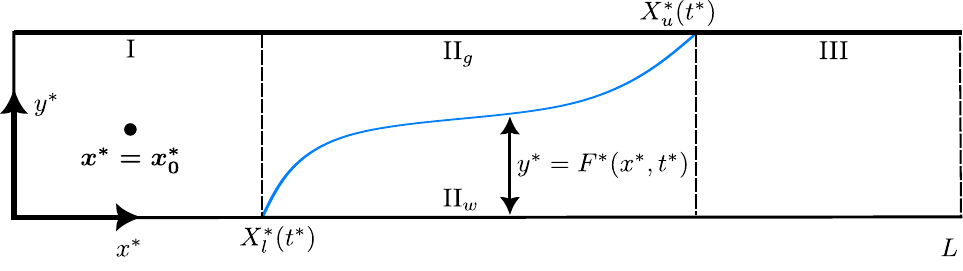}}
  \caption{Schematic showing the displacement of an ambient liquid in a porous medium (regions II$_w$ and III) due to the injection of a compressible gas (occupying regions I and II$_g$) from a line source at $\boldsymbol{x}^* = \boldsymbol{x}_0^*$. The interface separating the fluids is sharp and located at ${y}^* = {F}^*({x}^*, {t}^*)$. The pressure in the ambient brine is hydrostatic at the outlet at $x^*=L$.}
\label{fig:Diagram}
\end{figure}

In this paper, we extend the framework of \citet{pegler_fluid_2014} for injection of a gas into a confined porous channel containing brine (figure~\ref{fig:Diagram}) to investigate how gas compressibility interacts with buoyancy and viscous effects in both fluids to determine the spreading dynamics. Using a long-wave approximation, we derive two coupled nonlinear evolution equations governing the {gas/liquid} interface height and the gas pressure field.  These are mass conservation equations, with the gas pressure serving as a proxy for gas density via the equation of state.
The viscous pressure drop ahead of the upper contact line is captured by a boundary condition, while two kinematic conditions determine the motion of the contact lines. The model admits several distinguished asymptotic limits that reveal the roles of different mechanisms across parameter space; we derive a set of reduced models that capture dominant balances. In particular, we estimate key quantities of practical interest, including the pressure scale, pressure rise time, and the breakthrough time. In \S~\ref{sec:Model}, we present the governing equations and describe the scaling argument used to derive the long-wave model. The system is characterised by three primary dimensionless parameters, and we simplify the model in various sub-limits, identifying relevant scaling relationships. In \S~\ref{sec:results}, we present numerical solutions of the full model, illustrating how the key features of compressible flow emerge and comparing them with predictions from asymptotic analysis. Finally, in \S~\ref{sec:discussion}, we revisit the dominant balances in each regime, interpret the underlying physical mechanisms, and discuss the broader implications for hydrogen storage and related gas-injection technologies.

\section{Model formulation} 
\label{sec:Model}
\subsection{Governing equations}
We consider a two-dimensional model of fluid displacement driven by the injection of a compressible gas into a planar horizontally-confined porous medium with height $H$ and length $L$, as illustrated in figure \ref{fig:Diagram}. Initially, the medium is predominantly saturated with a liquid, with only a small amount of gas present. The length of the region initially occupied by gas is denoted by $L_{g0}$; the aspect ratio of this region is assumed small, \hbox{i.e.} $H/L_{g0} \ll 1$. (We do not seek here to model the initial formation of the gas bubble.) The permeability of the medium is taken to be uniform, with value $k_0$.

At time $t^* = 0$, gas is injected into the channel from a line source located at \(\boldsymbol{x}^* ~\equiv (x^*,y^*) = \boldsymbol{x}^*_0\).  Variations in the injection rate are governed by the function $q \, \mathcal{Q}(\omega {t^*})$, where $1/\omega$ represents the timescale of injection, $q$ denotes the source strength per unit width of the channel, and $\mathcal{Q}$ is a dimensionless function. The pressure at the outlet ($x^* = L$) is taken to be hydrostatic with a baseline value of $p_{g0}$. Using a sharp-interface model, disregarding capillary or miscibility 
effects, the fluids are segregated into two distinct phases and the domain is divided into four regions (figure~\ref{fig:Diagram}): the gas occupies regions \RomanNumeralCaps{1} and \RomanNumeralCaps{2}$_g$, while the liquid occupies regions \RomanNumeralCaps{2}$_w$ and \RomanNumeralCaps{3}.  We use the subscripts $g$ and $w$ to represent quantities in the gas and liquid (water) phase respectively. We revisit all modelling assumptions in \S~\ref{sec:discussion}.

\begin{figure}
    \centering
    \begin{subfigure}{0.49 \textwidth}
    \caption{}
        \includegraphics[width = \textwidth]{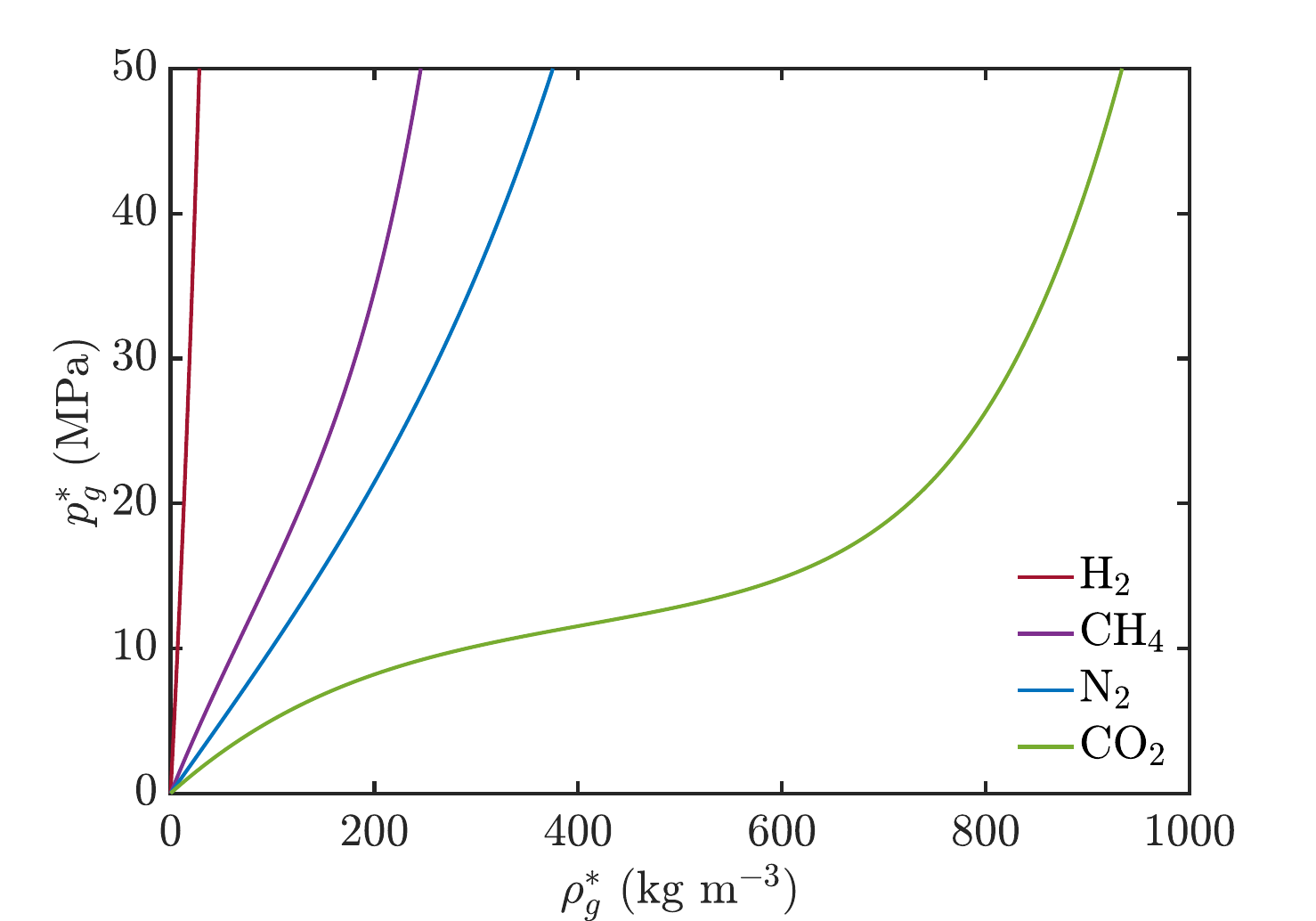}
    \end{subfigure}
    \hfill
    \begin{subfigure}{0.49 \textwidth}
    \caption{}
        \includegraphics[width = \textwidth]{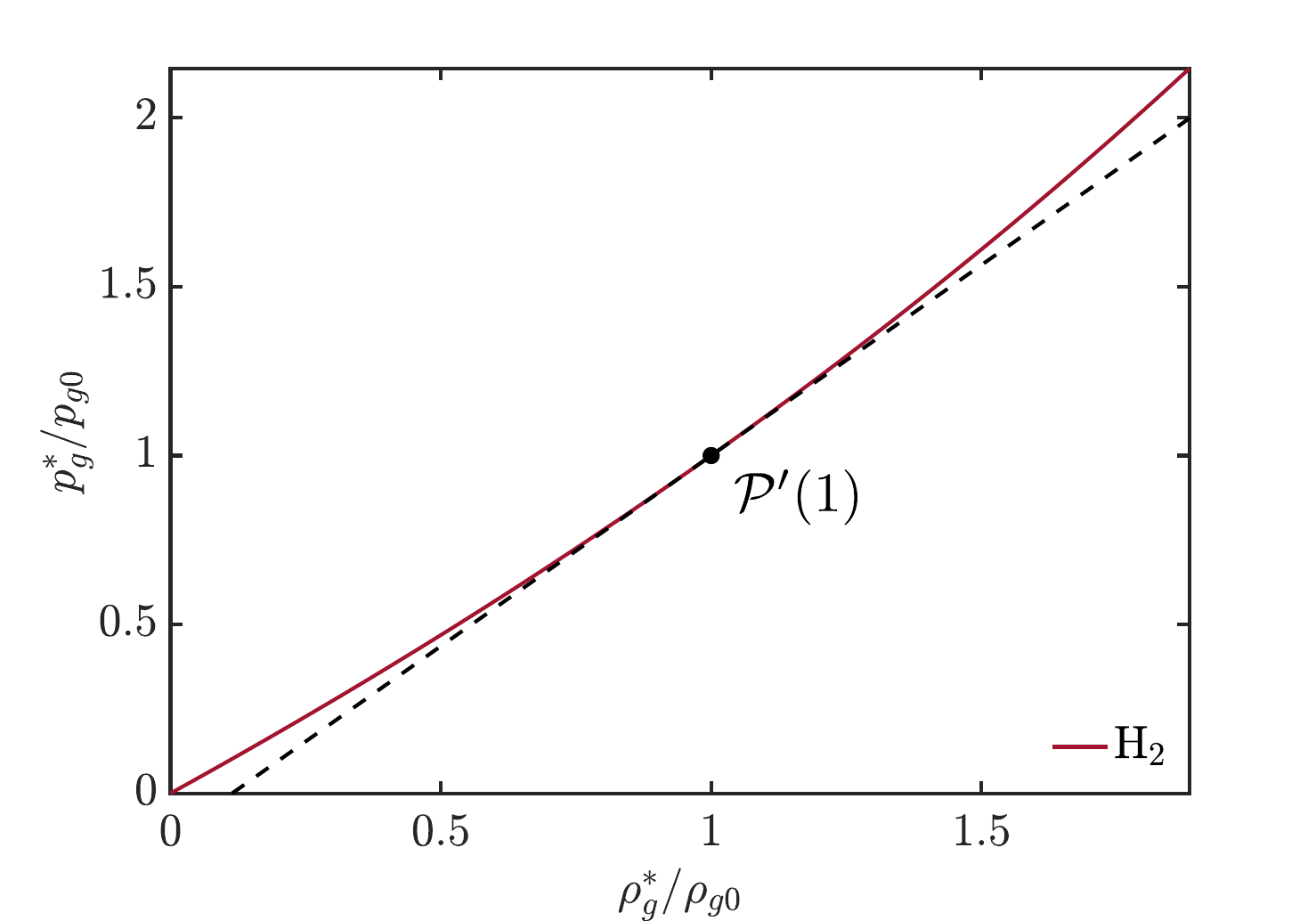}
    \end{subfigure}
        \caption{(a) Isothermal equations of state for hydrogen, methane, nitrogen, and carbon dioxide at 333\si{K}, sourced from \citet{NIST}. (b) The dimensionless function  $\mathcal{P}$ \eqref{eqState1} for hydrogen, computed using the reference pressure $p_{g0} = 23$~{MPa} and reference density $\rho_{g0} = 15$ $\si{kg.m^{-3}}$. The dashed line represents the tangent at the reference state, with slope  $\mathcal{P}^{\prime}(1)$. }
    \label{fig:EqState}
\end{figure}

We denote the height of the gas-liquid interface by $y^* = F^*(x^*, t^*)$ and the points of contact with the upper and lower boundaries by $X^*_u({t^*})$ and $X^*_l({t^*})$ respectively.  For later convenience, we extend the definition of $F^*$ to the whole domain such that 
\begin{equation}\label{eq:f}
    F^*(x^*,t^*)=\begin{cases} 0, & 0\leq x^*\leq X^*_l(t^*), \\
    H, & X^*_u(t^*) \leq  x^* \leq L. 
    \end{cases}
\end{equation}
The liquid is assumed to be incompressible with constant density ${\rho}_w$. {The initial gas and liquid volumes are assumed to be in thermal equilibrium with the surrounding pore matrix. Gas compression is treated as isothermal, an assumption that is valid when the thermal inertia of the solid pore matrix is much greater than that of the gas \citep{kushnir_temperature_2012}}.  At a fixed temperature, we define $\rho_{g0}$ to be the equilibrium gas density when at the outlet pressure $p_{g0}$. Then, using $p_{g0}$ and $\rho_{g0}$ as a reference pressure and a reference density, we write the equation of state for the gas (relating density $\rho^*_g$ to pressure $p^*_g$) and the equilibrium speed of sound in the gas ($c$) as
\begin{equation}\label{eqState1}
    p^*_g = p_{g0} \mathcal{P}\left(\frac{\rho^*_g}{\rho_{g0}}\right), \qquad c^2 \equiv \left. \PDfirst{p^*_g}{\rho^*_g} \right|_{\rho_{g0}} = \frac{p_{g0}}{\rho_{g0}} \mathcal{P}^{\prime}(1),
\end{equation}
respectively, for a suitable dimensionless function $\mathcal{P}$, for which $\mathcal{P}(1)=1$ and $\mathcal{P}'(1)>0$.  For hydrogen, $\mathcal{P}'(1)\geq 1$ and $\mathcal P''(1)\geq 0$ at temperatures and pressures of interest (figure~\ref{fig:EqState}$b$), so that $c^2$ rises with increasing pressure, although this may not be the case for gases such as CO$_2$ (figure~\ref{fig:EqState}$a$). 

The fluid velocities $\boldsymbol{u}_i^* = (u^*_i, v^*_i)$ in both phases are then given by
\begin{equation}\label{Darcy1}
      \boldsymbol{u}^*_i = -\frac{k_0}{\mu_i} \: ( \boldsymbol{\nabla}^* p^*_i -  \rho^*_i \boldsymbol{g}), \qquad (i = w,g),
\end{equation}
where $\boldsymbol{g}$ is the acceleration due to gravity, $\rho_w^*\equiv \rho_w$ and $\mu_g$ and $\mu_w$ are gas and liquid viscosities, assumed constant. The pore-scale Reynolds number, $Re = q d / (H\mu_g)$, is assumed sufficiently small to neglect inertia, where $d$ is a typical pore diameter.  Taking 
$H = 10\, \si{m}$, $\mu_g\approx 10^{-6}\,\si{kg.m^{-1}.s^{-1}}$ and $d\approx 10\,\mu\si{m}$ \citep{hashemi_pore-scale_2021}, the condition $Re\leq 1$ suggests that inertia can be neglected for injection rates up to $ q \approx 1\, \si{kg.m^{-1}.s^{-1}}$.

The continuity equations for the gas and liquid phases are
\addtocounter{equation}{1}
\begin{equation}\label{Continuity1}
\rho^*_{g, t^*} + \boldsymbol{\nabla}^* \cdot (\rho^*_g\boldsymbol{u}^*_g) = q \,\mathcal{Q}(\omega t^*) \, \delta^*(\boldsymbol{x}^* - \boldsymbol{x}_0^*), \quad
\boldsymbol{\nabla}^* \cdot \boldsymbol{u}_w^* = 0,
\tag{2.4a,b}
\end{equation}
respectively, where
$\delta^*(\boldsymbol{x}^*)$ is the Dirac delta function. The source at $\boldsymbol{x}^* = \boldsymbol{x}_0^*$ lies within a distance of order $H$ of an impermeable boundary at $x^*=0$. Equations \eqref{eqState1}, \eqref{Darcy1} and (\mainref{Continuity1}{2.4}$a$) form the governing equations for the gas and are applied in regions \RomanNumeralCaps{1} and \RomanNumeralCaps{2}$_g$; \eqref{Darcy1} and (\mainref{Continuity1}{2.4}$b$)
apply in regions \RomanNumeralCaps{2}$_w$ and \RomanNumeralCaps{3}. The kinematic condition
\begin{equation}\label{KinematicEq1}
        F^*_{t^*} + u^*_i F^*_{x^*} =  v^*_i,\quad \text{at} \quad  y^* = F^*, \quad (i = w,g; ~~0\leq x^*\leq L),
\end{equation}
relates the velocity of the interface to the fluid velocities for $X^*_l<x^*<X^*_u$. Subscripts $x^*$ and $t^*$ denote partial derivatives.  No-penetration at the walls implies $u^*_g = 0$ at $x^* = 0$, $v^*_i = 0 $ at $y^* = H$ and $v^*_i = 0 $ at $y^* = 0$. The relative strength of buoyancy to capillary forces can be characterised by a Bond number. We adopt the definition of the Bond number from \citet{GOLDING_2011} and \citet{zheng_self-similar_2019}, which measures the relative magnitude of buoyancy to the capillary entry pressure $p_e$ in the largest pores. Using $p_e = \gamma / k_0^{1/2}$ \citep{zheng_self-similar_2019}, the Bond number can be expressed as $Bo = \Delta \rho g H k_0^{1/2}/\gamma$, where $\Delta \rho = \rho_w - \rho_{g0} \approx 10^3\,\si{kg.m^{-3}}$ is the density difference between the fluids, $g$ is gravitational acceleration, $H \approx 10$ -- $50\,\si{m}$ is a characteristic aquifer height, $k_0 \approx 10^{-14}$ -- $10^{-12}\,\si{m^2}$ is the permeability, and $\gamma \approx 7 \times 10^{-2}\,\si{N.m^{-1}}$ is the hydrogen–liquid interfacial tension \citep{zivar_underground_2021}. An order-of-magnitude estimate gives $Bo \sim 0.1\,\text{--}\,10$, indicating that capillary effects may be important in shallow, low-permeability reservoirs. A detailed treatment of the capillary fringe is beyond the scope of this study, although we discuss its potential influence in \S~\ref{sec:discussion}. Accordingly, we assume a sharp interface and take the pressure to be continuous across it: 
\begin{equation}\label{ContinuityOfPressure1}
        p^*_g = p^*_w \quad \text{at} \quad y^* = F^* \quad\text{for}\quad X^*_l<x^*<X^*_u.
\end{equation}
The liquid pressure at the outlet is
\begin{equation}\label{OutletBC}
    p^*_w = p_{g0} - \rho_w g y^* \quad \text{at} \quad x^* = L.
\end{equation}
The initial bubble length specifies the initial gas volume (per unit width) via $V^*(0)=\int_0^{X^*_u} (H-F^*(x^*,0))\,\mathrm{d}x^*=HL_{g0}$. The interface is assumed to have a linear initial profile over a lengthscale $\Delta_{g0}$, so that 
\begin{equation}
F^*(x^*,0)=\frac{H}{{\Delta}_{g0}}(x^*-X_l^*(0))~~\mathrm{for}~~ X_l^*(0)\equiv L_{g0}-\tfrac{1}{2}{\Delta}_{g0}<x^*<X_u^*(0)\equiv L_{g0}+\tfrac{1}{2}{\Delta}_{g0}.
\label{eq:linic}
\end{equation}

We integrate (\mainref{Continuity1}{2.4}$a$) over $F^*<y^*<H$ for $0\leq x^*<X^*_u$, and (\mainref{Continuity1}{2.4}$b$) across $0<y^*<F^*$ for $X^*_u<x<L$ which, after applying \eqref{KinematicEq1} and the no-penetration conditions, respectively yield the transport equations
\begin{subequations}
\label{eq:mc}
\begin{alignat}{2}
    \left( \int_{F^*}^H \rho^*_g \: \mathrm{d} y^* \right)_{t^*} +  \left( \int_{F^*}^H  \rho^*_g  u^*_g \:\mathrm{d} y^* \right)_{x^*} &= q \mathcal{Q}(\omega t^*) \delta^*(x^* - x^*_0), \qquad  &&(0 < x^* < X^*_u),\\
    {F^*}_{t^*} +  \left(\int_0^{F^*} u^*_w \: \mathrm{d} y^* \right)_{x^*}  &= 0, &&(X^*_l < x^* < L).
\end{alignat}
\end{subequations}
The total masses (per unit width) of gas and liquid in the channel at a given instant are then expressed  using (\ref{eq:f}) as 
%\begin{center}
\begin{subequations}\label{TotalMass}
\begin{alignat}{1}
          M^*_g(t^*) &= 
          \int_{0}
          ^{X^*_u(t^*)} \int_{F^*(x^*,t^*)}^H \rho^*_g(x^*,y^*,t^*)  \: \mathrm{d} y^* \mathrm{d} x^* 
          ,\label{GasMass} \\
          M^*_w(t^*) &= \int_{X^*_l(t^*)}^{L}
          \int_0^{F^*(x^*,t^*)} \rho^*_w \: \mathrm{d} y^* \mathrm{d} x^* 
          \label{LiquidMass},
    \end{alignat}
\end{subequations}
%\end{center}
respectively. Integration of (\ref{eq:mc}) along the channel, { and using the no-penetration condition on the wall at $x^* = 0$}, gives the rates of gas accumulation in the channel and liquid outflow as
\addtocounter{equation}{1}
\begin{equation}
\label{eq:DiffMass}
\Dfirst{M^*_g}{t^*} = q \,\mathcal{Q}(\omega t^*) ,\quad
\Dfirst{M^*_w}{t^*}  = - \rho^*_w \int_0^H u^*_w(L,y^*,t^*) \mathrm{d} y^*.
\tag{2.11a,b}
\end{equation}
Equation (\mainref{eq:DiffMass}{2.11}$a$) shows that the change in mass of gas within the system is due to the influx from the source; (\mainref{eq:DiffMass}{2.11}$b$) shows that the change in mass of the liquid is equal to the negative of the flux at the outlet. Gas compressibility allows $\rho^*_g$ to vary in space and time; however in the incompressible limit, when $\rho^*_g$ is effectively constant, the fixed domain volume ensures that 
\begin{equation}
    \frac{\mathrm{d}}{\mathrm{d}t^*} \left(
    \frac{M^*_g}{\rho^*_g}+\frac{M^*_w}{\rho_w}\right)=0,
    \label{eq:bmass}
\end{equation}
relating the source strength to the outflow directly via (\mainref{eq:DiffMass}{2.11}).  In general, however, (\ref{eq:bmass}) will not hold while gas is compressed transiently in regions I and II$_g$.  

Input parameters are summarized in table~\ref{tab:kd}.  We seek outputs such as the breakthrough time $t^*_b$ (at which $X^*_u(t^*_b)=L$) and the total mass of gas delivered at this time $M_g^*(t_b^*)$.  {(The model applies only for $0<t^*\leq t_b^*$.)}  We now present an asymptotic reduction of (\ref{eq:mc}) to {a coupled system of evolution equations} ((\ref{eq:put2}$a$,$b$), below) for the interface {position} $F^*$ and the gas pressure $p^*_g$, beginning with a scaling argument.
\begin{table}
  \begin{center}
  \begin{tabular}{llll}
      Parameter  & Value/Range & Units & Source \\[3pt]
      
        Viscosity of hydrogen $\mu_g$ & $ 10^{-6} $ & \si{kg.m^{-1}.s^{-1}} & \cite{zivar_underground_2021}\\
        
        Viscosity of brine $\mu_w$ & $10^{-4}$ &\si{kg.m^{-1}.s^{-1}} & \cite{muhammed_review_2022} \\
        
        Density of hydrogen $\rho_{g0}$ & 10$^{-1}$ -- 10 & \si{kg.m^{-3}}& \cite{muhammed_review_2022} \\
        
        Density of reservoir brine ${\rho}_w$ & 10$^{3}$ & \si{kg.m^{-3}} & \cite{muhammed_review_2022} \\
        
        Pressure of hydrogen $p_{g0}$ & 10$^{-1}$ -- 10 & MPa & \cite{muhammed_review_2022}\\
        
        Typical aquifer height $H$ & {10 -- 50} &\si{m} & \cite{muhammed_review_2022} \\
        
        Permeability of reservoir $k_0$ & {10$^{-14}$ -- 10$^{-12}$} & \si{m^2} & \cite{okoroafor_toward_2022}\\

        Frequency of injection $\omega $ & $10^{-8}$ -- $10^{-7}$ & \si{s^{-1}} & 
        \cite{tarkowski_underground_2019} \\

        Length of channel $L$ &{10$^4$ -- 10$^5$} & \si{m}  & \cite{jenkins_impact_2019}\\
        
         Speed of sound in hydrogen $c$ & 1.5 $\times$ $10^{3}$ &\si{m.s^{-1}} &  \cite{NIST}\\

        Injection rate $q$ & 10$^{-2}$ -- 10$^{-1}$& \si{kg.m^{-1}s^{-1}} & Estimated \\
        
        Length of initial gas plume $L_{g0}$ & $10^2$ -- $10^3$ & \si{m} & Estimated\\
        
        Length of initial interface $\Delta_{g0}$ &$10^{-1}$ -- $10$ & \si{m} & Estimated \\
        
  \end{tabular}
  \caption{ 
  Parameters used in the governing equations, their approximate values and corresponding sources.}
  \label{tab:kd}
  \end{center}
\end{table}

\subsection{Scaling argument}
\label{sec:scaling}

The gas mass flux per unit {width} from the source $q$ generates a volume flux per unit length of order $q/\rho_{g0}$ and a horizontal velocity of order $q/(\rho_{g0} H)$.  The corresponding transit time along the channel is of order $\rho_{g0} H L/q$.  Assuming for the time being that $\mu_g$ and $\mu_w$ are comparable, the viscous pressure drop along the channel is of order $\mu_g q L/(k_0 \rho_{g0} H)$. Vertical hydrostatic pressure variations in each phase are of order $\rho_{g0} g H$ and $\rho_w gH$, with the difference $\Delta \rho g H$ 
% (where $\Delta \rho\equiv \rho_w-\rho_{g0}$)  
providing the buoyancy force that seeks to flatten the gas-liquid interface.  In the long-wave limit, vertical velocities are expected to be a factor $H/L_{g0}$ smaller than the horizontal velocity, allowing the dominant contributions to the gas and liquid pressure fields to be hydrostatic plus a field that varies with ${x}^*$ and ${t}^*$ to leading order, independent of ${y}^*$. Assuming $\rho_{g0} \ll \rho_w\approx \Delta \rho$, the hydrostatic field in the liquid dominates that in the gas. 

Suppose that the buoyant and compressible pressure variations are comparable to $p_{g0}$, but that the viscous pressure scale is larger, so that 
\begin{equation}
\frac{\mu_g}{\mu_w} {\,=\,} O(1),\quad \frac{q \mu_g}{k_0 p_{g0} \rho_{g0}}\frac{L}{H}\gg 1, \quad \frac{\rho_{g0} g H}{p_{g0}}\ll \frac{\Delta \rho g H}{p_{g0}} {\,=
\,} O(1), \quad \frac{\rho_{g0} c^2}{p_{g0}} \gtrsim 1.
\label{eq:parsc}
\end{equation}
We assume that the equation of state (\ref{eqState1}) can be approximated by the truncated Taylor series $\mathcal{P}(1+\vartheta)\approx 1+\vartheta\mathcal{P}'(1)+\tfrac{1}{2}\vartheta^2\mathcal{P}''(1)$, with $\mathcal{P}'(1)$ and $\mathcal{P}''(1)$ both being of order unity. (Figure~\ref{fig:EqState}(b) shows that, for hydrogen, we can assume $\mathcal{P}'(1)\approx 1$ and $\mathcal{P}''(1)\ll 1$).  Thus relative density variations {in (\ref{eqState1})}, $\rho^*_g/\rho_{g0}=1+\vartheta$, generate gas pressure variations of order $p_{g0}$.  For $\vartheta \ll 1$, for example, (\ref{eqState1}) {can be approximated by the linear relationship}
\begin{equation}
\label{eq:losX}
{p^*_g\approx p_{g0}+c^2(\rho^*_g-\rho_{g0})}.    
\end{equation}
We then use (\ref{eq:losX}) to write the transport equation for gas density (\ref{eq:mc}) as an evolution equation for gas pressure.  Anticipating no vertical gradients in $p^*_g$ to leading order (the hydrostatic component being subdominant), {so that $p_g^*=p_g^*(x^*,t^*)$,} the transported quantity {$\int \rho_g^* \,\mathrm{d}y^*$} in (\ref{eq:mc}$a$) is proportional to $(H-F^*)(\rho_{g0} + (p^*_g-p_{g0})/c^2)$.  The time derivative {(in (\ref{eq:mc}$a$))} then includes $-\rho_{g0}F^*_{t^*}$ and $Hp^*_{g,t^*}/c^2$. 

Eliminating $p_{g0}$ from {ratios in} the second, third and fourth terms of (\ref{eq:parsc}) identifies the dimensionless parameters
\begin{equation}
\mathcal{M}=\frac{\mu_g}{\mu_w}\lesssim 1, \quad \mathcal{L} \equiv \frac{q \mu_g L}{k_0 \rho_{g0} \Delta \rho g H^2}\gg 1,\quad 
\zeta\equiv \frac{\Delta \rho g H}{\rho_{g0} c^2} \lesssim 1, \quad 
\beta=\frac{p_{g0}}{\rho_{g0} c^2}\equiv\frac{1}{\mathcal{P}'(1)}\lesssim 1.
\label{Eq:DimParams}
\end{equation}
We take {the viscosity ratio} $\mathcal{M}$ to be of order unity for now, specialising later to the limit $\mathcal{M}\ll 1$.  Buoyancy-driven flattening of the interface takes place over a lengthscale $L_b$ for which the pressure gradient $\Delta \rho g H/L_b$ generates a horizontal velocity $k_0 \Delta \rho g H/(\mu_g L_b)$ that we assume balances the source-driven velocity $q/(\rho_{g0} H)$.  It follows that 
\begin{equation}
    L_b=L/\mathcal{L} \ll L.
\end{equation}  
{Thus we can interpret $\mathcal{L}$ as a measure of the channel length relative to the lengthscale over which buoyancy forces flatten the interface.} The corresponding timescale {for buoyancy-driven flow} is $T_b=L_b \rho_{g0} H/q$, which is smaller than the transit time by a factor $1/\mathcal{L} \ll 1$.  Taking the pressure scale $\Delta \rho g H$, the two time derivatives in the gas transport equation ($-\rho_{g0}F^*_{t^*}$ and $Hp^*_{g,t^*}/c^2$) differ in magnitude by $\zeta$, {a parameter that measures hydrostatic pressure changes relative to those associated with gas compressibility.} Thus for $\zeta \lesssim O(1)$ with $\mathcal{M}=O(1)$, we expect compressible effects to operate on shorter timescales than source- and buoyancy-driven spreading, which in turn may take a long time to reach the end of the channel. The parameter $\beta$ in (\ref{Eq:DimParams}) measures the global convexity of the equation of state at the baseline state: as illustrated in figure~\ref{fig:EqState}(b) for hydrogen, $\beta$ is slightly below unity because the equation of state is almost linear but with small positive curvature.

To model these multiple physical effects, we therefore scale $x^*$ by $L_b$, $y^*$ and $F^*$ by $H$, $t^*$ by $T_b$, $p^*_g$ by $\Delta \rho g H$ and $\rho^*_g$ by $\rho_{g0}$.  The dimensionless domain length is then $\mathcal{L}$.  Assuming $\mathcal{L}\gg 1$ allows us to track the evolution of the interface over long distances and long times. At the early stages of spreading, when compressible effects are likely most dominant, the dimensionless viscously-generated gas pressure is of order $\mathcal{L}$, which yields dimensionless density variations ($\vartheta$) of order $\zeta \mathcal{L}$. To capture compressible effects, we seek to accommodate the distinguished limit
\begin{equation}
\zeta\rightarrow 0, \quad \mathcal{L}\rightarrow \infty \quad \mathrm{with}\quad \theta\equiv~ 
\zeta \mathcal{L} \equiv \frac{q \mu_g L}{k_0 \rho_{g0}^2 H c^2} {\,=\,} O(1),\label{eq:dlim}
\end{equation}
where $\theta$ measures the relative magnitude of viscously-generated pressure compared with that due to compressibility; this is analogous to the compressibility number defined in \citet{cuttle_compression-driven_2023}. Here the long domain length ($\mathcal{L}\gg 1$) generates large pressure deviations that cause appreciable density variations, even in a weakly compressible gas ($\zeta \ll 1$). 

Shortly, we will extend the model to incorporate the limit $\mathcal{M}\ll 1$, representing the motion of a gas into a liquid.  In this case, a thin film of gas can spread rapidly over the liquid, reducing the time taken for the gas to travel to the channel outlet.  We will show how compressible effects can remain dominant throughout this spreading process.  For now, we proceed assuming $\mathcal{M}=O(1)$.

\subsection{Model equations}
\label{Sec:Model}
\subsubsection{\color{black}{Nondimensionalization}}
Adopting the proposed scalings ($x^*=L_bx$, $y^*=Hy$, $t^*=T_bt$, $\rho^*_g=\rho_{g0} \rho_{g}$, $p^*_g=(\Delta \rho g H) p_g$, $p^*_w=\Delta \rho g H(p_w-y)$, $V^*=HL_bV$, $M_g^*=\rho_{g0}HL_b M_g$), $(u_g^*,v_g^*)=(q/H\rho_{g0})(u_g,\epsilon v_g)$, $(u_w^*,v_w^*)=(q/H\rho_{g0})(u_w,\epsilon v_w)$ with $\epsilon\equiv H/L_b\ll 1$, the Darcy equations (\ref{Darcy1}) become
\begin{subequations}
    \begin{align}
    u_g&=-p_{g,x}, & u_w&=-\mathcal{M} p_{w,x}, \\
\epsilon^2 v_g&=-p_{g,y}+(\rho_{g0}/\Delta \rho), & \epsilon^2 v_w&=\mathcal{M}\left[-p_{w,y}+(\rho_{g0}/\Delta\rho)\right].
\end{align}
\end{subequations}
Thus assuming $\rho_{g0}\ll \rho_w$, ensuring $\rho_{g0}\ll \Delta\rho$, $p_g$ and $p_w$ are independent of $y$ at leading order.  We also assume for simplicity that $\beta \mathcal{P}''(1) \ll 1$: with a linearised equation of state (\ref{eq:losX}), the gas density $\rho_g$ is represented {in terms of gas pressure} by $1-\beta+\zeta p_g$. {With the flow configured as in figure~\ref{fig:Diagram},} we {then} recover from (\ref{ContinuityOfPressure1}, \ref{OutletBC}, \ref{eq:mc}) 
% [Appendix~\ref{app:a}]
\begin{subequations}
\label{eq:put1}
\begin{align}
\left[(1-F)(1-\beta+\zeta p_g )\right]_t &= \left[ (1-F)(1-\beta+\zeta p_g)p_{g,x}\right]_x & (0<x<X_u),\\
F_t &= \mathcal{M} \left[F p_{w,x}
%(p_{g,x}+F_x)
\right]_x, & (X_l<x<X_u),\\
 0 &= \mathcal{M} \left[p_{w,x}\right]_x, & (X_u<x<\mathcal{L}),\\
- (1-\beta+\zeta p_g)p_{g,x}&=\mathcal{Q}, & (x=0), \\
p_w&=p_g+F, & (X_l<x\leq X_u),\\
(p_{g,x}+F_x)\big\vert_{X_u-}&=p_{w,x}\big\vert_{X_u+}, \\
p_w&=\beta/\zeta, & (x=\mathcal{L}, y=0).
\end{align}
\end{subequations}
The problem is governed by two coupled evolution equations (\ref{eq:put1}$a$,$b$), unlike the incompressible problem in which a single evolution equation describes the dynamics.  These equations describe respectively conservation of mass in the gas (of thickness $1-F$) and of the liquid beneath it (of thickness $F$).  Equation~(\ref{eq:put1}$c$) is a statement of mass conservation where the liquid fully fills the channel; (\ref{eq:put1}$d$) balances the mass flux of gas near the inlet (assuming $F(0,t)=0$) with the imposed source; (\ref{eq:put1}$e$) ensures continuity of pressure across the gas-liquid interface; (\ref{eq:put1}$f$) ensures continuity of liquid flux across the upper contact line; and (\ref{eq:put1}$g$) imposes the hydrostatic pressure constraint at the channel outlet.
Integrating (\ref{eq:put1}$c$) to find $p_w$, and applying (\ref{eq:put1}$e$,$f$) at $x=X_u$ and (\ref{eq:put1}$g$), gives
\begin{equation} 
p_g+ (p_{g,x}+F_x) (\mathcal{L} - X_u)=({\beta}/{\zeta}) -1,\quad\quad\quad(x=X_u-).
    \label{eq:pgu}
    \end{equation}
{Referring to Fig.~\ref{fig:Diagram}, two boundary conditions are required in Region I (for gas alone), two boundary conditions are required in Region III (for liquid alone) and four are required for Region II (for gas and liquid).  Thus we have a single inlet condition (\ref{eq:put1}$d$), a single outlet condition (\ref{eq:put1}$g$) plus continuity and kinematic conditions at internal free boundaries.  Specifically,} combining (\ref{eq:put1}$a$,$b$) with the constraints $F(X_l,t)=0$, $F(X_u,t)=1$ requires that
\begin{align}
\label{eq:xint}
    X_{l,t}&=-\mathcal{M} (p_{g,x}+F_x)\big\vert_{X_l+}, &
    X_{u,t}&=- p_{g,x}\big\vert_{X_u-}.
\end{align}

We define the interface length as $\Delta(t) \equiv X_u(t) - X_l(t)$, with initial value $\Delta(0)=\Delta_0~\equiv {\Delta}_{g0}/L_b$. The initial condition (\ref{eq:linic}) becomes
\begin{equation}\label{eq:ini}
    F(x,0)%=F_0(x)
    \equiv \begin{cases}0, & 0\leq x<X_l(0)\equiv L_0-\tfrac{1}{2}\Delta_0, \\
    (x-X_l(0))/\Delta_0, & X_l(0)\leq x<X_u(0), \\ 1, & L_0+\tfrac{1}{2}\Delta_0\equiv X_u(0)\leq x\leq \mathcal{L},\end{cases}
\end{equation}
where $L_0\equiv L_{g0}/L_b$.  An initial condition is also required for $p_g$; we assume for now that it is large enough in magnitude for the gas volume to increase.  The volume of gas in the channel (per unit width) is given by the integral
\begin{equation}\label{eq:V}
    V(t) = \int_0^{X_u(t)} (1 - F) \, \mathrm{d} x;
\end{equation}
thus (\ref{eq:ini}) defines the initial gas volume $V(0)=\tfrac{1}{2} [X_u(0) + X_l(0)]=L_0$.
The mass of gas (\mainref{eq:DiffMass}{2.11}a) evolves according to 
\begin{equation}
\frac{\mathrm{d}M_g}{\mathrm{d}t}=\mathcal{Q}(\Omega t) \quad\mathrm{where}\quad \Omega=\omega T_b.
\label{eq:massg}
\end{equation}

{The full dimensionless problem 
%(\ref{eq:put1})--(\ref{eq:massg}) 
is governed by {seven} dimensionless parameters ($\beta$, $\Omega$, $L_0$, $\Delta_0$, $\mathcal{L}$, $\mathcal{M}$, $\zeta$){\color{black}, which are summarized in table~\ref{tab:Dimensionless}}.  We will show shortly how the convexity parameter $\beta$ can be eliminated, and will focus attention primarily on steady injection with $\mathcal{Q}=1$, although we retain $\mathcal{Q}$ in the following in order to comment later on time-varying injection rates parametrized by $\Omega$.  In many instances we expect details of the initial conditions $L_0$, $\Delta_0$ to be unimportant at large times.  This leaves the domain length $\mathcal{L}$, viscosity ratio $\mathcal{M}$ and compressibility $\zeta$ as parameters of primary interest.}

\begin{table}
  \centering
  \color{black}
  \begin{tabular}{l@{\hspace{0.5cm}}l@{\hspace{0.5cm}}}
      Parameter  & Description\\[4pt]
      $\beta  = p_{g0}/(\rho_{g0} c^2)$& Convexity of equation of state \\[1pt]
      
      $\Omega = \omega T_b$ & Buoyancy vs injection time scales \\[1pt]
      $L_0 = L_{g0}/L_b$ & Initial length of gas \\[1pt]
      
      $\Delta_0 = \Delta_{g0}/L_b$ & Initial length of interface\\[1pt]
      
      $\mathcal{L} = q \mu_g L/(k_0 \rho_{g0} \Delta \rho g H^2)$ & Channel vs buoyancy length scales \\[1pt]
      
      $\mathcal{M} = \mu_g/\mu_w$ & Viscosity ratio \\[1pt]
      
       $\mathcal{\zeta} = \Delta \rho g H/(\rho_{g0} c^2)$ & Hydrostatic vs compressive pressure \\
      \end{tabular}
  \caption{{\color{black}A summary of the seven dimensionless parameters in the governing equations (\ref{eq:put1}).}
  }
  \label{tab:Dimensionless}
\end{table}
\subsubsection{\color{black}{Incompressible limit}}
The incompressible limit of (\ref{eq:put1}) with $\mathcal{L}=O(1)$ is recovered by taking $\zeta\rightarrow 0$, $\beta\rightarrow 0$ with $\beta/\zeta=p_{g0}/(\Delta \rho g H)  \,{=\,} O(1)$. {Using (\ref{eq:put1}$e$) to eliminate $p_w$, the two evolution equations (\ref{eq:put1}$a$,$b$) become} 
\begin{subequations}\label{eq:put1inc}
\begin{align}
-F_t &= \left[ (1-F)p_{g,x}\right]_x, & (0<x<X_u),\\
F_t &= \mathcal{M} \left[F (p_{g,x}+F_x)\right]_x, & (X_l<x<X_u),
\end{align}
\end{subequations}
which are subject to the boundary conditions (\ref{eq:put1}$d$), (\ref{eq:pgu}), (\ref{eq:xint}), from which the flux constraint 
\begin{align}
    (1-F)p_{g,x}+\mathcal{M}F(p_{g,x}+F_x)&=-{\mathcal{Q}}, & (0<x<X_u),
    \label{eq:incflux}
\end{align}
emerges.   
Elimination of $p_{g,x}$ between (\ref{eq:put1inc}$a$, \ref{eq:incflux}) yields a single evolution equation for $F$,
\begin{align}
    F_t&=\mathcal{M}\left[\frac{F(1-F)F_x-F\mathcal{Q}}{1-F+\mathcal{M}F}\right]_x, & (X_l<x<X_u),
    \label{eq:incF}
\end{align}
a system addressed by \cite{pegler_fluid_2014} and others (with $\mathcal{Q}=1$). The term proportional to $F_x$ captures the role of buoyancy; the term proportional to $\mathcal{Q}$ represents advection driven by injection.  The source strength $\mathcal{Q}$ appears only in a boundary condition in the compressible problem (\ref{eq:put1}), but incompressibility enables its presence to be felt throughout the domain in (\ref{eq:incF}).  Assuming the (diffusive) buoyancy term is subdominant to (advective) viscous effects, so that $F_t+\mathcal{M}\mathcal{Q} (1-F+\mathcal{M}F)^{-2}F_x=0$, a solution using characteristics shows how the source drives both contact lines directly:
\begin{subequations}
\begin{equation}
    X_l(t)= X_l(0) + \mathcal{M} \int_0^t \mathcal{Q}\,\mathrm{d}t, \quad X_u(t)= X_u(0) + \frac{1}{\mathcal{M}}\int_0^t\mathcal{Q}\,\mathrm{d}t.
    \label{eq:fullcharsol}
\end{equation}
For $\mathcal{Q}=1$ this becomes, in the large-time limit \citep{pegler_fluid_2014},
\begin{equation}
    X_l(t)\approx % X_l(0) + 
    {\mathcal{M} t}, \quad X_u(t)\approx % X_u(0) + 
    {t}/{\mathcal{M}}.
    \label{eq:clsp}
\end{equation}
\end{subequations}
In this formulation, $t=\mathcal{L}$ corresponds to the transit time of source-driven flow. The breakthrough time of $X_u$ at the outlet predicted by (\ref{eq:clsp}), $t_b=\mathcal{M}\mathcal{L}$, lies below the transit time for $\mathcal{M}<1$, because a thin film of gas spreads rapidly over the top of the liquid. 

\subsubsection{\color{black}{Compact formulation of the model}}
A more compact formulation of (\ref{eq:put1}), that is independent of $\beta$, is obtained by writing {the gas density as} $\zeta P=1-\beta+\zeta p_g$ (or $p^*_g=p_{g0}-\rho_{g0} c^2+(\Delta \rho g H) P$, so that the equation of state (\ref{eq:losX}) reduces to $\zeta P =\rho_{g}$), and (\ref{eq:put1}) simplifies to 
\begin{subequations}
\label{eq:put2}
\begin{align}
\left[(1-F)P\right]_t &= \left[ (1-F) P P_{x}\right]_x, & (0<x<X_u),\\
F_t &= \mathcal{M} \left[F(P_{x}+F_x)\right]_x, & (X_l<x<X_u),\\
- \zeta P P_x&=\mathcal{Q}, & (x=0),\\
P+(P_{x}+F_x) (\mathcal{L} - X_u) &=(1/\zeta)-1,  &(x=X_u),\\
    X_{l,t}&=-\mathcal{M}  (P_{x}+F_x), & (x={X_l+}), \\
    X_{u,t}&=- P_{x},& (x={X_u-}).
\end{align}
\end{subequations}
Taking $\mathcal{Q}(0) = 1$, the initial pressure field is chosen to be
\begin{equation}\label{eq:PIC}
        P(x,0) = P_c - \frac{1}{ \zeta P_c} x, \qquad (0 < x < X_u),
\end{equation}
where the constant $P_c$ satisfies the quadratic equation
\begin{equation}
\label{eq:pquad}
    P_c^2 + P_c \left( \frac{\mathcal{L} - X_u(0)}{\Delta_0} 
    + 1 - \frac{1}{\zeta} \right) - \frac{\mathcal{L}}{\zeta }  =0,
\end{equation}
where $X_u(0)=L_0+\tfrac{1}{2}\Delta_0$.  Equation (\ref{eq:pquad}) results from substituting (\ref{eq:PIC}) into (\ref{eq:put2}$d$) so that the initial pressure field is consistent with the boundary conditions. The initial interface slope $F_x(x,0) = 1/\Delta_0$ in (\ref{eq:ini}) must be consistent with the inequality $P_x \equiv -1/\zeta P_c < (1/\zeta - 1)/(\mathcal{L} - X_u(0)) - 1/\Delta_0$ to ensure compatibility with the boundary condition (\ref{eq:put2}$d$). Otherwise, the initial condition (\ref{eq:PIC}) would lead to negative densities ($P<0$), rendering (\ref{eq:put2}$a$) ill-posed. 

Having eliminated $p_w$ and $\beta$ from the model, we can interpret (\ref{eq:put2}$a$) as mass conservation for gas, (\ref{eq:put2}$b$) as mass conservation for liquid, (\ref{eq:put2}$c$) as the mass flux driving the flow, (\ref{eq:put2}$d$) as a pressure condition at the upper contact line (capturing the viscous resistance of the liquid-filled region of the aquifer) and (\ref{eq:put2}$e$,$f$) as kinematic conditions for the contact-line locations where $F=0$ and $F=1$ respectively.

If, at any point during the flow, the slope of the interface at the lower contact line exceeds the magnitude of the pressure gradient at that location, buoyancy will cause the lower contact line to move backwards through (\ref{eq:put2}$e$). As a result, the lower contact line may reach the boundary at $x = 0$ before the upper contact line reaches the outlet. To accommodate this, once $X_l = 0$ we replace the boundary condition (\ref{eq:put2}$c$) with
\begin{equation}\label{eq:OriginBC}
    \zeta (1-F) P P_x = -\mathcal{Q}, \qquad F_x = -P_x, \qquad (x = 0).
\end{equation}
This modification sets the liquid flux at the origin to zero and reduces the effective area through which the gas flux enters the channel by a factor of $1 - F$.

It can be verified from (\ref{eq:put2}) that the mass of gas 
\begin{equation}
\label{eq:InMas}
M_g(t)=\zeta\int_0^{X_u}(1-F)P\,\mathrm{d}x,
\end{equation}
satisfies (\ref{eq:massg}).  In particular, when $\mathcal{Q}=1$, this implies that $M_g(t_b)=M_g(0)+t_b$, \hbox{i.e.} the breakthrough time $t_b$ reveals the total mass of gas delivered by the source.

In (\ref{eq:put2}), the incompressible limit (\ref{eq:put1inc}, \ref{eq:incflux}) with $\mathcal{L}=O(1)$ and small $\zeta$ is recovered using the expansion $P=(1/\zeta)+\hat{P}+\dots$ for $\zeta \ll 1$, with the large mean pressure arising from the $1/\zeta$ term in (\ref{eq:put2}$d$).

\begin{figure}
\centering
\includegraphics[width = \linewidth]{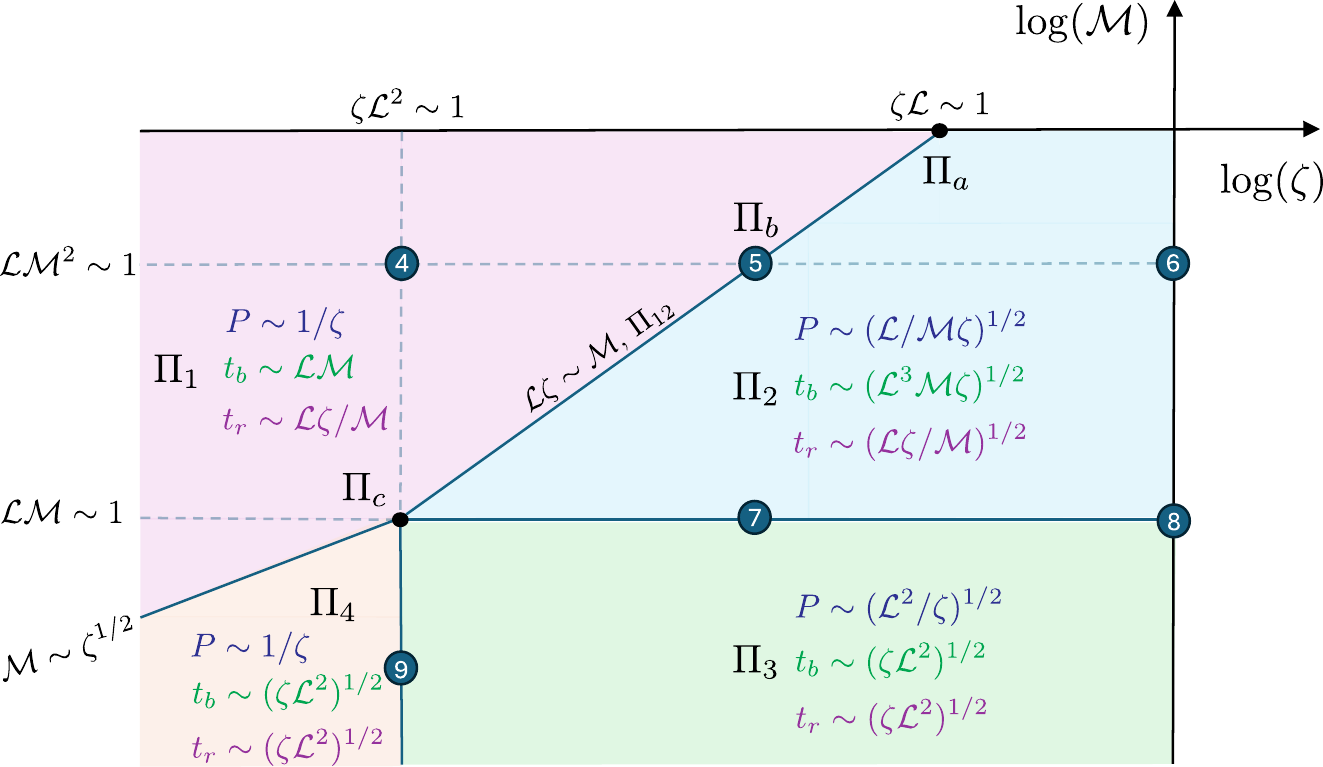}
\caption{A schematic map of $(\zeta, \mathcal{M})$-parameter space.  The full problem (\ref{eq:put2}) is derived for $\mathcal{M}\sim \mathcal{L} \sim \zeta\sim 1$.  The map specialises to the case $\mathcal{L}\gg 1$, focusing on high gas mobility and weak compressibility ($\mathcal{M}\ll 1$, $\zeta \ll 1$); features of the model in the shaded regions in the map become increasingly distinct as $\mathcal{L}\rightarrow \infty$.  The distinguished limit $\mathcal{M}\sim \mathcal{L} \zeta \sim 1$ (problem $\Pi_a$) is given by (\ref{eq:put2t}) below.  An inner/outer structure given by (\ref{eq:put2ti}, \ref{eq:put2to}) emerges from this system along $\mathcal{L}^{-1}\ll\mathcal{L} \zeta\sim \mathcal{M} \ll 1$ (problem $\Pi_{12}$).  The outer problem simplifies to (\ref{eq:put2toa}) for $\max(\mathcal{L} \zeta,\zeta^{1/2})\ll \mathcal{M} \ll 1$ (shaded pink, region $\Pi_1$) and (\ref{eq:put2toc}) for $\mathcal{L}^{-1}\ll \mathcal{M}\ll \min(1,\mathcal{L} \zeta)$ (shaded blue, region $\Pi_2$). Buoyancy effects influence the inner region for $\mathcal{L} \mathcal{M}^2\lesssim 1$, allowing $X_l$ to recede.  Problem $\Pi_b$ arises at $\mathcal{L}\mathcal{M}^2\sim 1$ and $\mathcal{L}\zeta\sim\mathcal{M}$. Ultra-low viscosity effects emerge via (\ref{eq:put21}) (problem $\Pi_c$) at $\mathcal{M}\sim \mathcal{L}^{-1}$ and $\zeta\sim \mathcal{L}^{-2}$, from which emerge sub-limits shown in green (region $\Pi_3$; $\mathcal{LM}\ll 1$, $\mathcal{L}^{-2}\ll\zeta\ll 1$) and orange (region $\Pi_4$; $\mathcal{M}\ll \zeta^{1/2}$, $\mathcal{L}^{2}\zeta\ll 1$). Scales for the pressure at the origin $P$, breakthrough time $t_b$ and pressure rise time $t_r$ are shown in blue,  green and magenta respectively in the coloured regions; $t_r\ll t_b$ above the line $\mathcal{L} \mathcal{M} \sim 1$ for $\mathcal{L}\zeta \gtrsim 1$.  Relative locations within parameter space of results shown in figures \ref{fig:M_0.1_zeta_10-4}--\ref{fig:M_10-3_zeta_10-4} below are indicated with symbols.}
    \label{fig:overview}
\end{figure}

\subsubsection{\color{black} Summary of model formulation}
To summarise, we have taken a long-wave limit to derive evolution equations (\ref{eq:put1})--(\ref{eq:xint}) for the interface shape and gas pressure, generalising an existing model to incorporate gas compressibility.  This system has been reformulated as (\ref{eq:put2}), the core model of interest, which shows how the transport involves competing nonlinear diffusion (with fluxes proportional to $(1-F)PP_x$ and $FF_x$) and advection ($FP_x$).  Numerical solutions are given in \S~\ref{sec:results} below, obtained using methods described in Appendix~\ref{app:nummeth}.  The system (\ref{eq:put2}) is parametrized by $\mathcal{M}$, $\zeta$, $\mathcal{L}$, $L_0$ and $\Delta_0$ (plus the parameter $\Omega\equiv \omega T_b$ relating to the source variation $\mathcal{Q}$). We assume henceforth that the flow domain is long ($\mathcal{L}\gg 1$).  For fixed initial condition parameters $\Delta_0$, $L_0$, we will henceforth focus on behaviour across $(\zeta,\mathcal{M})$-parameter space, in particular the quadrant $\zeta\lesssim 1$ and $\mathcal{M}\lesssim 1$ of relevance to weakly compressible gases.

The map in figure~\ref{fig:overview} will be helpful in navigating further reductions of this model. The boundaries of the map will be justified as we proceed; at present the reader is asked just to consider its overall organisation.  As is typical of multi-parameter problems, distinguished limits arise in which different combinations of physical effects determine the dynamics.  Figure~\ref{fig:overview} identifies codimension-0 regions (coloured areas) of parameter space, labelled $\Pi_1$--$\Pi_4$; these are separated by codimension-1 boundaries (lines we shall label $\Pi_{ij}$, bounding $\Pi_i$ and $\Pi_j$), which intersect codimension-2 points (labelled $\Pi_a$--$\Pi_c$).  Reduced forms of the model can be derived in each case, with the number of competing physical effects, and the number of relevant independent parameters, increasing with codimension. We will start in \S~\ref{sec:pia} at the codimension-2 point $\Pi_a$ (the limit $\mathcal{M}\sim 1$, $\zeta\sim 1/\mathcal{L}\ll 1$, parametrised by $\mathcal{M}$ and $\theta\equiv\zeta \mathcal{L}$, as in (\ref{eq:dlim})).  Problem $\Pi_a$ {encompasses as a special case} the codimension-1 problem $\Pi_{12}$, which we address in \S~\ref{sec:pi12} (the specialised limit $\mathcal{M}\sim\theta\ll 1$, parametrized by $\mathcal{M}/\theta$).  {Contained in} problem $\Pi_{12}$ are sub-limits $\Pi_1$ ($\mathcal{M}/\theta \gg 1$) and $\Pi_2$ ($\mathcal{M}/\theta\ll 1$), represented respectively by the pink and blue regions of parameter space in figure~\ref{fig:overview}. We will also highlight the {codimension-2 problems} $\Pi_b$ ($\mathcal{M}\sim \mathcal{L}^{-1/2}$, $\zeta\sim \mathcal{L}^{-3/2}$) to assess the role of buoyancy (Appendix~\ref{app:buoyancy}) and $\Pi_c$ ($\mathcal{M}\sim \mathcal{L}^{-1}$, $\zeta\sim \mathcal{L}^{-2}$) to assess flows at ultra-low viscosity ratio (Appendix~\ref{app:verylowviscosity}). Within problem $\Pi_c$ we identify additional codimension-1 problems $\Pi_{23}$, $\Pi_{34}$ and $\Pi_{14}$ that bound two remaining {codimension-0} regions of parameter space ($\Pi_3$ and $\Pi_4$, orange and green respectively in figure~\ref{fig:overview}).  Many of the sub-problems will also involve consideration of distinct asymptotic regions in time (distinguishing early from late spreading) and space (near-source and far field). {\color{black}We will see that buoyancy effects (represented by the term $F_x$ in (\ref{eq:put2}b)) are often subdominant to pressure gradients, and we will identify conditions under which spreading is driven directly by the volume flux $\mathcal{Q}$ in (\ref{eq:put2}c), or indirectly by the evolving pressure field near the source.} We now address the different physical balances within each limit; for later reference, these are summarised in table~\ref{tab:regionsummary} below.

\subsection{Problem $\Pi_a$: weak compressibility}
\label{sec:pia}
We first consider the distinguished limit (\ref{eq:dlim}) with $\mathcal{M}=O(1)$, large $\mathcal{L}$, small $\zeta$ and $\theta\equiv \zeta\mathcal{L}=O(1)$. We distinguish the initial phase of spreading, when $X_u=O(1)$, from the later phase when $X_u$ is comparable to $\mathcal{L}\gg 1$.  

\subsubsection{Early times}
\label{sec:EarlyTimes}
At early times, we pose the expansion $P=P_0(t)/\zeta+P_1(x,t)+\dots$, $F=F_0{(x,t)}+O(1/\zeta)$, in which case (\ref{eq:put2}) yields, at leading order,
\begin{subequations}
\label{eq:put3}
\begin{align}
P_0^{-1} \left[(1-F_0)P_0\right]_t &= \left[ (1-F_0)  P_{1x}\right]_x, & (0<x<X_u),\\
F_{0t} &= \mathcal{M} \left[F_0(P_{1x}+F_{0x})\right]_x, & (X_l<x<X_u),\\
- P_0 P_{1x}&=\mathcal{Q}, & (x=0),\\
P_0+(P_{1x}+F_{0x}) \theta &=1,  &(x=X_u), \\
    X_{l,t}&=-\mathcal{M}  (P_{1x}+F_{0x}), & (x={X_l+}), \\
    X_{u,t}&=- P_{1x}, & (x={X_u-}).
\end{align}
Functions in expansions will be assumed to remain $O(1)$ in the relevant limit; here, $P_0$, $P_1$ and $F_0$ are assumed to be $O(1)$ as $\mathcal{L}\rightarrow \infty$ with $\theta=O(1)$ and $\mathcal{M}=O(1)$.
Initial conditions for $P_0$ and $P_1$ are obtained by expanding the initial condition for the full pressure field \eqref{eq:PIC}, yielding
\begin{align}
    P_0(0) &= \frac{1}{2} - \frac{1}{{2 \Delta_0} }\left[ \theta - \left( (1+4\theta) \Delta_0^2 + \theta^2 - 2 \theta \Delta_0\right)^{1/2} \right].
    \label{eq:IC2}
\end{align}
\end{subequations}
The formulation (\ref{eq:put3}) couples compressible effects, captured by the unsteady but spatially-uniform pressure component $P_0$, with buoyancy and spreading effects captured by gradients of the smaller pressure component $P_{1}$. 

To interpret (\ref{eq:put3}), it is helpful to integrate (\ref{eq:put3}$a$) across regions I and II$_g$ of figure~\ref{fig:Diagram}, using boundary conditions (\ref{eq:put3}$c$,$f$) to give the mass balance {for the gas phase} $[P_0 V(t)]_t=\mathcal{Q}$; the gas volume $V(t)$ is defined in (\ref{eq:V}). Integrating (\ref{eq:put3}$b$) across region II$_w$ and using (\ref{eq:put3}$d$,$e$) gives a mass flux balance for the incompressible liquid phase $V_t=\mathcal{M}(P_0-1)/\theta$.  Together, these yield 
\begin{equation}
    \frac{P_{0,t}}{P_0} V(t) = \frac{\mathcal{Q}}{P_0} - \frac{\mathcal{M} ( P_0 - 1)}{\theta}.
    \label{eq:Boyles}
\end{equation}
This is an analogue of the differentiated form of Boyle’s law, which for a closed system with a fixed mass of gas would be $(P_0 V)_t = 0$. In contrast, (\ref{eq:Boyles}) applies to an open system in which gas enters the channel; the first term on the right-hand side accounts for the volume flux of gas entering the channel, while the second term captures changes in gas volume due to liquid leaving the channel. Consequently, the gas compresses when the inflow exceeds the outflow.
Eq.~(\ref{eq:Boyles}) shows how $P_0$ can equilibrate to a constant value $P_\infty$, say, as spreading initiates (while $X_u$ remains small compared to $\mathcal{L}$), satisfying
\begin{equation}
    P_\infty-\frac{\mathcal{Q} \mathcal{L}\zeta}{P_\infty \mathcal{M} } =1
    \label{eq:pinf}
\end{equation}
(provided $\mathcal{Q}$ is sufficiently slowly varying). Setting $P_0=P_\infty$ in (\ref{eq:put3}) recovers the incompressible flux constraint (\ref{eq:incflux}) with $p_g$ replaced by $P_1$ and $\mathcal{Q}$ replaced by $\mathcal{Q}/P_\infty$.  As anticipated in \S\ref{sec:scaling}, (\ref{eq:Boyles}) shows how compressibility may lead to a rise in gas pressure prior to appreciable expansion of the gas domain. 

Although derived for $\mathcal{M}\sim 1$ and $\zeta \mathcal{L} \sim 1$, (\ref{eq:Boyles}) extends to the limit $\mathcal{M}\sim \zeta\mathcal{L} \ll 1$, allowing $P_0$ to equilibrate over a timescale $t_r {\,=\,} O(1) $ while $X_u$ remains small compared to $\mathcal{L}$. We note for later reference that when $\mathcal{Q}=1$ for $\mathcal{M}\ll \mathcal{L} \zeta \ll 1$ (region $\Pi_2$), (\ref{eq:Boyles}) predicts that $P$ rises to $(\mathcal{L} /\mathcal{M} \zeta )^{1/2}$ over a timescale $t_r\sim (\mathcal{L} \zeta/\mathcal{M})^{1/2}$; in contrast, for $ \mathcal{L} \zeta \ll \mathcal{M}\ll 1$ (region $\Pi_1$), (\ref{eq:Boyles}) predicts that $P$ rises to $1/\zeta$ over a timescale $t_r\sim \mathcal{L} \zeta /\mathcal{M}$.  Below, we will compare $t_r$ to the breakthrough time $t_b$ at which the upper contact line first reaches the outlet.

\subsubsection{Late times}

Staying in the distinguished limit (\ref{eq:dlim}) of large $\mathcal{L}$ and small $\zeta$ with $\mathcal{M}\sim 1$, we now turn to spreading at later times when $X_u$ becomes comparable to $\mathcal{L}$. Setting $x=\mathcal{L} \bar{x}$, $t=\mathcal{L} \bar{t}$, $P=\mathcal{L} \bar{P}$, $F(x,t)=\bar{F}(\bar{x},\bar{t})$, (\ref{eq:put2}) becomes, at leading order with $\theta\sim 1$ and $\mathcal{M}\sim 1$,
\begin{subequations}
\label{eq:put2t}
\begin{align}
\left[(1-\bar{F)}\bar{P}\right]_{\bar{t}} &= \left[ (1-\bar{F}) \bar{P} \bar{P}_{\bar{x}}\right]_{\bar{x}}, & (0<\bar{x}<\bar{X}_u),\\
\bar{F}_{\bar{t}} &= \mathcal{M} \left[ \bar{F}\bar{P}_{\bar{x}}\right]_{\bar{x}}, & (\bar{X}_l<\bar{x}<\bar{X}_u),\\
- \theta \bar{P} \bar{P}_{\bar{x}}&=\mathcal{Q}, & (\bar{x}=0),\\
\theta\left[ \bar{P}+\bar{P}_{\bar{x}} (1-\bar{X}_u)\right] &=1,  &(\bar{x}=\bar{X}_u),\\
    \bar{X}_{l,\bar{t}}&=-\mathcal{M} \bar{P}_{\bar{x}}, & (\bar{x}={\bar{X}_l+}), \\
    \bar{X}_{u,\bar{t}}&=- \bar{P}_{\bar{x}},& (\bar{x}={\bar{X}_u-}).
\end{align}
Over long spatial scales, buoyancy terms that would appear in (\ref{eq:put2t}$b${,$d$,$e$}), associated with tilting of the interface, are subdominant {(of order $1/\mathcal{L})$}, but may influence dynamics close to the lower contact line where the interface is likely to be steepest. {Otherwise (\ref{eq:put2t}) retains the viscous and compressible effects of the full model (\ref{eq:put2}) throughout the region occupied by gas.} Associated mass and volume constraints are
\begin{equation}
\left(\int_0^{\bar{X}_u} (1-\bar{F})\bar{P}\,\mathrm{d}\bar{x}\right)_{\bar{t}}=\frac{\mathcal{Q}}{\theta}, \quad
\left(\int_0^{\bar{X}_u} (1-\bar{F})\,\mathrm{d}\bar{x}\right)_{\bar{t}}=\frac{\mathcal{M}}{1-\bar{X}_u}\left[\bar{P}\big\vert_{\bar{X}_u}-\frac{1}{\theta}\right],
\end{equation}
\end{subequations}
implying that $\bar{P}(\bar{X}_u,\bar{t})\rightarrow 1/\theta$ as $\bar{X}_u\rightarrow 1$. We next seek a reduction of this system when $\mathcal{M}$ and $\theta$ are both small, in which (\ref{eq:put2t}) develops a spatial inner/outer structure.

\subsection{Problem $\Pi_{12}$: high gas mobility}
\label{sec:pi12}

We now refine the $\Pi_a$ problem in the limit in which the viscosity ratio becomes small ($\mathcal{L}^{-1}\ll \mathcal{M}\sim \zeta\mathcal{L}\ll 1$), focussing on the late-time structure (\ref{eq:put2t}) in which $X_u$ moves over distances comparable to the channel length. Low $\mathcal{M}$ can be expected to promote rapid spreading of a thin layer of gas above the liquid \citep{pegler_fluid_2014}; likewise, (\ref{eq:put2t}$e$) shows how motion of the lower contact line is suppressed.  In the distinguished limit $\theta\sim \mathcal{M} \ll 1$, we therefore consider motion over short times $\bar{t}=O(\mathcal{M})$, { during which a thin film extends from a shorter region in which the interface moves more slowly.  Accordingly,} we split the domain into an inner region over distances $\bar{x}=O(\mathcal{M}^2)$ for which $\bar{F}=O(1)$ (discussed in \S~\ref{sec:251}), and an outer region over distances $\bar{x}=O(1)$ for which $1-\bar{F}=O(\mathcal{M})$ (discussed in \S~\ref{sec:252}). The contact-line locations $\bar{X}_l$ and $\bar{X}_u$ lie in the inner and outer regions respectively. Compressibility turns out to have a sustained effect in the outer thin-film region.

\subsubsection{Inner region}
\label{sec:251}

In the inner region, we set $\bar{t}=\mathcal{M}\tilde{t}$, $\bar{x}=\mathcal{M}^2\tilde{x}$ and $\bar{F}=\tilde{F}(\tilde{x},\tilde{t})+O(\mathcal{M})$, $\bar{P}=\mathcal{M}^{-1}\tilde{P}_0(\tilde{t})+\mathcal{M}^2\tilde{P}_1(\tilde{x},\tilde{t})+\dots$. Then (\ref{eq:put2t}) becomes
\begin{subequations}
    \label{eq:put2ti}
\begin{align}
0 &= \left[ (1-\tilde{F}) \tilde{P}_0 \tilde{P}_{1\tilde{x}}\right]_{\tilde{x}}, & \\
\tilde{F}_{\tilde{t}} &=  \left[ \tilde{F}\tilde{P}_{1\tilde{x}}\right]_{\tilde{x}}, & \\
-  \tilde{P}_0 \tilde{P}_{1\tilde{x}}&=\mathcal{Q}(\mathcal{M}/\theta), & (\tilde{x}=0),\\
    \tilde{X}_{l,\tilde{t}}&=-\tilde{P}_{1\tilde{x}}, & (\tilde{x}={\tilde{X}_l+}).
\end{align}
This is an unsteady incompressible viscous flow problem involving a so-far undetermined leading-order gas pressure $\tilde{P}_0$.  The gas and liquid share the same pressure gradient $\tilde{P}_{1\tilde{x}}$ in (\ref{eq:put2ti}$a$,$b$); $\tilde{P}_0$ in (\ref{eq:put2ti}$a$,$c$), a proxy for gas density, converts volume fluxes in the gas into mass fluxes. It follows from (\ref{eq:put2ti}$a$,$c$) that $(1-\tilde{F}) \tilde{P}_0 \tilde{P}_{1\tilde{x}}=-\mathcal{Q}(\mathcal{M}/\theta)$ throughout the inner region (the gas mass flux is uniform in this incompressible limit) and that 
\begin{equation}
\tilde{F}_{\tilde{t}}+\frac{\mathcal{Q}}{\tilde{P}_0} \frac{\mathcal{M}}{\theta}\frac{\tilde{F}_{\tilde{x}}}{(1-\tilde{F})^2}=0.
\label{eq:innerhyp}
\end{equation}
The interface height $\tilde{F}$ can be integrated on characteristics.  There will be some transient adjustment of $\tilde{F}$ at early times via (\ref{eq:put3}); for simplicity, we assume this is modest and impose the initial condition (\ref{eq:linic}), giving
\begin{equation}
    \tilde{F}= 1-\left[\frac{\mathcal{M}}{\theta}\frac{1}{\tilde{x}-\xi} \int_0^{\tilde{t}} \frac{\mathcal{Q}}{\tilde{P}_0}\,\mathrm{d}\tilde{t}'\right]^{1/2}, \quad \tilde{F}=\frac{\xi-\tilde{X}_l(0)}{\tilde{X}_u(0)-\tilde{X}_l(0)}
\end{equation}
for $\tilde{X}_l(0)<\xi<\tilde{X}_u(0)$, which can be expressed as 
\begin{equation}
(1-\tilde{F})^2 \left[\tilde{x}-\{\tilde{X}_l(0)+(\tilde{X}_u(0)-\tilde{X}_l(0))\tilde{F}\}\right]=
\frac{\mathcal{M}}{\theta}\int_0^{\tilde{t}} \frac{\mathcal{Q}}{\tilde{P}_0}\,\mathrm{d}\tilde{t}
\label{eq:charsol}
\end{equation}
for $0<\tilde{F}<1$.
At the outer end of the inner region, $\tilde{F}\rightarrow 1$ with 
\begin{equation}
\label{eq:innerout}
    \tilde{F}\approx 1-\left[\frac{\mathcal{M}}{\theta}\frac{1}{\tilde{x}} \int_0^{\tilde{t}} \frac{\mathcal{Q}}{\tilde{P}_0}\,\mathrm{d}\tilde{t}'\right]^{1/2}.
\end{equation}
\end{subequations}
This profile describes spreading of incompressible gas, from a source with time-dependent volume flux proportional to $\mathcal{Q}/\tilde{P}_0$, into a thin gas film at the top of the aquifer. The inner solution depends on the unknown $\tilde{P}_0(\tilde{t})$, which must be determined by matching to the outer {(thin-film)} problem, but is independent of the precise initial condition.

\subsubsection{Outer region}
\label{sec:252}

In the outer region, we set $\bar{P}=\mathcal{M}^{-1}\hat{P}(\bar{x},\tilde{t})$ and $\bar{F}=1-\mathcal{M}\hat{F}(\bar{x},\tilde{t})$.  Then (\ref{eq:put2t}) becomes, to leading order in $\mathcal{M} \ll 1$,
\begin{subequations}
\label{eq:put2to}
\begin{align}
\left[\hat{F}\hat{P}\right]_{\tilde{t}} &= \left[\hat{F} \hat{P} \hat{P}_{\bar{x}}\right]_{\bar{x}}, & (0<\bar{x}<\tilde{X}_u),\\
-\hat{F}_{\tilde{t}} &= \hat{P}_{\bar{x}\bar{x}}, & (0<\bar{x}<\tilde{X}_u),\\
\hat{P}+\hat{P}_{\bar{x}} (1-\tilde{X}_u) &=\mathcal{M}/\theta,  &(\bar{x}=\tilde{X}_u),\\
    \tilde{X}_{u,\tilde{t}}&=- \hat{P}_{\bar{x}},& (\bar{x}={\tilde{X}_u-}).
\end{align}
The transport equation (\ref{eq:put2to}$a$) demonstrates that compressibility is significant along the thin gas film.  Matching conditions at $\bar{x}\rightarrow 0$ are provided by the outer limit (\mainref{eq:innerout}{2.38}$h$) of the inner problem. Continuity of pressure, gas mass flux and of interface shape requires that 
\begin{align}
\hat{P}&\rightarrow \tilde{P}_0, & 
\hat{F}\hat{P}\hat{P}_{\bar{x}}&\rightarrow -\frac{\mathcal{M}}{\theta}\mathcal{Q}, &
\hat{F}&\approx \left[\frac{\mathcal{M}}{\theta}\frac{1}{\bar{x}} \int_0^{\tilde{t}} \frac{\mathcal{Q}}{\tilde{P}_0}\,\mathrm{d}\tilde{t}'\right]^{1/2}, & (\bar{x}\rightarrow 0).
\end{align}
The associated mass and volume constraints are 
\begin{equation}
\left(\int_0^{\tilde{X}_u} \hat{F}\hat{P}\,\mathrm{d}\bar{x}\right)_{\tilde{t}}=\frac{\mathcal{Q}
\mathcal{M}}{\theta}, \quad
\left(\int_0^{\tilde{X}_u} \hat{F}\,\mathrm{d}\bar{x}\right)_{\tilde{t}}=\frac{1}{1-\tilde{X}_u}\left[\hat{P}\big\vert_{\tilde{X}_u}-\frac{\mathcal{M}}{\theta}\right].
\end{equation}
\end{subequations}
To match with the early time evolution (\ref{eq:Boyles}, \ref{eq:pinf}), we impose $\hat{P}(0,0)=(\mathcal{M}/\theta)P_\infty$ in instances when $P_0$ has had time to equilibrate.

\subsection{Problems $\Pi_1$ and $\Pi_2$}

The codimension-2 problem (\ref{eq:put2t}) (problem $\Pi_a$) is parametrized by $\mathcal{M}$ and $\theta$; this problem retains compressibility but buoyancy becomes subdominant. The reduced codimension-1 problem $\Pi_{12}$ emerging from $\Pi_a$, represented by (\ref{eq:put2ti}, \ref{eq:put2to}), is parametrized by the single parameter $\mathcal{M}/\theta$, which is formally of order unity in Problem $\Pi_{12}$; here the low viscosity ratio promotes formation of a thin film in which compressible effects remain significant, connected to an inner region where the flow is incompressible. We can recover two further simplifications, when $\theta\ll \mathcal{M}\ll 1$ and $\mathcal{M}\ll\theta \ll 1$; these are the codimension-0 problems in regions $\Pi_1$ and $\Pi_2$ in figure~\ref{fig:overview}. The problems differ in the outer region but share the same inner region structure, as described in \S~\ref{sec:251}.  We now address each outer region in turn. 

\textbf{Region $\Pi_1$}.  For $\theta\ll \mathcal{M}\ll 1$, we expect compressibility effects to be suppressed. We set $\hat{P}=(\mathcal{M}/\theta)+\hat{P}_1+\dots$, consistent with (\ref{eq:pinf}). The outer problem (\ref{eq:put2to}) becomes, to leading order in $\theta/\mathcal{M}\ll 1$, 
\begin{subequations}
\label{eq:put2toa}
\begin{align}
\hat{F}_{\tilde{t}} &= \left[\hat{F} \hat{P}_{1\bar{x}}\right]_{\bar{x}}, & (0<\bar{x}<\tilde{X}_u),\\
-\hat{F}_{\bar{t}} &= \hat{P}_{1\bar{x}\bar{x}}, & (0<\bar{x}<\tilde{X}_u),\\
 \hat{P}_1+\hat{P}_{1\bar{x}} (1-\tilde{X}_u) &=0,  &(\bar{x}=\tilde{X}_u),\\
    \tilde{X}_{u,\tilde{t}}&=- \hat{P}_{1\bar{x}},& (\bar{x}={\tilde{X}_u-}), \\
\hat{F}\hat{P}_{1\bar{x}}&=-\mathcal{Q}, \quad 
\hat{F}\approx \left[{\tilde{\mathcal{V}}}/{\bar{x}} \right]^{1/2}, & (\bar{x}\rightarrow 0).
\end{align}
Here we have introduced $\tilde{\mathcal{V}}(\tilde{t})\equiv \int_0^{\tilde{t}} \mathcal{Q}\,\mathrm{d}\tilde{t}'$; (\ref{eq:put2toa}$e$) matches to (\mainref{eq:innerout}{2.38}$h$).  The mass and volume constraints are 
\begin{equation}
\left(\int_0^{\tilde{X}_u} \hat{F}\,\mathrm{d}\bar{x}\right)_{\tilde{t}}=\mathcal{Q}, \quad
\left(\int_0^{\tilde{X}_u} \hat{F}\,\mathrm{d}\bar{x}\right)_{\tilde{t}}=\frac{1}{1-\tilde{X}_u}\hat{P}_1\big\vert_{\tilde{X}_u},
\end{equation}
\end{subequations}
implying that $\hat{P}_1\big\vert_{\tilde{X}_u}=\mathcal{Q}(1-\tilde{X}_u)$.  It follows from (\ref{eq:put2toa}$a$,$b$,$e$) that $\hat{P}_{1\bar{x}}=-\mathcal{Q}/(1+\hat{F})$ and
\begin{equation}
    \label{eq:hyp}
    \hat{F}_{\tilde{t}}+\frac{\mathcal{Q} \hat{F}_{\tilde{x}}}{(1+\hat{F})^2}=0.
\end{equation}
Solving using characteristics, assuming that at early times the interface is confined to a region $\bar{x}=o(1)$, gives
\begin{align}
\hat{F}&=\left[{\tilde{\mathcal{V}}}/{\bar{x}}\right]^{1/2}-1, & \left(0<\bar{x}<\tilde{X}_u=\tilde{\mathcal{V}}\right).
    \label{eq:Peg}
\end{align}
In (\ref{eq:put2toa})--(\ref{eq:Peg}), we have recovered the incompressible spreading problem described by \cite{pegler_fluid_2014} and others, extended to account for time-dependent forcing, consistent with (\ref{eq:fullcharsol}).  The volume flux at the source has an instantaneous effect on the leading-order location of the upper contact line.  The corresponding pressure field is
\begin{align}
    \bar{P}(\bar{x},\tilde{t})&=\frac{1}{\theta}+\frac{\mathcal{Q}}{\mathcal{M}}\left[1-\frac{\tilde{\mathcal{V}}}{3}\left(1+2\left(\frac{\bar{x}}{\tilde{\mathcal{V}}}\right)^{3/2}\right) \right],& (0<\bar{x}<\tilde{\mathcal{V}} \leq 1).
    \label{eq:poutinc}
\end{align}
Here, with $\theta\ll \mathcal{M}\ll 1$, the large but uniform leading-order pressure $1/\theta$ (obtained via rapid compression of the gas) effectively decouples from the pressure field driving spreading.  However (\ref{eq:poutinc}) suggests that once $\mathcal{M}$ and $\theta$ are of comparable magnitude, viscous and compressible effects will together influence spreading.

\textbf{Region $\Pi_2$}.  The other simpler case arising from (\ref{eq:put2to}) arises for $\mathcal{M}\ll \theta$, {a limit in which we expect compressibility to be promoted}.  Here we 
set $\hat{P}=(\mathcal{M}/\theta)^{1/2}\check{P}(\bar{x},\check{t})$, $\tilde{t}=(\theta/\mathcal{M})^{1/2}\check{t}$, $\hat{F}=\check{F}(\bar{x},\check{t})$, $\bar{X}_u=\check{X}_u(\check{t})$, leading to
\begin{subequations}
\label{eq:put2toc}
\begin{align}
\left[\check{F}\check{P}\right]_{\check{t}} &= \left[\check{F} \check{P} \check{P}_{\bar{x}}\right]_{\bar{x}}, & (0<\bar{x}<\check{X}_u),\\
-\check{F}_{\check{t}} &= \check{P}_{\bar{x}\bar{x}}, & (0<\bar{x}<\check{X}_u),\\
\check{P}+\check{P}_{\bar{x}} (1-\check{X}_u) &=0,  &(\bar{x}=\check{X}_u),\\
    \check{X}_{u,\check{t}}&=- \check{P}_{\bar{x}},& (\bar{x}={\check{X}_u-}), \\
\check{F}\check{P}\check{P}_{\bar{x}}&\rightarrow -\mathcal{Q}, \quad
\check{F}\approx \left[\frac{1}{\bar{x}} \int_0^{\check{t}} \frac{\mathcal{Q}}{\check{P}(0,\check{t}^{\prime})}\,\mathrm{d}\check{t}'\right]^{1/2}, & (\bar{x}\rightarrow 0).
\end{align}
The gas transport equation (\ref{eq:put2toc}$a$) balances viscous and compressible effects; it communicates with an incompressible inner region via (\ref{eq:put2toc}$e$), which matches to (\mainref{eq:innerout}{2.38}$h$).  The pressure field is smaller in magnitude and the spreading rate is slower than in problem $\Pi_{1}$.  The system (\ref{eq:put2toc}) satisfies the mass and volume constraints
\begin{equation}
    \left(\int_0^{\check{X}_u}\check{F}\check{P}\,\mathrm{d}\check{x} \right)_{\check{t}}=\mathcal{Q},\quad
    \check{V}_{\check{t}}\equiv \left(\int_0^{\check{X}_u}\check{F}\,\mathrm{d}\check{x} \right)_{\check{t}}=\frac{1}{1-\check{X}_u} \check{P}\big\vert_{\check{X}_u}.
\end{equation}
\end{subequations}
From (\ref{eq:put2toc}$c$,$d$,$f$) the volume constraint can be written as $\check{V}_{\check{t}}= \check{X}_{u,\check{t}}$.  Integrating up to the breakthrough time $\check{t}_b$ gives $\check{V}(\check{t}_b) = \check{V}_0 + 1$, where $\check{V}_0=L_0/(\mathcal{LM})$; $\check{V}_0\ll 1$ in region $\Pi_2$.  Thus provided $\mathcal{L}^{-1}\ll \mathcal{M}\ll 1$, the dimensional delivered gas volume (per unit width)
\begin{equation}
\label{eq:dimVol}
    V^*(t_b^*) \approx H L \mathcal{M},
\end{equation}
matches the volume delivered in the incompressible case $\Pi_1$, although the delivered mass $\int_0^{t_b} \mathcal{Q}\,\mathrm{d}t$ will differ.

For $\mathcal{Q}=1$ the problem (\ref{eq:put2toc}) is parameter-free, yielding a solution that depends only on details of the initial condition.  Numerical solutions of (\ref{eq:put2toc}) are compared to solutions of the full problem (\ref{eq:put2}) below.

\subsection{Overview of asymptotic regimes}

A summary of the asymptotic regions described so far is provided in figure~\ref{fig:overview}.  In the pink shaded region ($\Pi_1$), compressibility regulates rapid initial adjustment of the pressure  via (\ref{eq:Boyles}) but {thereafter mass-flux-driven} spreading is controlled by incompressible effects via (\ref{eq:put2toa}).  In the blue region ($\Pi_2$), viscous and compressible effects combine to determine the pressure scale and the spreading rate via (\ref{eq:put2toc}).  There is again a rapid adjustment of the pressure before spreading initiates.  Further analysis at lower viscosity ratios reveals new balances.  First, as explained in Appendix~\ref{app:buoyancy}, buoyancy effects allow the lower contact line to recede for $\mathcal{L}\mathcal{M}^2\lesssim 1$, without having an appreciable effect on spreading rates. Second, as explained in Appendix~\ref{app:verylowviscosity}, at even lower values of $\mathcal{M}$, the film of gas escaping from the bubble becomes so thin that it carries very low flux relative to that provided by the source.  The bubble pressure rises at a rate determined by the source strength, and compressive effects determine the breakthrough time.  {This is addressed in Appendix~\ref{app:verylowviscosity} by formulating the reduced problem at codimension-2 point $\Pi_c$, from which the codimension-1 problem $\Pi_{12}$ is recovered, while revealing new codimension-0 regions $\Pi_3$ and $\Pi_4$ (shaded orange and green respectively in figure~\ref{fig:overview}).  In $\Pi_3$, compressible effects persist in the thin film and dynamic changes to the pressure of the main gas bubble influence spreading dynamics.  In $\Pi_4$, the thin-film flow is incompressible but compressive effects in the main gas bubble ensure that spreading is pressure-driven rather than directly driven by the imposed mass flux.}  The pressure rises to levels comparable to the incompressible problem in region $\Pi_4$ (see (\ref{eq:put24})), but to a level depending on $\zeta$ when the gas is more compressible in region $\Pi_3$ (see (\ref{eq:put22})).  These asymptotic predictions will now be evaluated against simulations of the full model (\ref{eq:put2}).

\section{Results}
\label{sec:results}

\begin{figure}
    \centering
    \begin{subfigure}{0.49\textwidth}
        \caption{}
        \includegraphics[width = \textwidth]{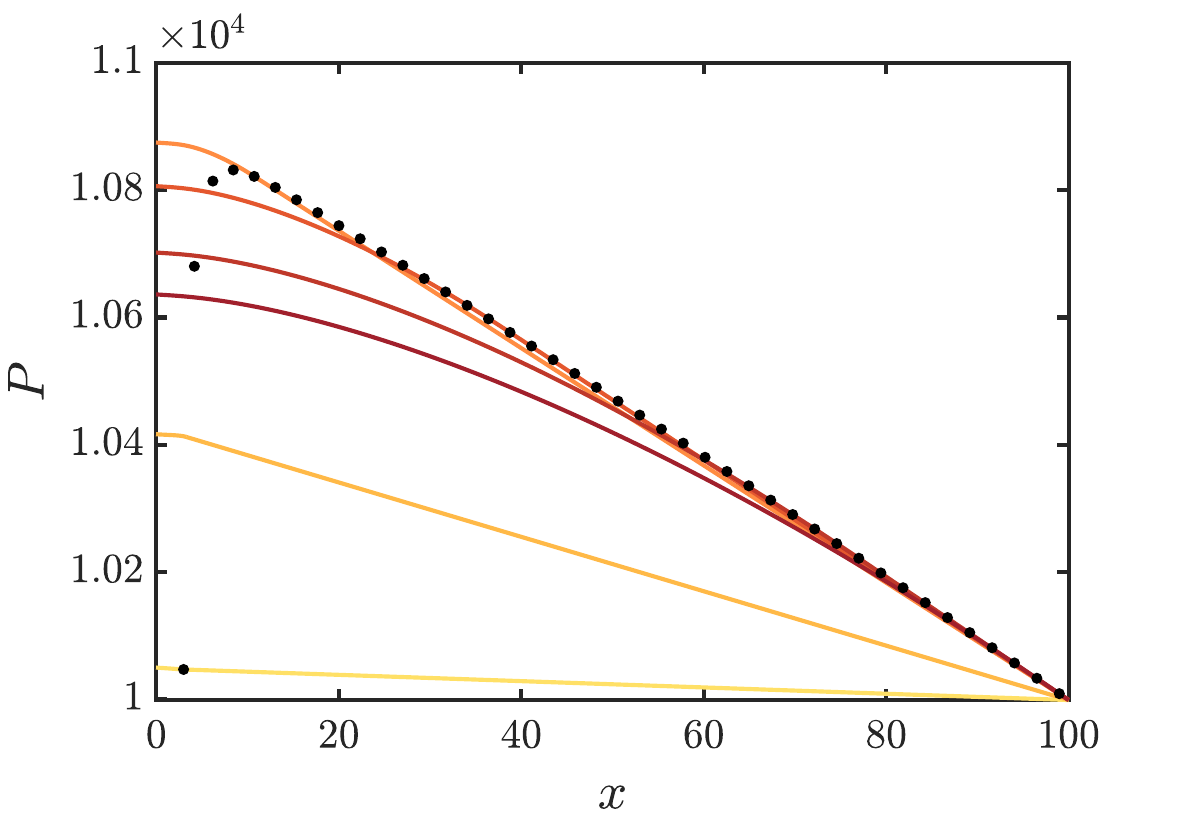}
    \end{subfigure}
    \hfill
    \begin{subfigure}{0.49\textwidth}
        \caption{}
        \includegraphics[width = \textwidth]{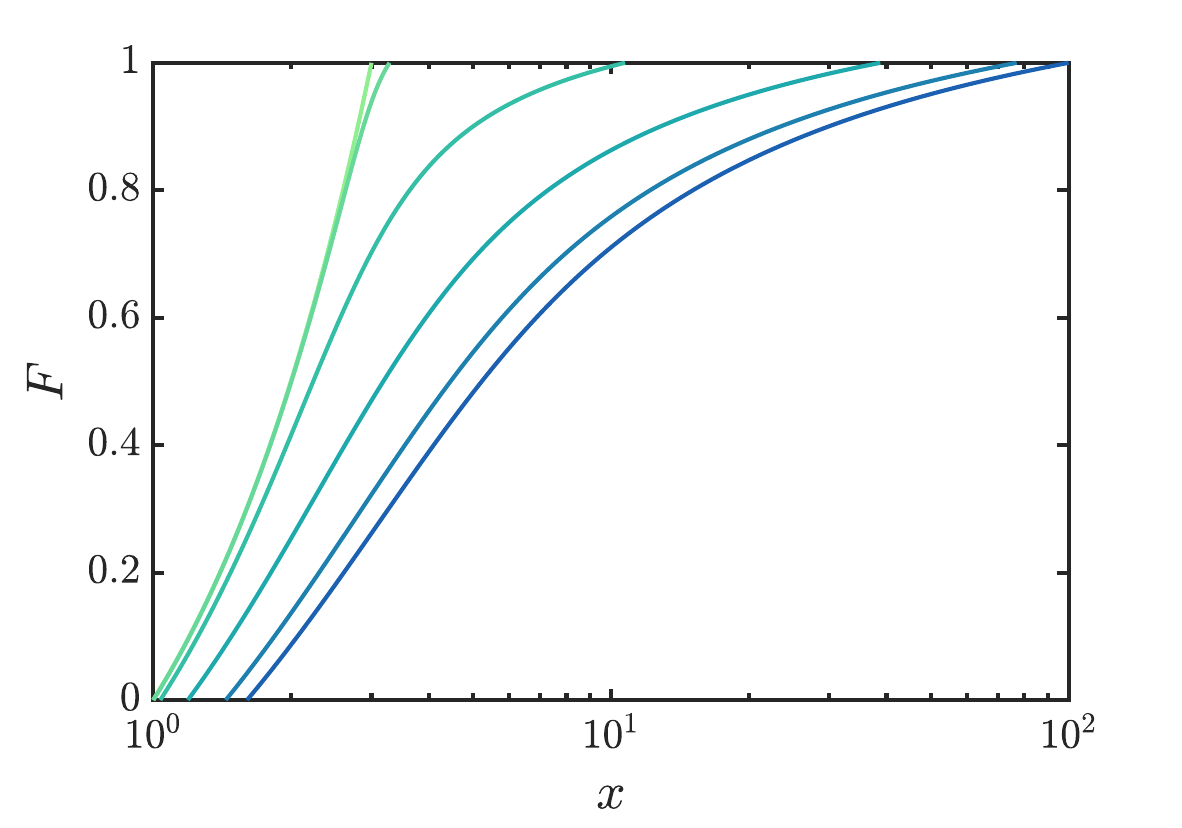}
    \end{subfigure}
    \\[-2\baselineskip]  
    \begin{subfigure}{0.49\textwidth}
        \caption{}
        \includegraphics[width = \textwidth]{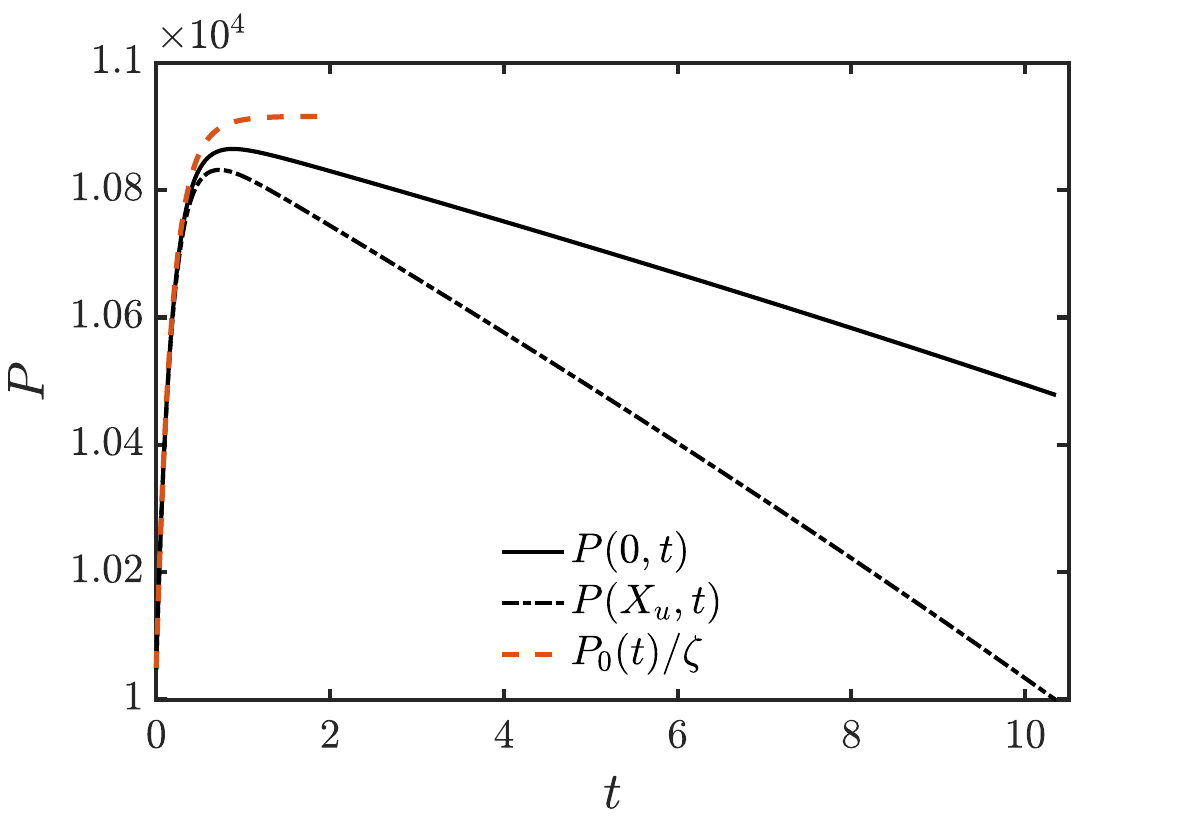}
    \end{subfigure}
    \hfill
    \begin{subfigure}{0.49\textwidth}
        \caption{}
        \includegraphics[width = \textwidth]{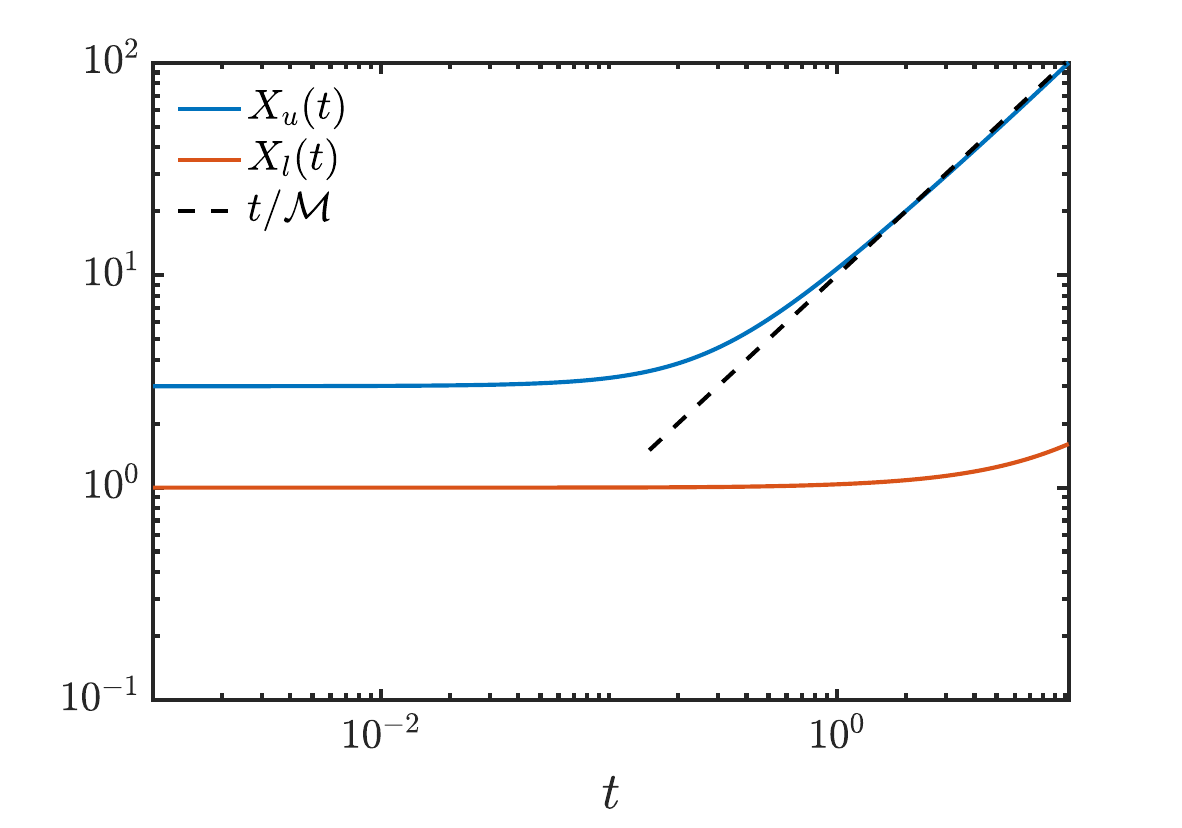}
    \end{subfigure}
     \caption{A simulation of (\ref{eq:put2}) for $\zeta = 10^{-4}$, $\mathcal{M} = 0.1$, and $\mathcal{L} = 100$, within the pink region of  figure~\ref{fig:overview}. (a) and (b) show the pressure field $P(x,t)$ and the interface height $F(x,t)$, respectively, at $t = \{0, 0.1, 1, 4, 8, 10.4\}$, following the light-to-dark colour gradient. Black dots in (a) show samples of the pressure at the upper contact line $P(X_u,t)$. (c) shows the pressure evolution at the origin and at the upper contact line; the early-time approximation given by \eqref{eq:Boyles} (dashed orange) captures the transient rise in pressure due to gas compressibility. (d) shows contact-line locations; $X_u$ approaches the late-time solution (\ref{eq:Peg}) (dashed black line). }
     \label{fig:M_0.1_zeta_10-4}
\end{figure}

For a dimensionless domain length of $\mathcal{L} = 100$, we performed extensive simulations of (\ref{eq:put2}) across a broad range of $( \zeta, \mathcal{M})$ values. Throughout, we take $L_0= \Delta_0 = 2$ in (\ref{eq:ini}). In \S~\ref{sec:Steady}, we present results for a steady injection rate ($\mathcal{Q} = 1$). To illustrate key flow behaviours, we show six representative simulations of \eqref{eq:put2}: one (figure~\ref{fig:M_0.1_zeta_10-4}) from region $\Pi_1$ in figure~\ref{fig:overview}, one on the boundary $\Pi_{12}$ (figure~\ref{fig:M_0.1_zeta_0.001}), three from region $\Pi_2$ or its boundary with region $\Pi_3$ (figures~\ref{fig:M_0.1_zeta_1}–\ref{fig:M_0.01_zeta_1}), and one from the ultra-low viscosity regime near the $\Pi_{3}/\Pi_4$ boundary (figure~\ref{fig:M_10-3_zeta_10-4}). In \S~\ref{sec:unsteady}, we compare results for increasing and decreasing injection rates.

\subsection{Steady injection}
\label{sec:Steady}

\begin{figure}
    \centering
    \begin{subfigure}{0.49\textwidth}
        \caption{}
        \includegraphics[width = \textwidth]{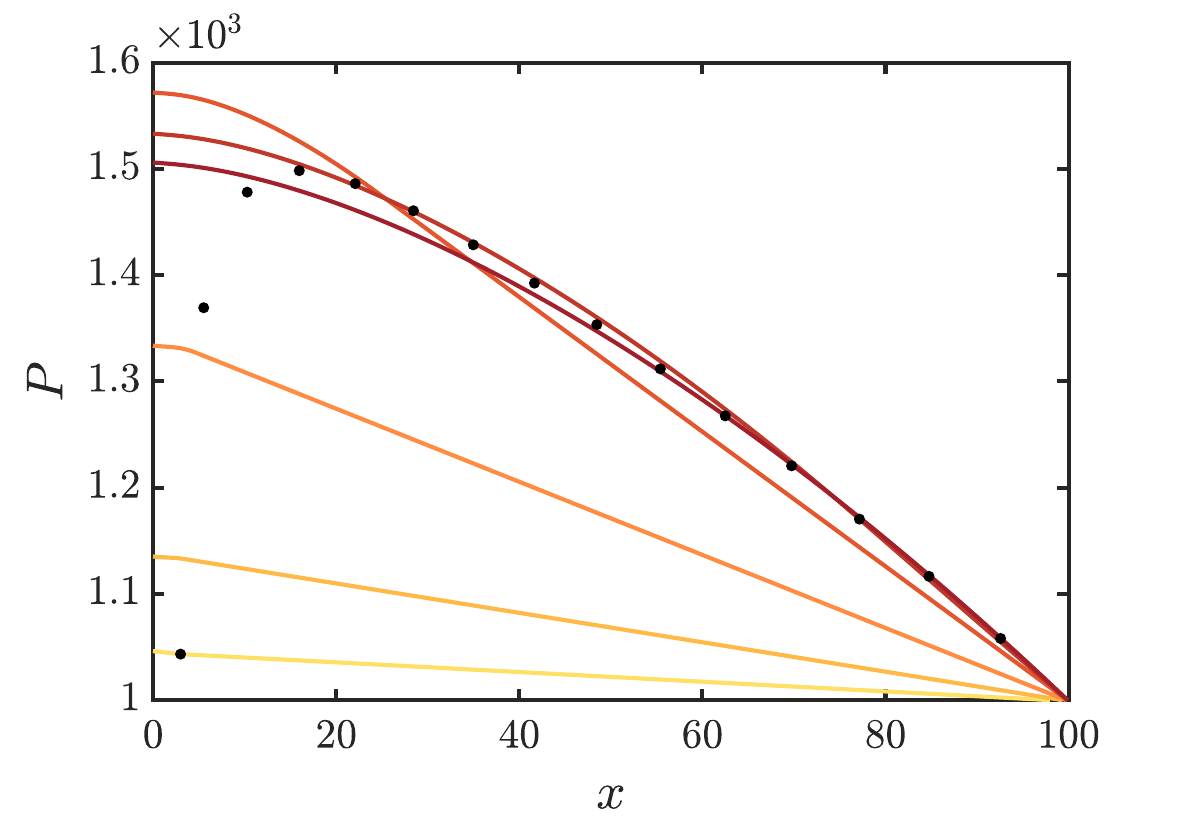}
    \end{subfigure} 
    \hfill
    \begin{subfigure}{0.49\textwidth}
        \caption{}
        \includegraphics[width = \textwidth]{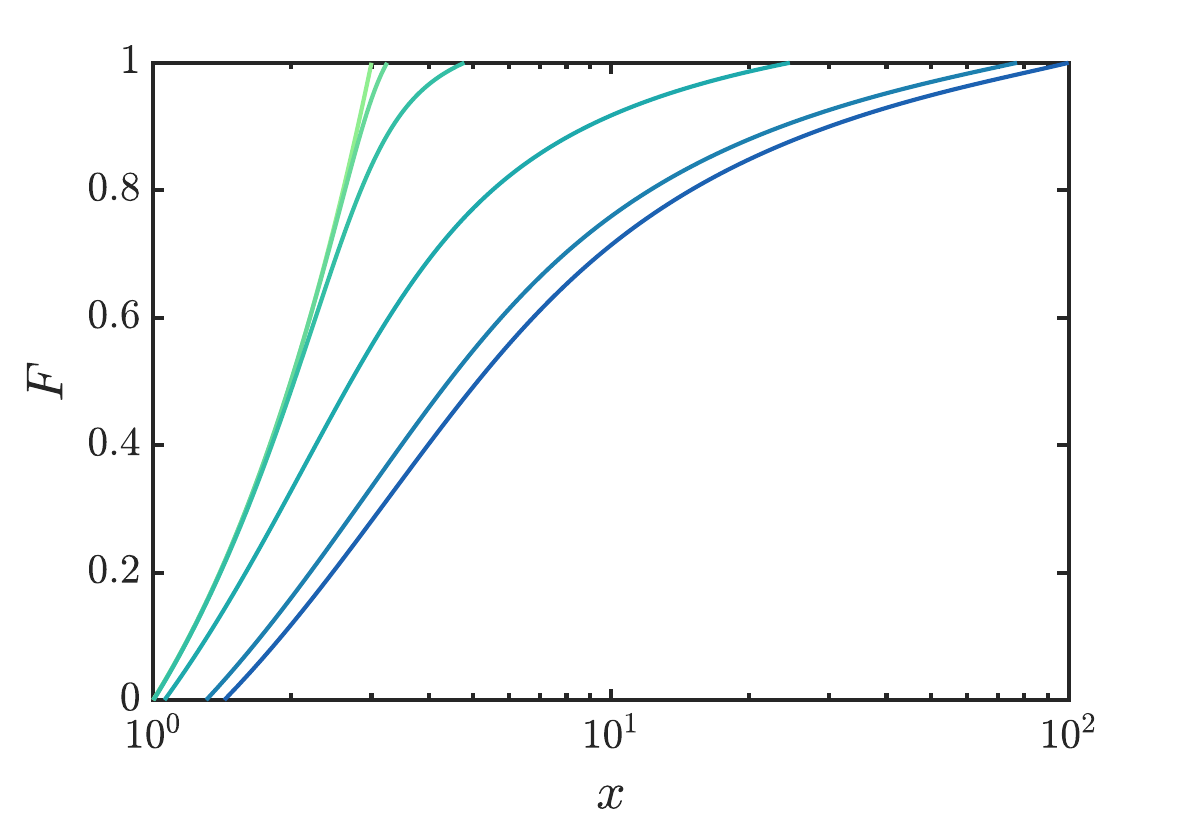}
    \end{subfigure}
    \\[-2\baselineskip]    
    \begin{subfigure}{0.49\textwidth}
        \caption{}
        \includegraphics[width = \textwidth]{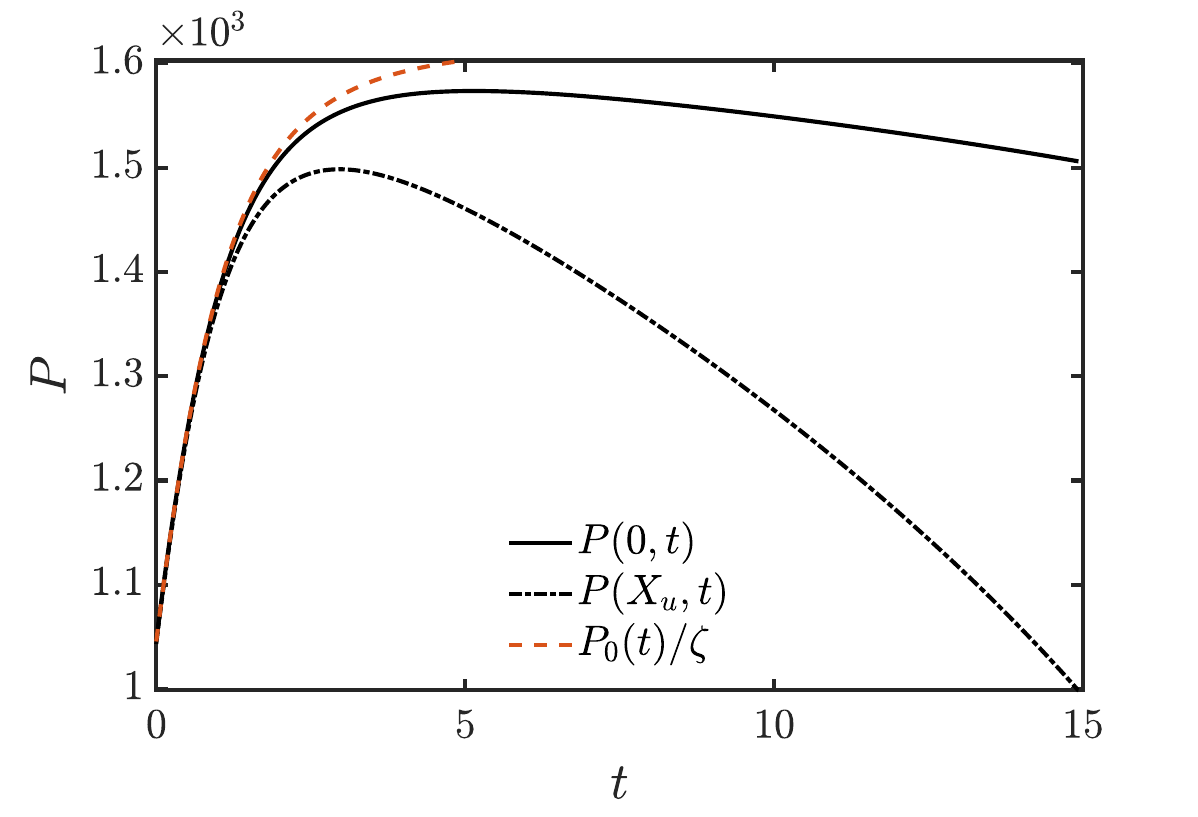}
    \end{subfigure}
    \hfill
    \begin{subfigure}{0.49\textwidth}
        \caption{}
        \includegraphics[width = \textwidth]{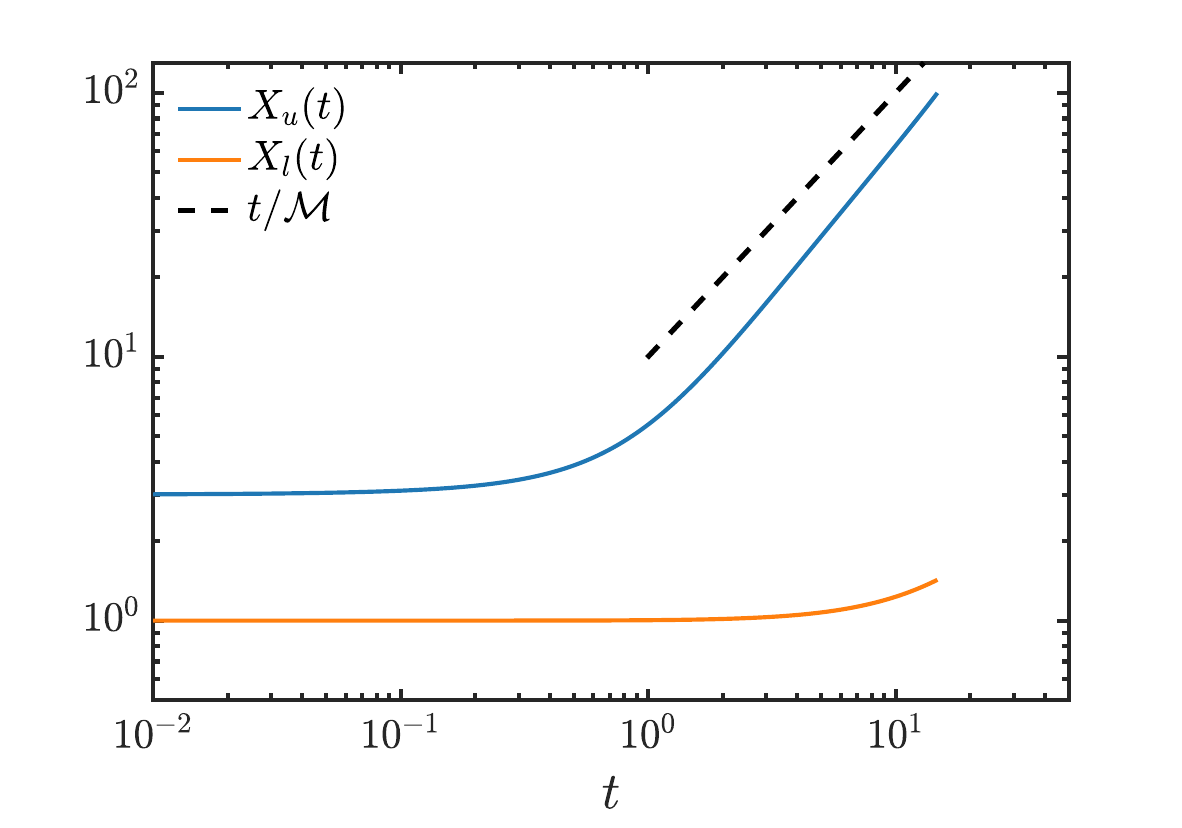}
    \end{subfigure}
     \caption{
     A simulation of (\ref{eq:put2}) for $\zeta = 10^{-3}$, $\mathcal{M} = 0.1$, and $\mathcal{L} = 100$.    
     (a) and (b) show $P(x,t)$ and $F(x,t)$ at $t = \{0, 0.2, 0.8, 4.4, 12, 15\}$.
     Solutions in (a-d) are plotted using the format described in figure~\ref{fig:M_0.1_zeta_10-4}.
}
     \label{fig:M_0.1_zeta_0.001}
\end{figure}

To set the scene, we first illustrate spreading in the almost incompressible regime with low viscosity ratio ($\mathcal{M}=0.1$, $\zeta=10^{-4}$; figure~\ref{fig:M_0.1_zeta_10-4}); {as indicated in figure~\ref{fig:overview}, this example lies in asymptotic region $\Pi_1$}.  As expected from previous studies of the incompressible problem with $\mathcal{M}<1$ \citep{pegler_fluid_2014}, a finger of gas advances rapidly along the top of the channel (figure~\ref{fig:M_0.1_zeta_10-4}b).  $X_u$ asymptotes promptly to $t/\mathcal{M}$ (figure~\ref{fig:M_0.1_zeta_10-4}d), advancing at a rate determined by the strength of the source, as predicted by (\ref{eq:clsp}); the lower contact line advances only marginally before the gas phase reaches the outlet.  $P(0,t)$, which measures changes in gas pressure near the source relative to the hydrostatic scale $\Delta \rho g H$, is non-monotonic (figure~\ref{fig:M_0.1_zeta_10-4}c): its early rise can be attributed to transient compressible effects, following (\ref{eq:Boyles}); its later fall is due to the shortening length of the liquid-filled domain, across which there is the primary viscous pressure drop.  The pressure field at any time is monotonic in $x$ (figure~\ref{fig:M_0.1_zeta_10-4}a), and is linear across $X_u<x<\mathcal{L}$.  The pressure at the upper contact line rises rapidly before falling slowly (figure~\ref{fig:M_0.1_zeta_10-4}a,c).

Figure~\ref{fig:M_0.1_zeta_0.001} illustrates the impact of increasing the compressibility parameter to $\zeta = 10^{-3}$, a case lying on the line $\mathcal{M} = \zeta \mathcal{L}$ in the parameter map shown in figure~\ref{fig:overview}.  The time taken for the pressure to rise is extended in comparison to figure~\ref{fig:M_0.1_zeta_10-4}; the early rise is still well captured by (\ref{eq:Boyles}) (figure~\ref{fig:M_0.1_zeta_0.001}c). 
Transient compression of the gas slows the rate of spreading relative to the incompressible case predicted by (\ref{eq:clsp}) (figure~\ref{fig:M_0.1_zeta_0.001}$d$). The spreading rate is further inhibited by the pressure reduction associated with increasing $\zeta$ (compare figure~\ref{fig:M_0.1_zeta_0.001}a to figure~\ref{fig:M_0.1_zeta_10-4}a), which reduces the viscous pressure gradient along the liquid column. A more compressible gas (larger $\zeta$) has higher density, enabling it to sustain similar mass fluxes at lower pressures. The lower contact line in this example advances marginally (figure~\ref{fig:M_0.1_zeta_0.001}b, d).  This example corresponds to problem $\Pi_b$, discussed in Appendix~\ref{app:buoyancy}, so that buoyancy effects are expected to influence the motion of $X_l$.  Otherwise the pressure and interfacial profiles broadly resemble the incompressible example in figure~\ref{fig:M_0.1_zeta_10-4}.  

\begin{figure}
    \centering
    \begin{subfigure}{0.49\textwidth}
        \caption{}
        \includegraphics[width = \textwidth]{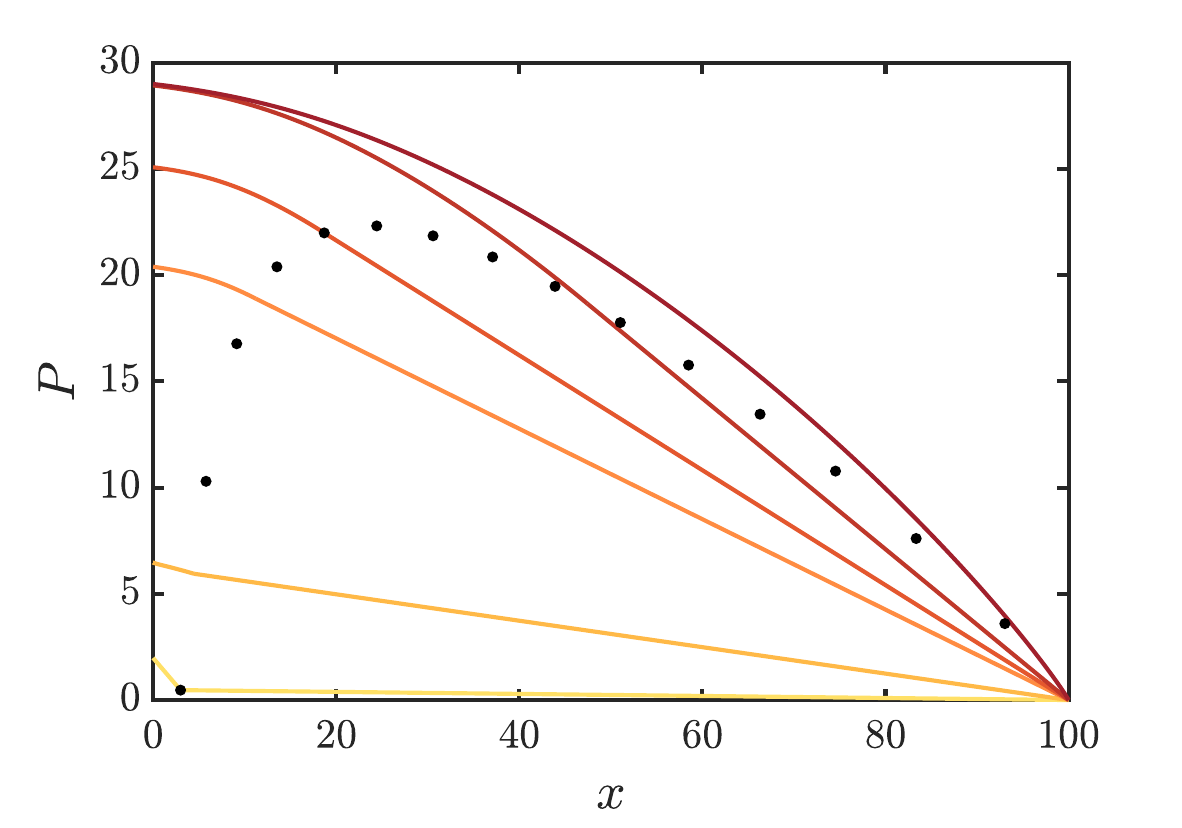}
    \end{subfigure} 
    \hfill
    \begin{subfigure}{0.49\textwidth}
        \caption{}
        \includegraphics[width = \textwidth]{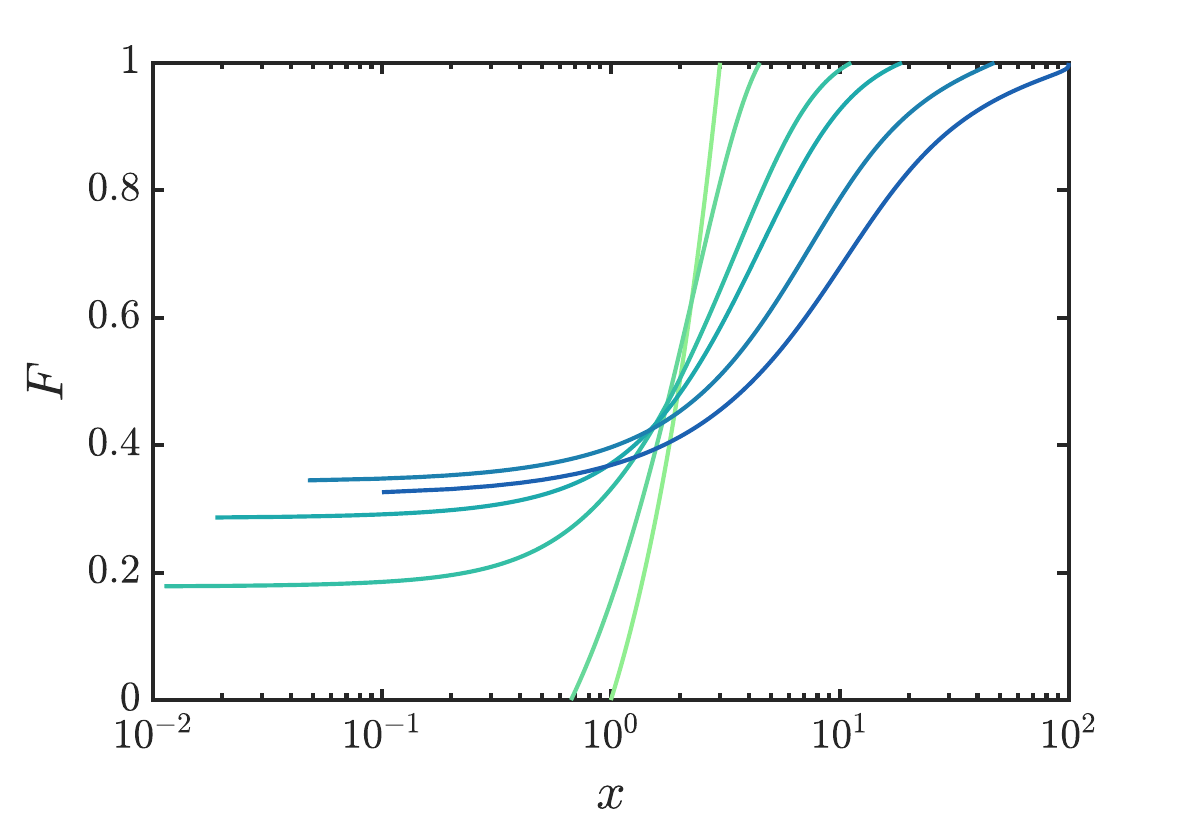}
    \end{subfigure}
    \\[-2\baselineskip]    
    \begin{subfigure}{0.49\textwidth}
        \caption{}
        \includegraphics[width = \textwidth]{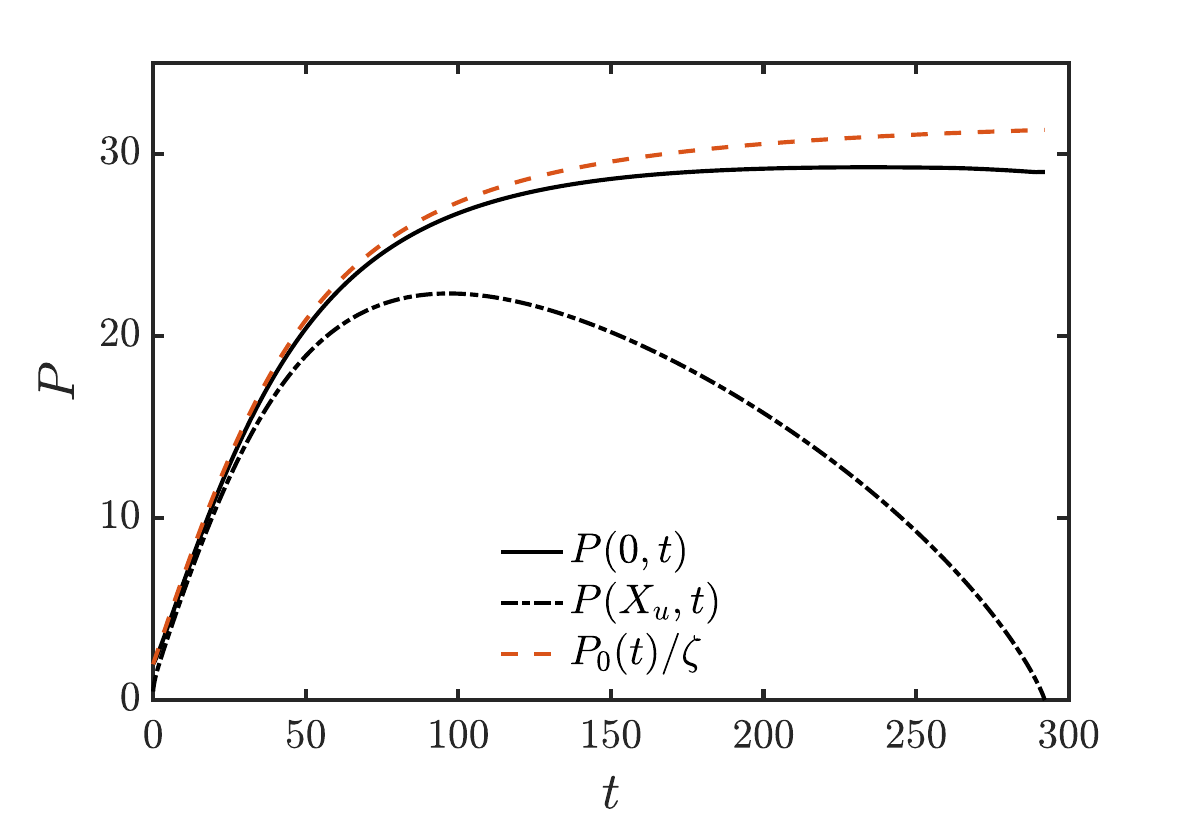}
    \end{subfigure}
    \hfill
    \begin{subfigure}{0.49\textwidth}
        \caption{}
        \includegraphics[width = \textwidth]{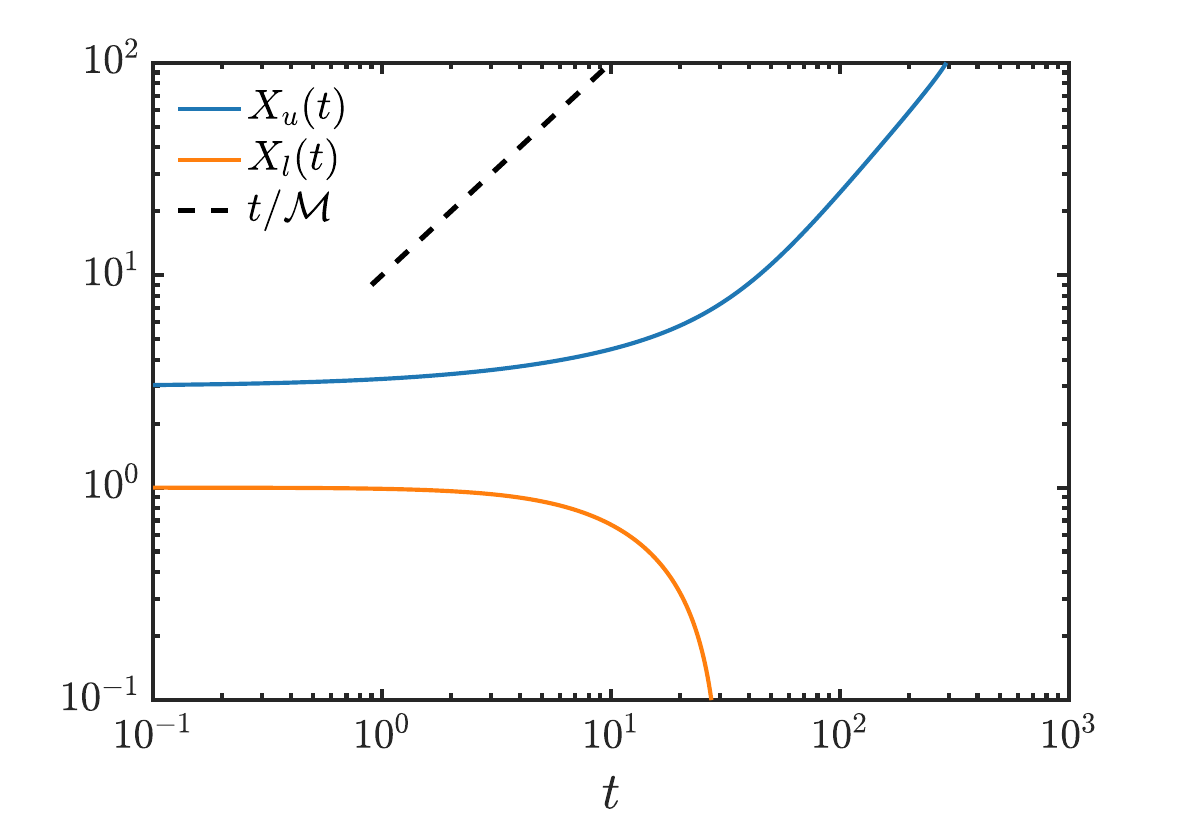}
    \end{subfigure}
     \caption{
     A simulation of (\ref{eq:put2}) for $\zeta = 1$, $\mathcal{M} = 0.1$, and $\mathcal{L} = 100$.
     (a) and (b) show $P(x,t)$ and $F(x,t)$ at $t = \{0,10,50,80,170,292.2\}$.  Solutions in (a-d) are plotted using the format described in figure~\ref{fig:M_0.1_zeta_10-4}.
     In (b, d), buoyancy forces drive the lower contact line backwards towards the origin and then up the wall at $x = 0$. 
     }
     \label{fig:M_0.1_zeta_1}
\end{figure}

A further increase in the compressibility parameter, to $\zeta=1$, brings more dramatic changes  (figure~\ref{fig:M_0.1_zeta_1}).  First, it significantly extends the proportion of spreading time in which $P(0,t)$ rises as the gas is compressed (figure~\ref{fig:M_0.1_zeta_1}c), almost until the breakthrough time.  Spreading is substantially delayed relative to the incompressible case, giving time for buoyancy effects to drive the lower contact line backwards towards $x=0$ (figure~\ref{fig:M_0.1_zeta_1}b, d).  

\begin{figure}
    \centering
    \begin{subfigure}{0.49\textwidth}
        \caption{}
        \includegraphics[width = \textwidth]{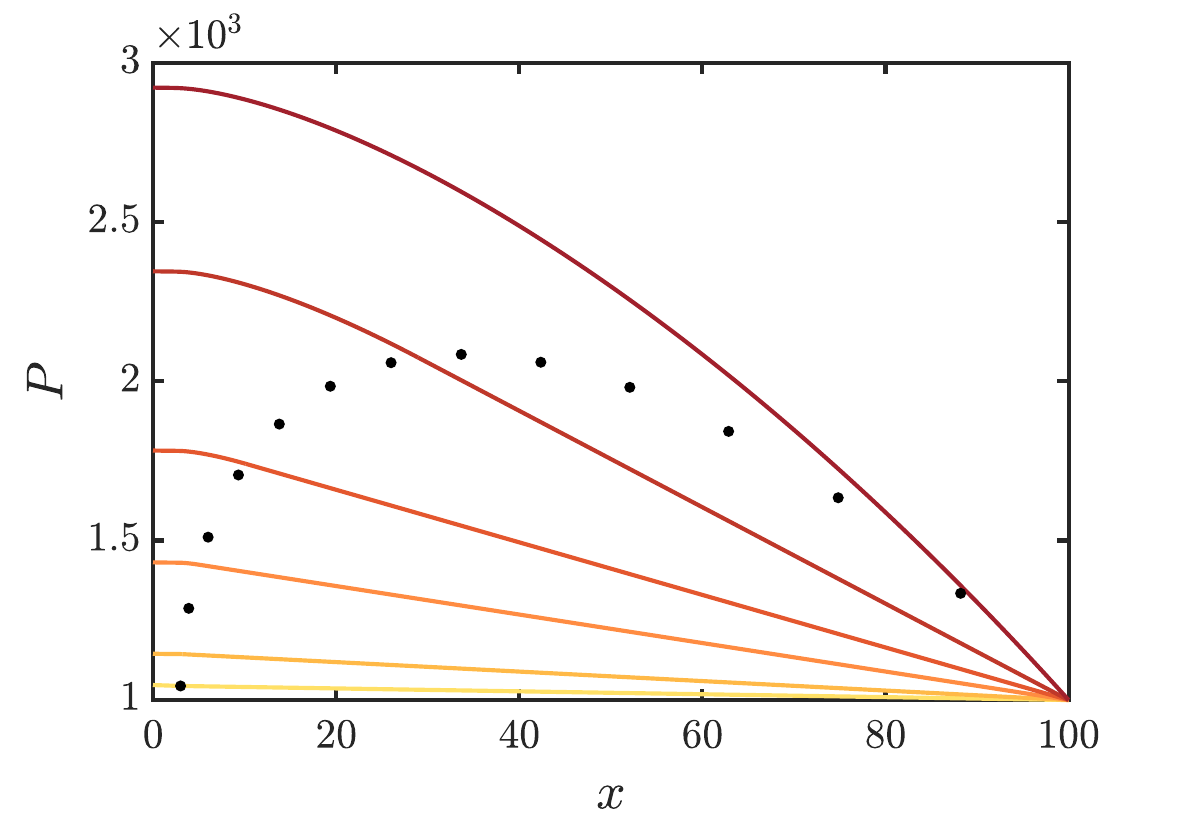}
    \end{subfigure} 
    \hfill
    \begin{subfigure}{0.49\textwidth}
        \caption{}
        \includegraphics[width = \textwidth]{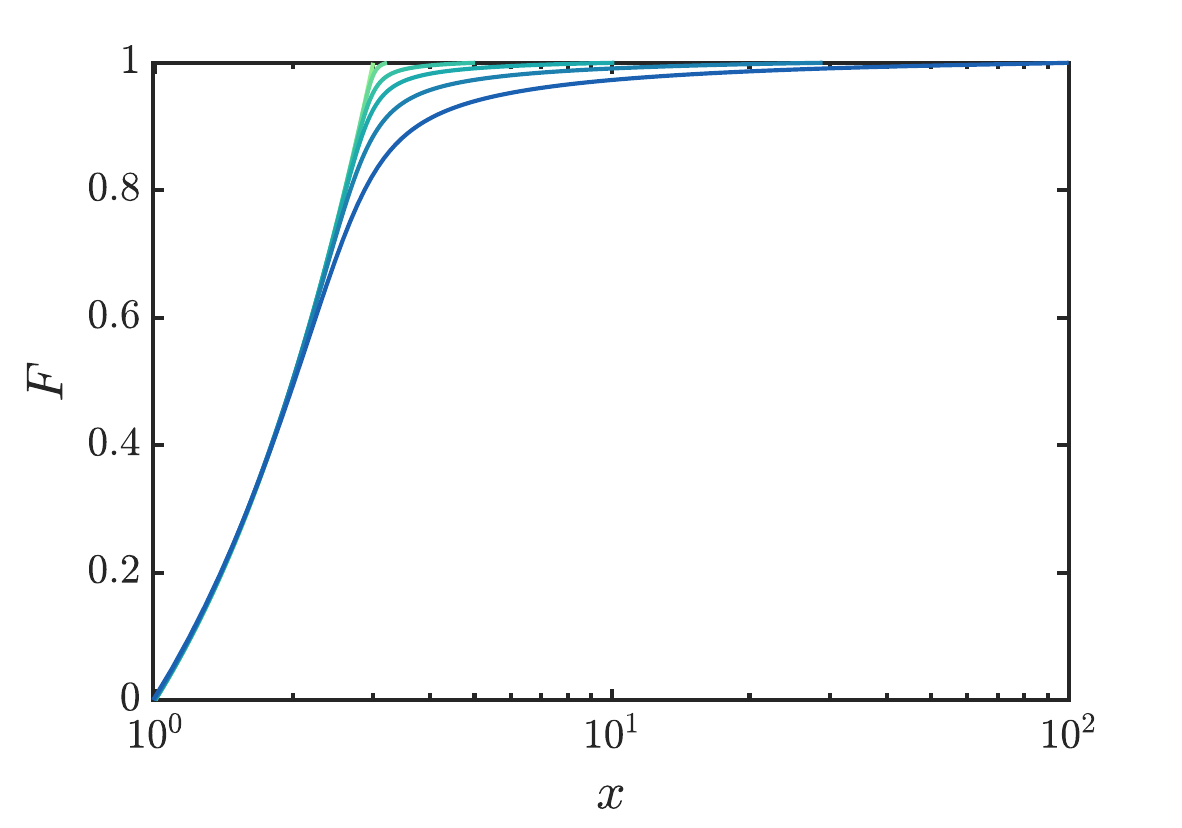}
    \end{subfigure}
    \\[-2\baselineskip]    
    \begin{subfigure}{0.49\textwidth}
        \caption{}
        \includegraphics[width = \textwidth]{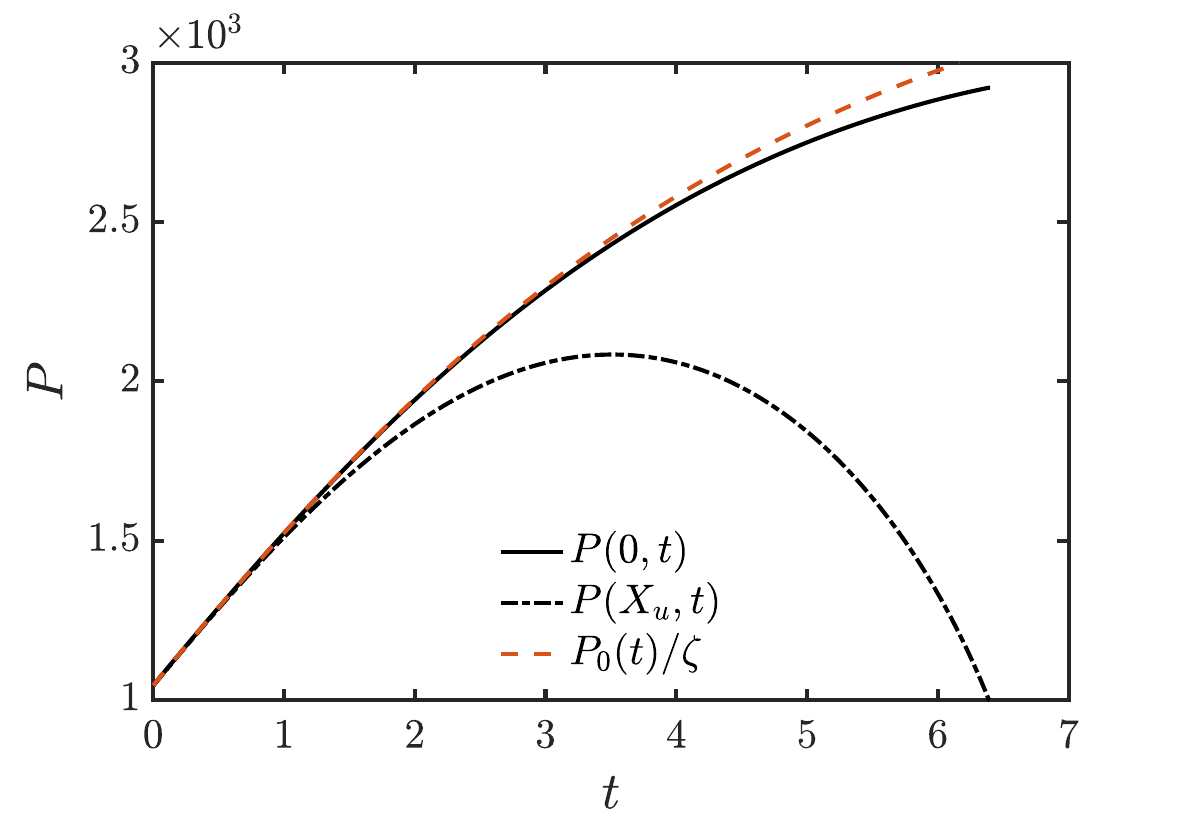}
    \end{subfigure}
    \hfill
    \begin{subfigure}{0.49\textwidth}
        \caption{}
        \includegraphics[width = \textwidth]{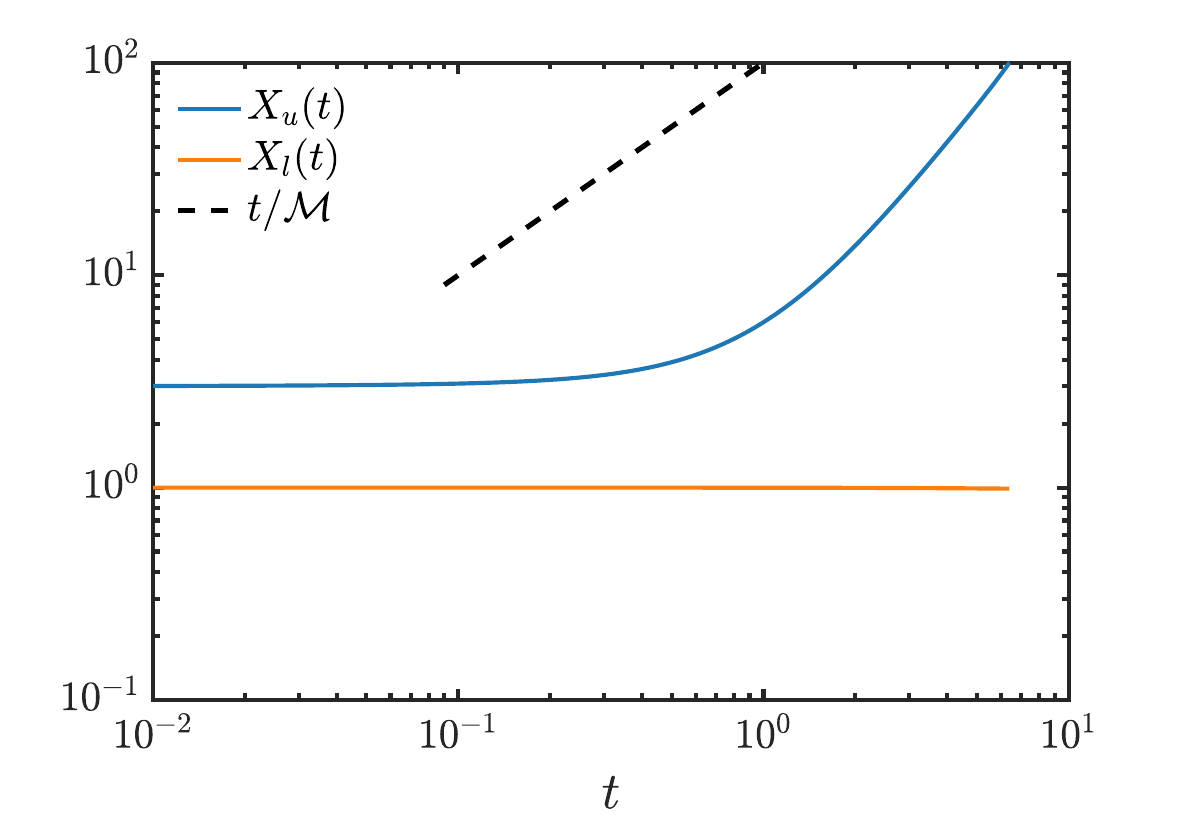}
    \end{subfigure}
     \caption{A simulation of (\ref{eq:put2}) for $\zeta = 10^{-3}$, $\mathcal{M} = 10^{-2}$, and $\mathcal{L} = 100$. 
               (a) and (b) show $P(x,t)$ and $F(x,t)$ at $t = \{0,0.2,0.8,1.6,3.2,6.4\}$.  Solutions in (a-d) are plotted using the format described in figure~\ref{fig:M_0.1_zeta_10-4}.\
}
     \label{fig:M_0.01_zeta_0.001}
\end{figure}

Reducing the viscosity ratio to $\mathcal{M}=0.01$ with $\zeta=10^{-3}$ also slows the rate at which the gas pressure initially rises (relative to the breakthrough time; compare figure~\ref{fig:M_0.01_zeta_0.001}$c$ to figure~\ref{fig:M_0.1_zeta_0.001}$c$).  In this example, with very low gas viscosity, spreading is sufficiently rapid for the initial interface profile to remain almost stationary as the gas film advances over the liquid (figure~\ref{fig:M_0.01_zeta_0.001}$b$), consistent with region-$\Pi_3$ analysis in Appendix~\ref{app:verylowviscosity}.  The thickness of the gas finger is reduced as a result of its low viscosity: the two layers have comparable fluxes driven by a shared pressure gradient, so that a thinner gas layer balances the smaller flux of the much more viscous liquid layer.  
Increasing $\zeta$, while retaining $\mathcal{M}=0.01$, again slows spreading substantially, giving buoyancy time to drive the lower contact line backwards (see figure~\ref{fig:M_0.01_zeta_1}(d) for $\zeta=1$), as anticipated in Appendix~\ref{app:buoyancy}.  Rather than equilibrating rapidly, $P(0,t)$ rises continuously as the gas bubble elongates (figure~\ref{fig:M_0.01_zeta_1}c), a signature of sustained compressive effects. 

\begin{figure}
    \centering
    \begin{subfigure}{0.49\textwidth}
        \caption{}
        \includegraphics[width = \textwidth]{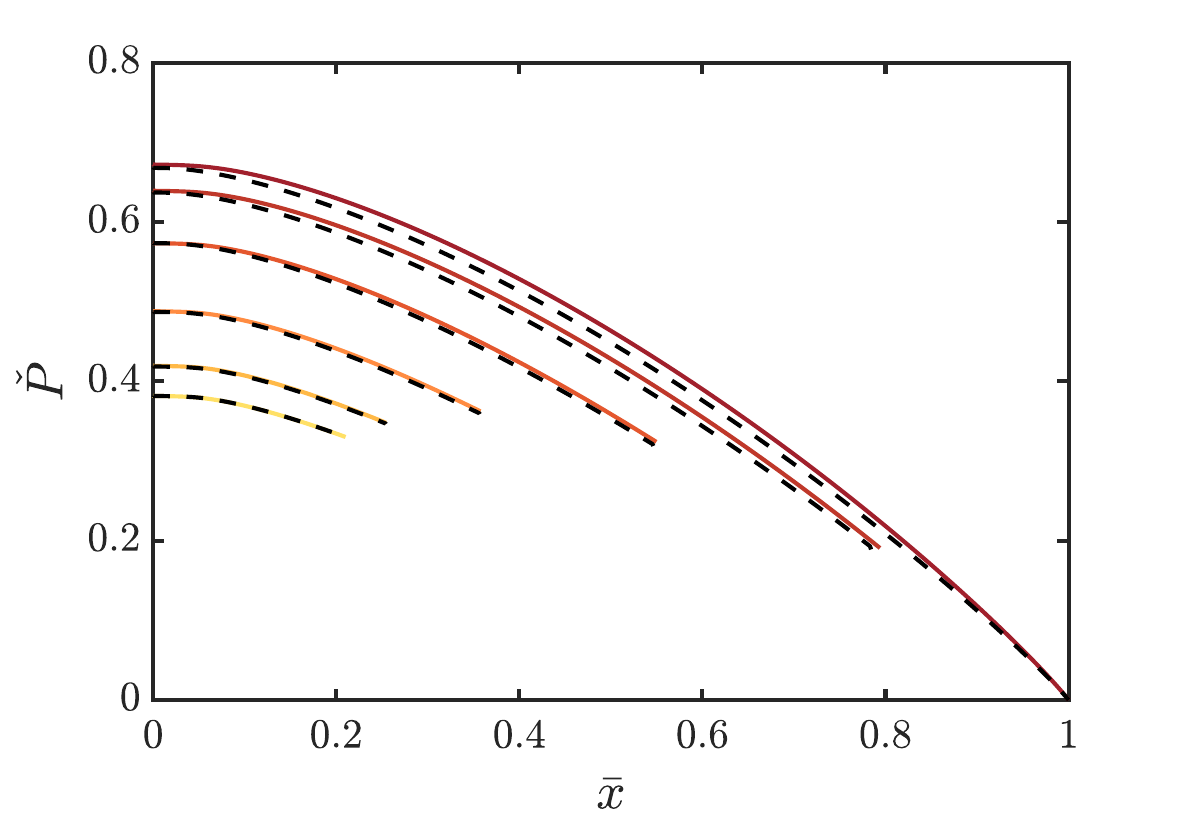}
    \end{subfigure} 
    \hfill
    \begin{subfigure}{0.49\textwidth}
        \caption{}
        \includegraphics[width = \textwidth]{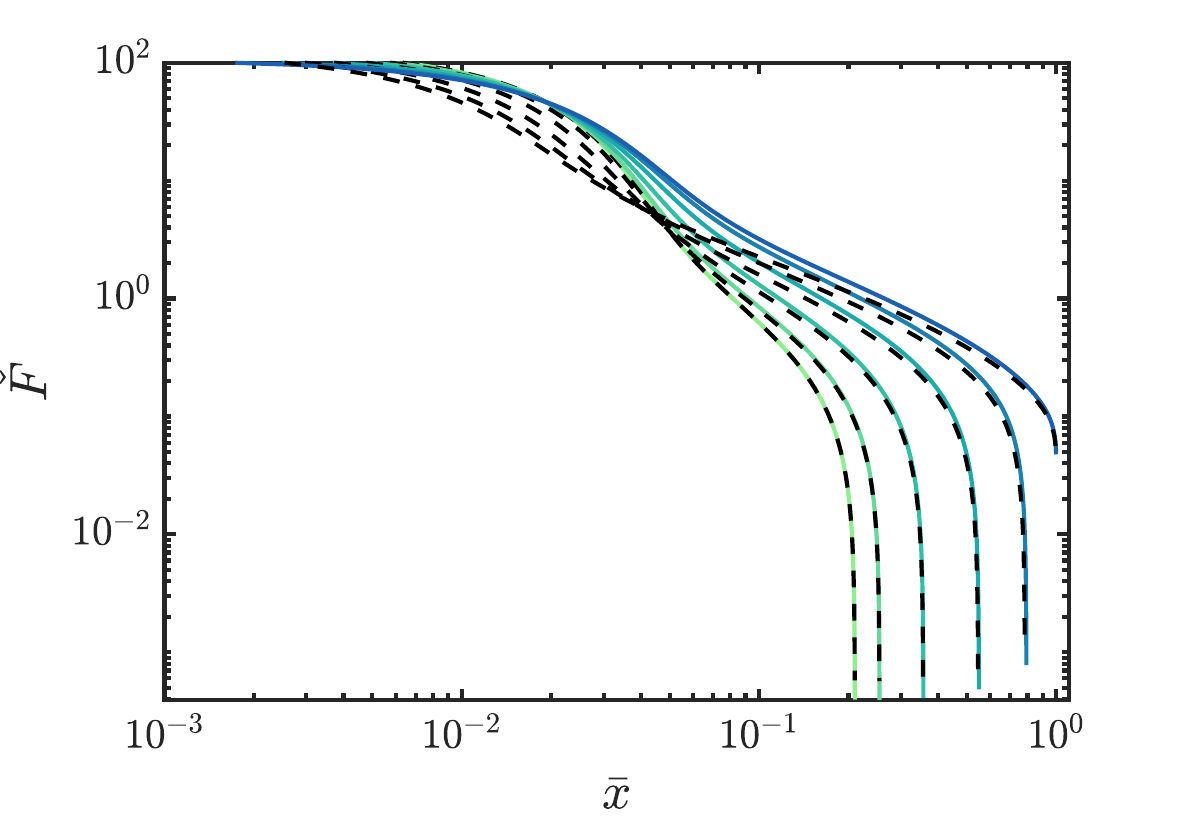}
    \end{subfigure}
    \\[-2\baselineskip]    
    \begin{subfigure}{0.49\textwidth}
        \caption{}
        \includegraphics[width = \textwidth]{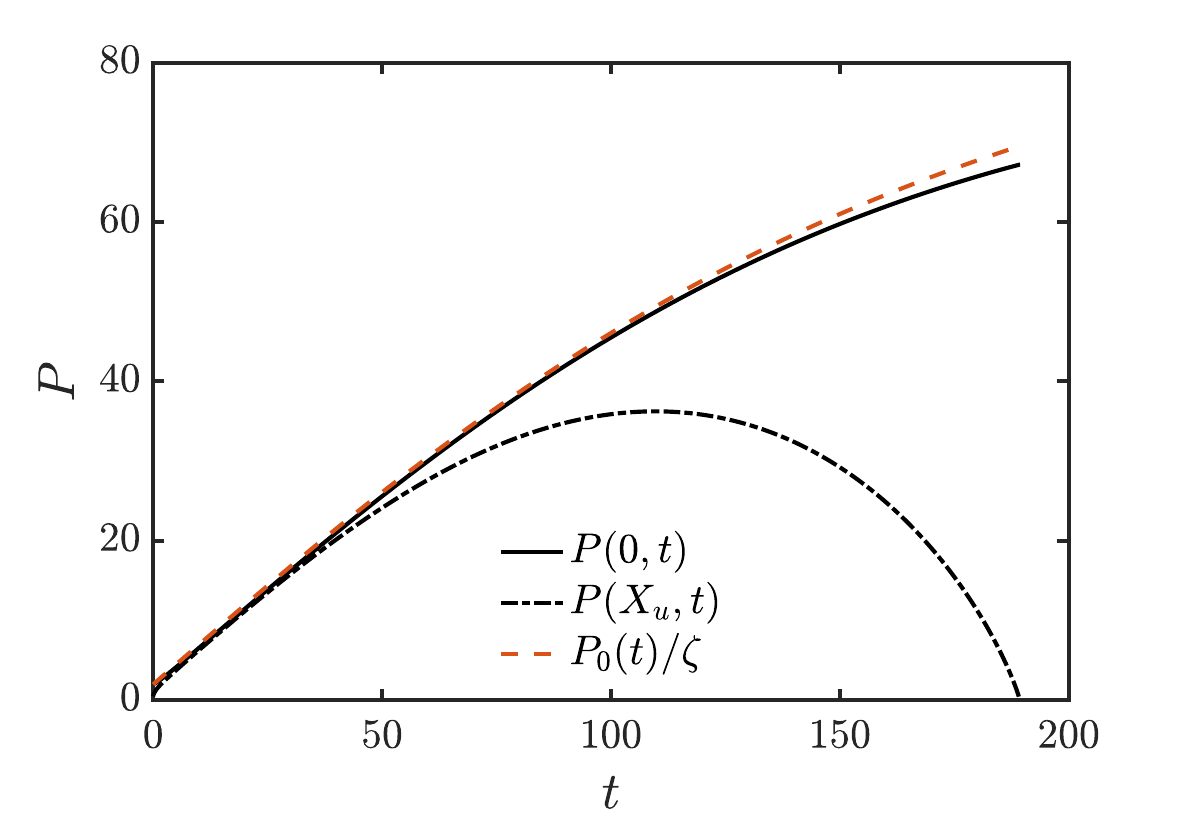}
    \end{subfigure}
    \hfill
    \begin{subfigure}{0.49\textwidth}
        \caption{}
        \includegraphics[width = \textwidth]{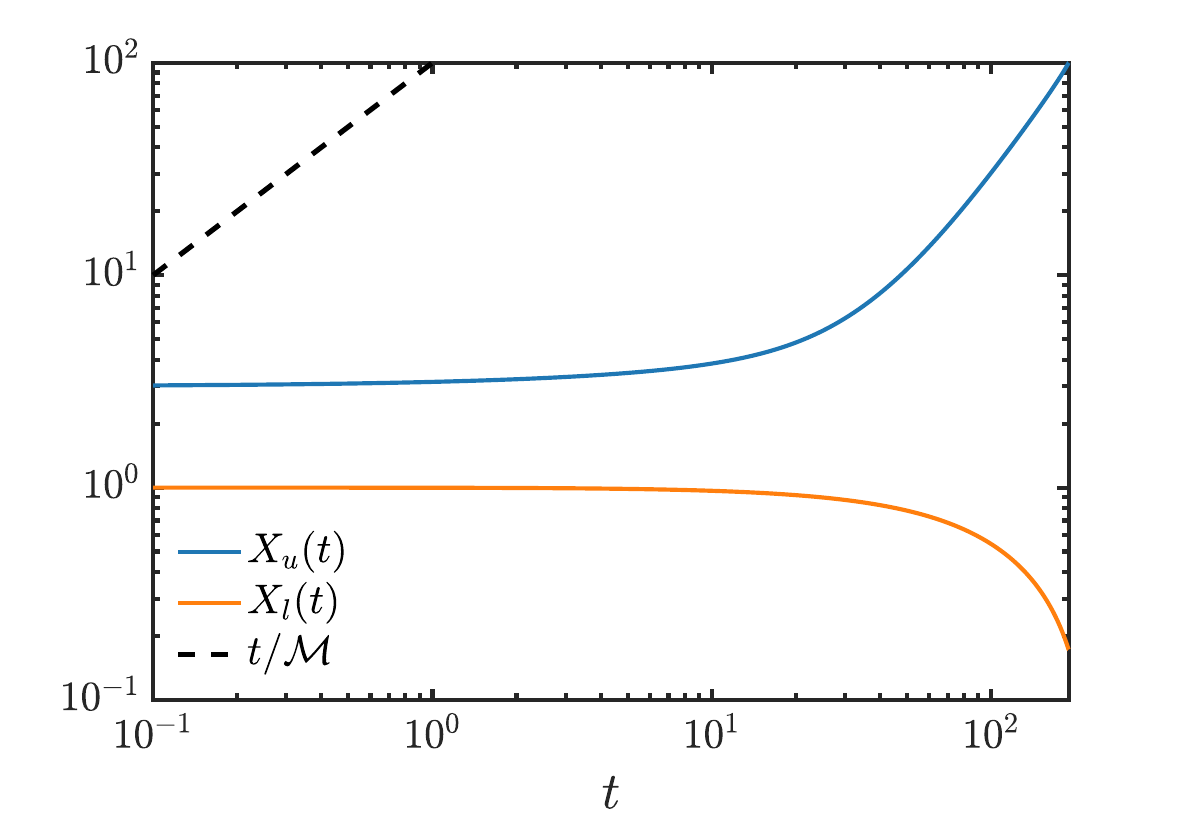}
    \end{subfigure}
     \caption{A simulation of (\ref{eq:put2}) and the outer problem (\ref{eq:put2toc}), for $\zeta = 1$, $\mathcal{M} = 10^{-2}$, and $\mathcal{L} = 100$. The solid curves in (a) and (b) show the full solution of \eqref{eq:put2} plotted in the outer coordinates of problem $\Pi_2$, $\check{P}(\bar{x},\check{t})$ and $\check{P}(\bar{x},\check{t})$, the dashed curves are numerical solutions of the outer equations \eqref{eq:put2toc}. The solutions are shown at times $\check{t} = \{0.8, 0.9, 1.1, 1.4, 1.7, 1.9\}$.  Solutions in (c) and (d) are from (\ref{eq:put2}) and are plotted using the format described in figure~\ref{fig:M_0.1_zeta_10-4}.
 }
     \label{fig:M_0.01_zeta_1}
\end{figure}

A quantitative comparison with the predictions of problem $\Pi_2$ is illustrated in figure~\ref{fig:M_0.01_zeta_1}, where solutions of (\ref{eq:put2}) are compared to solutions of the reduced model (\ref{eq:put2toc}).   The close agreement illustrates that the reduced model captures the dominant processes. It is helpful at this point to review the elements of the reduced model.  Eq.~(\ref{eq:put2toc}$a$) describes the evolution of the mass of gas, integrated across the thin film and represented by the density $\check{F}\check{P}$, advected by the pressure gradient $\check{P}_{\bar{x}}$.  The gas layer and the thicker liquid beneath share the same pressure gradient.  Eq.~(\ref{eq:put2toc}$b$) describes the evolution of the mass of liquid below the gas film. {Across this region $\check{P}_{\bar{x}\bar{x}}<0$}, indicting stretching of the liquid column that allows the gas thickness to increase locally.  Eq.~(\ref{eq:put2toc}$c$) captures the viscous pressure drop of the liquid column ahead of the gas bubble; as the column shortens, the pressure at the bubble tip falls.  Eq.~(\ref{eq:put2toc}$d$) is the kinematic relationship between contact line speed and pressure gradient: as the liquid column shortens, the pressure gradient across it rises and the contact line accelerates.  Finally, (\ref{eq:put2toc}$e$) couples mass input from the source to compression of gas in the bulk of the bubble. This is awkward to implement numerically: we implemented a local expansion that fails to capture the precise shapes of $\check{F}$ curves for smaller $\bar{x}$ in figure~\ref{fig:M_0.01_zeta_1}. 

Further reducing the viscosity ratio to $\mathcal{M} = 10^{-3}$ with $\zeta = 10^{-4} $ (figure~\ref{fig:M_10-3_zeta_10-4}) reveals the characteristic features of the ultra-low-viscosity regime described in Appendix~\ref{app:verylowviscosity}. In this regime, most of the interface remains stationary throughout the evolution, except for the thin film along the upper boundary; this is so narrow (figure~\ref{fig:M_10-3_zeta_10-4}b) that the source supplies a negligible flux to this region. Consequently, the pressure {rises} within the main gas bubble at a rate that is approximately linear in time when $\mathcal{Q}=1$ (figure~\ref{fig:M_10-3_zeta_10-4}c); the slight deviation of $P(0,t)$ from linearity reflects the slight expansion of the gas domain as the thin film elongates.  An analytical solution of the reduced-order model (\ref{eq:put24}) for $\mathcal{Q}=1$ implies that
\begin{equation}
\label{eq:xulv}
    {X_u}(t)=3\mathcal{L}\left[1-\left(1-\frac{t^2}{3L_0 \zeta \mathcal{L}^2}\right)^{1/2}\right],\quad \left(0<t<\mathcal{L}\sqrt{5L_0\zeta/3}\right).
\end{equation}
In this limit, spreading is driven primarily by the build-up of pressure in the gas bubble rather than directly from the mass flux delivered by the source. As a result, the position of the upper contact line $X_u$ initially grows quadratically in time. The approximation (\ref{eq:xulv}) agrees well with the full numerical simulation of (\ref{eq:put2}) across the entire evolution (figure~\ref{fig:M_10-3_zeta_10-4}d).

\begin{figure}
    \centering
    \begin{subfigure}{0.49\textwidth}
        \caption{}
        \includegraphics[width = \textwidth]{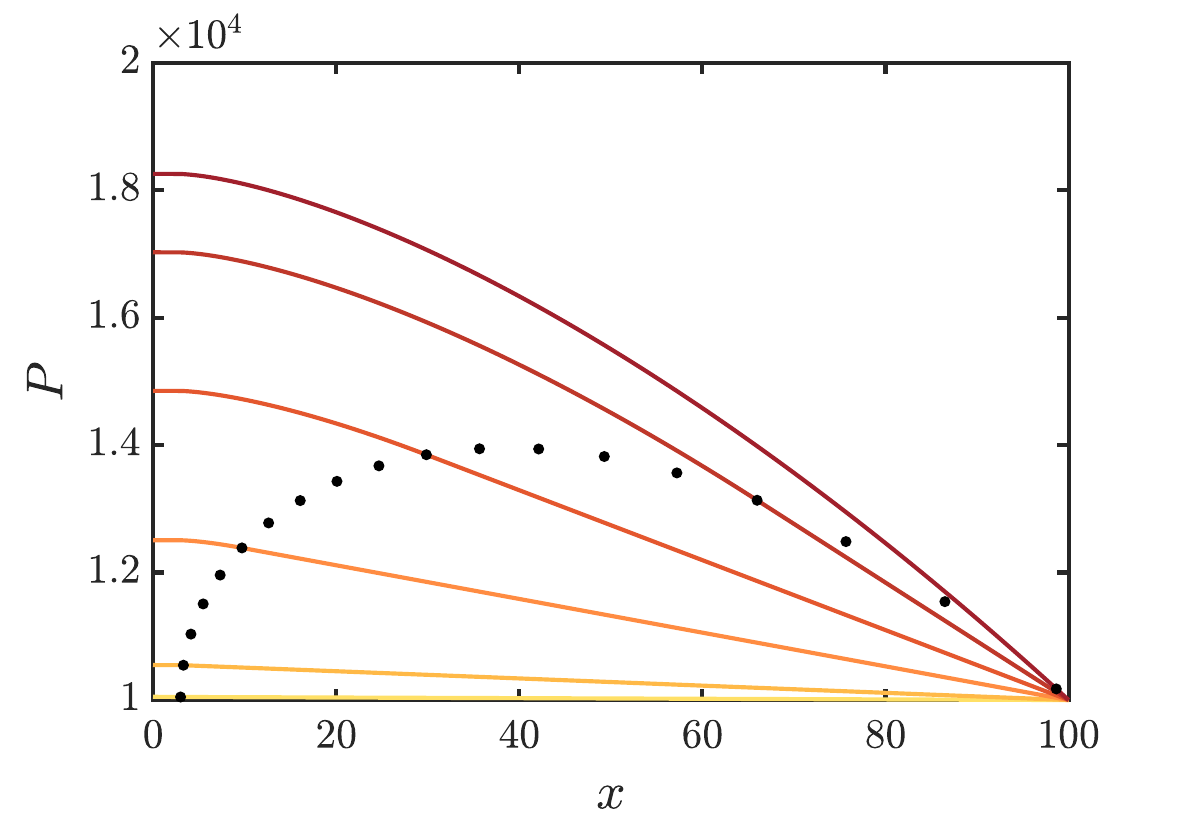}
    \end{subfigure} 
    \hfill
    \begin{subfigure}{0.49\textwidth}
        \caption{}
        \includegraphics[width = \textwidth]{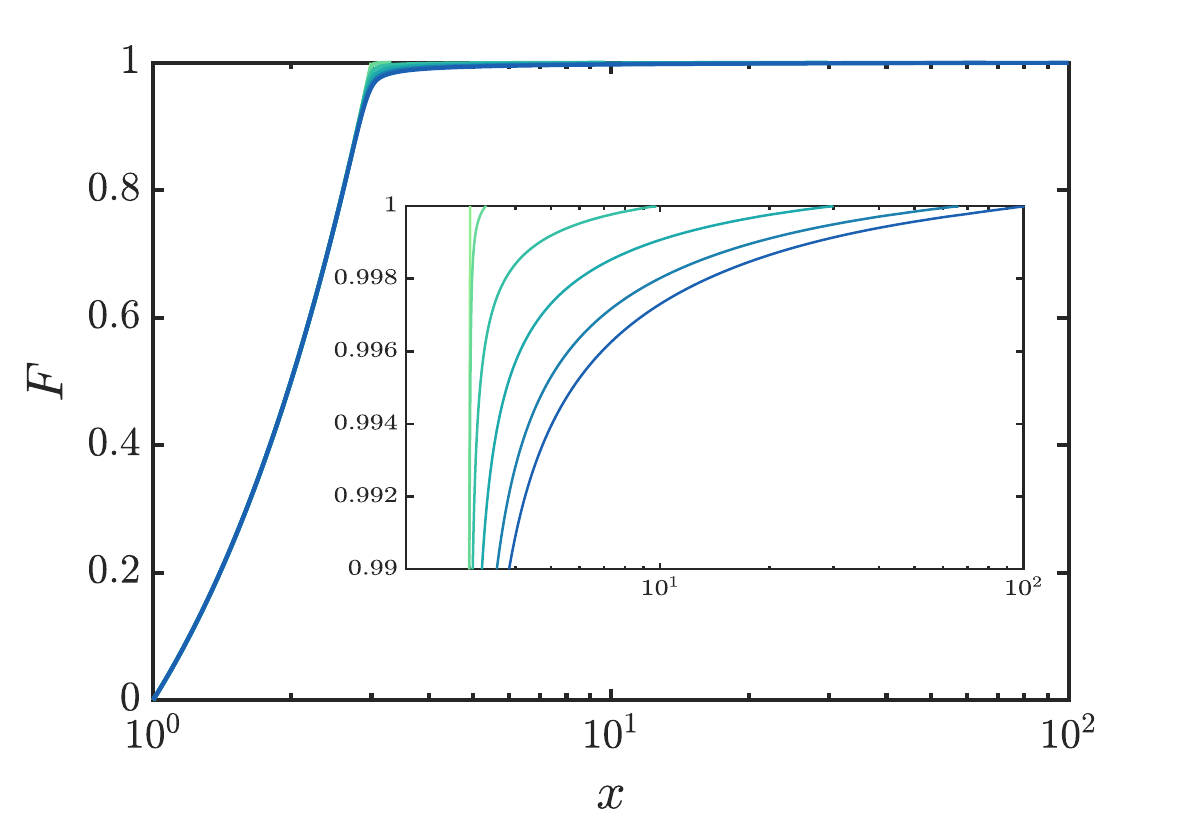}
    \end{subfigure}
    \\[-2\baselineskip]    
    \begin{subfigure}{0.49\textwidth}
        \caption{}
        \includegraphics[width = \textwidth]{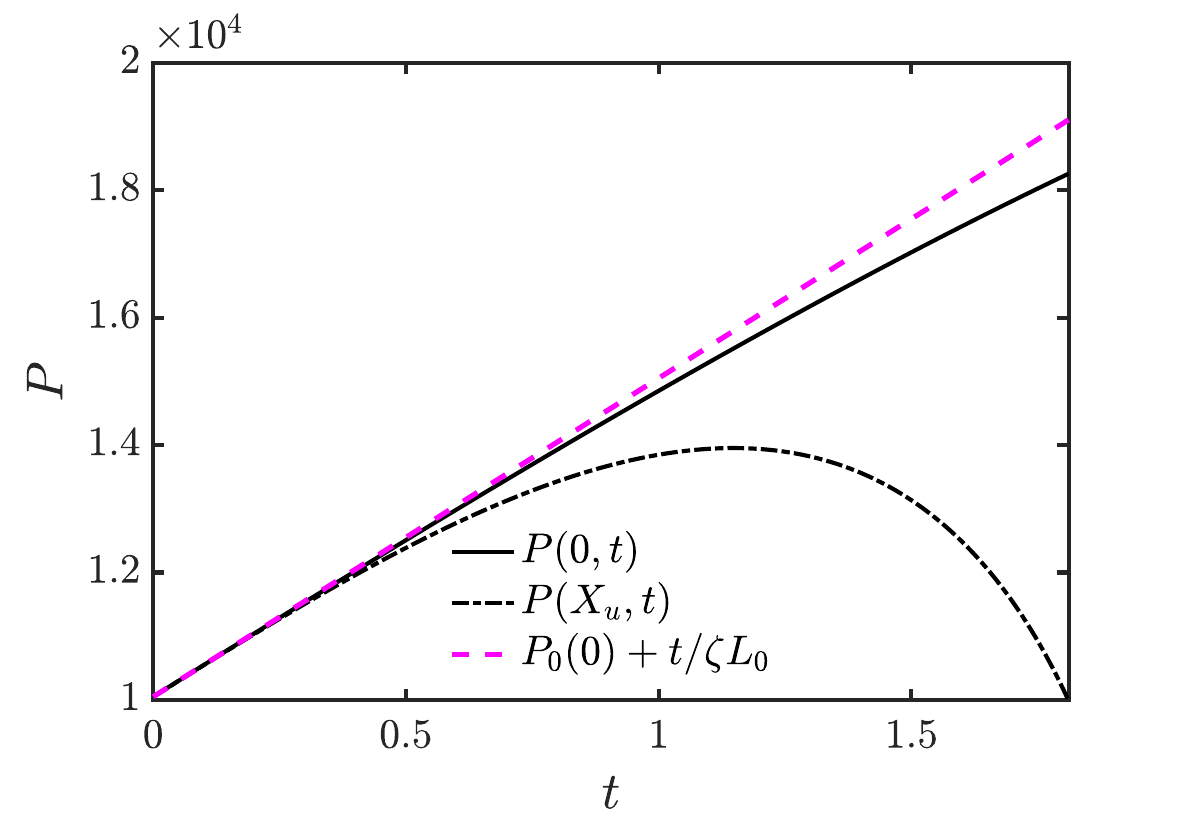}
    \end{subfigure}
    \hfill
    \begin{subfigure}{0.49\textwidth}
        \caption{}
        \includegraphics[width = \textwidth]{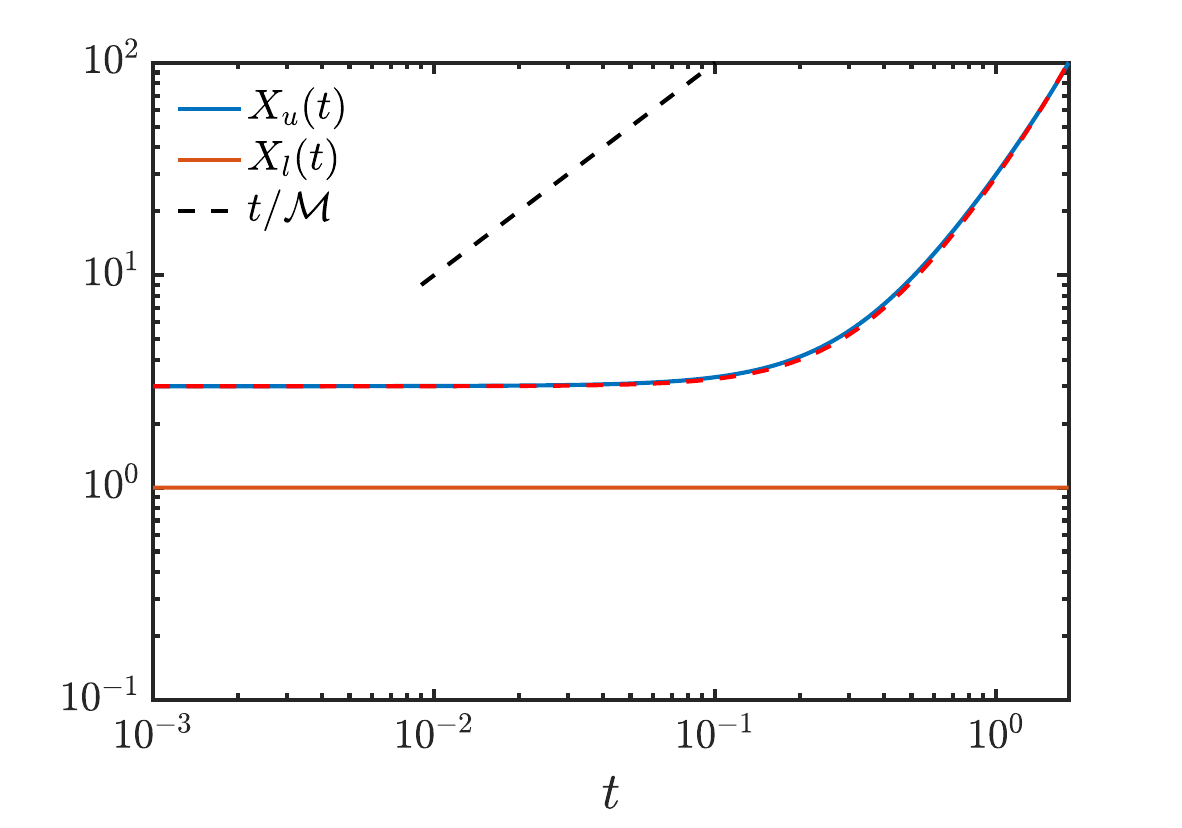}
    \end{subfigure}
     \caption{
A simulation of (\ref{eq:put2}) in the ultra-low viscosity regime, with $\zeta = 10^{-4}$, $\mathcal{M} = 10^{-3}$, and $\mathcal{L} = 100$. Solutions are plotted using the same format as figure~\ref{fig:M_0.1_zeta_10-4}. (a,b) Evolution of the pressure $P(x,t)$ and interface height $F(x,t)$ at times $t = \{0, 0.1, 0.5, 1, 1.5, 1.8\}$. The inset in (b) shows $F$ where it is close to unity.  The dashed pink curve in (c) is the approximation of the linear pressure rise within the main gas bubble, given by (\ref{eq:put24}$e$).  The dashed red line in (d) is the approximation given by (\ref{eq:xulv}).}
     \label{fig:M_10-3_zeta_10-4}
\end{figure}
\begin{figure}
\begin{subfigure}{0.49\textwidth}
        \caption{}
        \includegraphics[width = \textwidth]{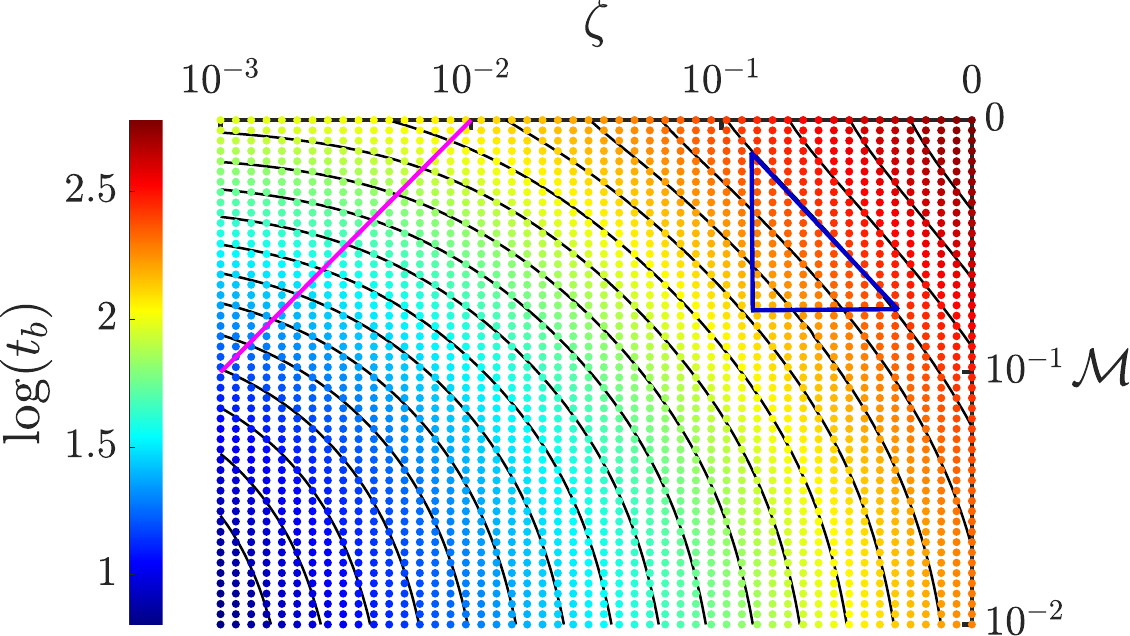}
\end{subfigure}
\hfill
\begin{subfigure}{0.49\textwidth}
        \caption{}
        \includegraphics[width = \textwidth]{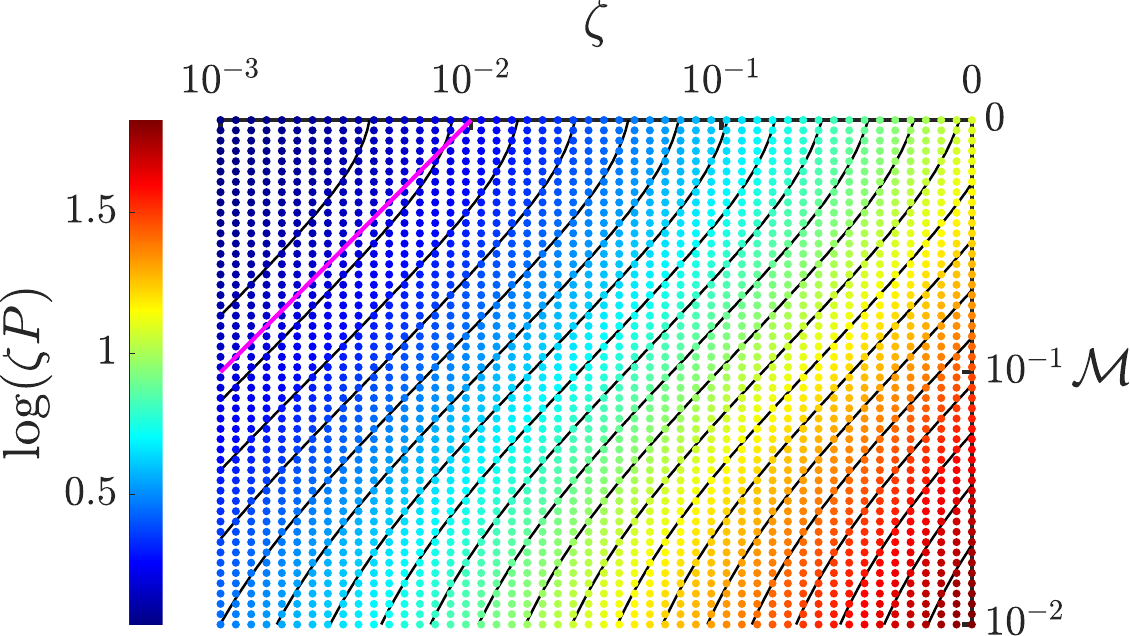}
\end{subfigure}
\caption{Contour plots showing (a) breakthrough times $t_b$ and (b) density $\zeta P$ at the origin at the breakthrough time, in $(\zeta, \mathcal{M} )$-parameter space. Dots capture data from 2500 simulations conducted with $\mathcal{L} = 100$.  Black contour lines, derived from interpolated data, are evenly spaced on a logarithmic scale. The magenta line illustrates the approximate location of the asymptotic boundary $\Pi_{12}$ separating region $\Pi_1$ to the left from region $\Pi_2$ to the right (regions are illustrated in figure~\ref{fig:overview}).The blue triangle in (a) indicates a slope of $-1$, illustrating the predicted scaling $t_b \sim (\mathcal{L}^3 \mathcal{M} \zeta )^{1/2}$. The asymptotic boundary $\Pi_{23}$ lies along $\mathcal{M}\sim \mathcal{L}^{-1}$ at the base of map.  $t_b$ and $\zeta P$ are predicted to be independent of $\mathcal{M}$ in region $\Pi_3$, beneath this boundary (Appendix~\ref{app:verylowviscosity}).}
    \label{fig:Contour_tb}
\end{figure}

Computations across parameter space show how the breakthrough time $t_b$ increases with both $\zeta$ and $\mathcal{M}$ (figure~\ref{fig:Contour_tb}a).  To the left of the line $\zeta \mathcal{L} = \mathcal{M}$, the flow approaches the incompressible limit (problem $\Pi_1$), making the breakthrough time independent of compressibility. In this regime, $t_b\approx \mathcal{L} \mathcal{M}$, which is reflected in the horizontal levelling off of the contours. To the right, problem $\Pi_2$ predicts that compressible and viscous effects determine the breakthrough time, with $t_b \sim (\mathcal{L} / \mathcal{M} \zeta)^{1/2}$. This scaling implies that the contours should have a slope of $-1$, as demonstrated in the top right corner of the figure. Closer to the lower boundary of region $\Pi_2$ (\hbox{i.e.} the lower boundary of figure~\ref{fig:Contour_tb}a), the contours become distorted due to the influence of very small viscosity effects. In the limit of vanishing gas viscosity (region $\Pi_3$ in figure~\ref{fig:overview}), the breakthrough timescale $t_b \sim (\zeta \mathcal{L}^2)^{1/2}$ becomes independent of $\mathcal{M}$ (Appendix~\ref{app:verylowviscosity}), manifesting as the contours becoming vertical.

Figure~\ref{fig:Contour_tb}(b) shows how $\zeta P$, the gas density measured relative to $\rho_{g0}$, increases with $\zeta$ and decreases with $\mathcal{M}$ across region $\Pi_2$.  The dominant density in the incompressible region ($\Pi_1$, to the left of the figure) is $O(1)$, with a correction of $O(\zeta \mathcal{L}/\mathcal{M})$ (arising from the pressure $\hat{P}_1$ in (\ref{eq:put2toa})).  Accordingly, the density contours share the slope of the line $\mathcal{L}\zeta =\mathcal{M}$ near this region.  Across region $\Pi_2$, $\zeta P$ is $O(\zeta \mathcal{L}/\mathcal{M})^{1/2}$ (see (\ref{eq:put2toc})) and therefore the contours maintain unit slope.  However near the lower boundary of the figure, the contours become vertical, because $\zeta P=O(\mathcal{L}\zeta^{1/2})$ in the underlying region $\Pi_3$ (figure~\ref{fig:overview}), which is independent of $\mathcal{M}$ .

\begin{figure}
    \centering
    \begin{subfigure}{0.49\textwidth}
        \caption{}
        \includegraphics[width = \textwidth]{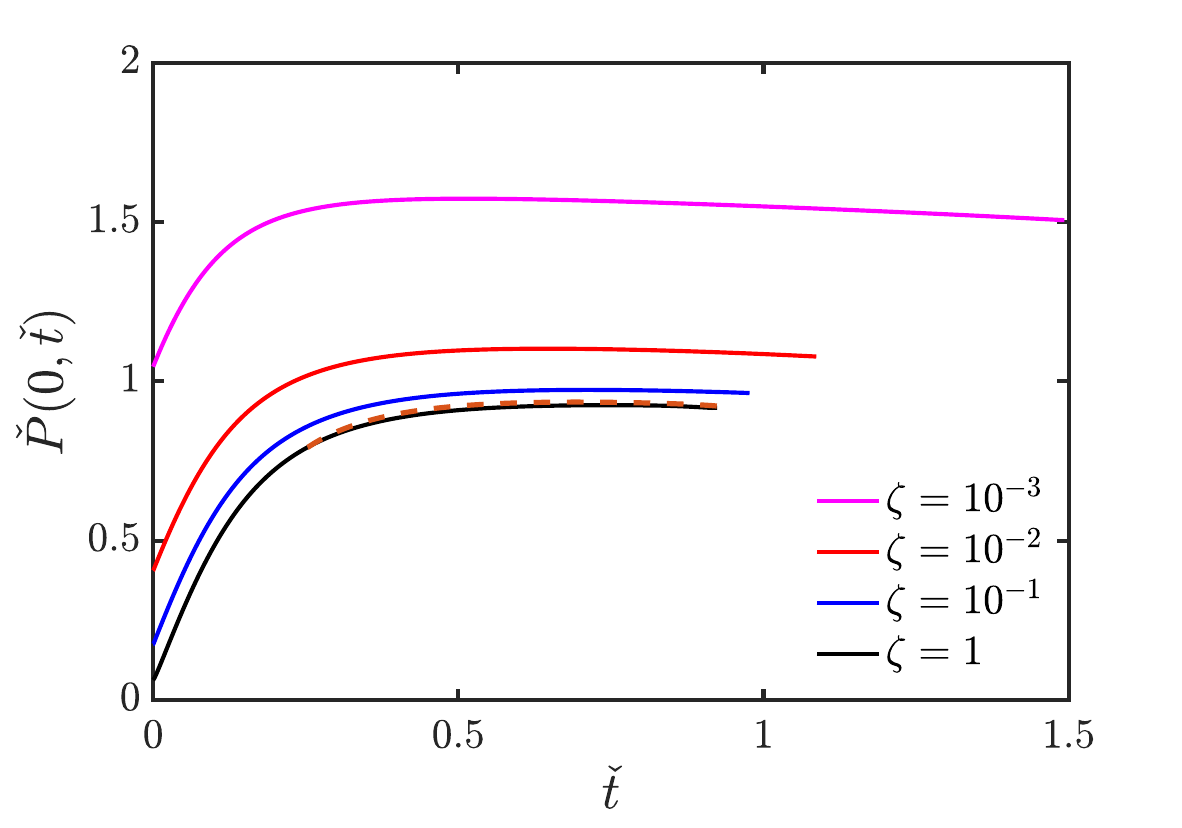}
    \end{subfigure} 
    \hfill
    \begin{subfigure}{0.49\textwidth}
        \caption{}
        \includegraphics[width = \textwidth]{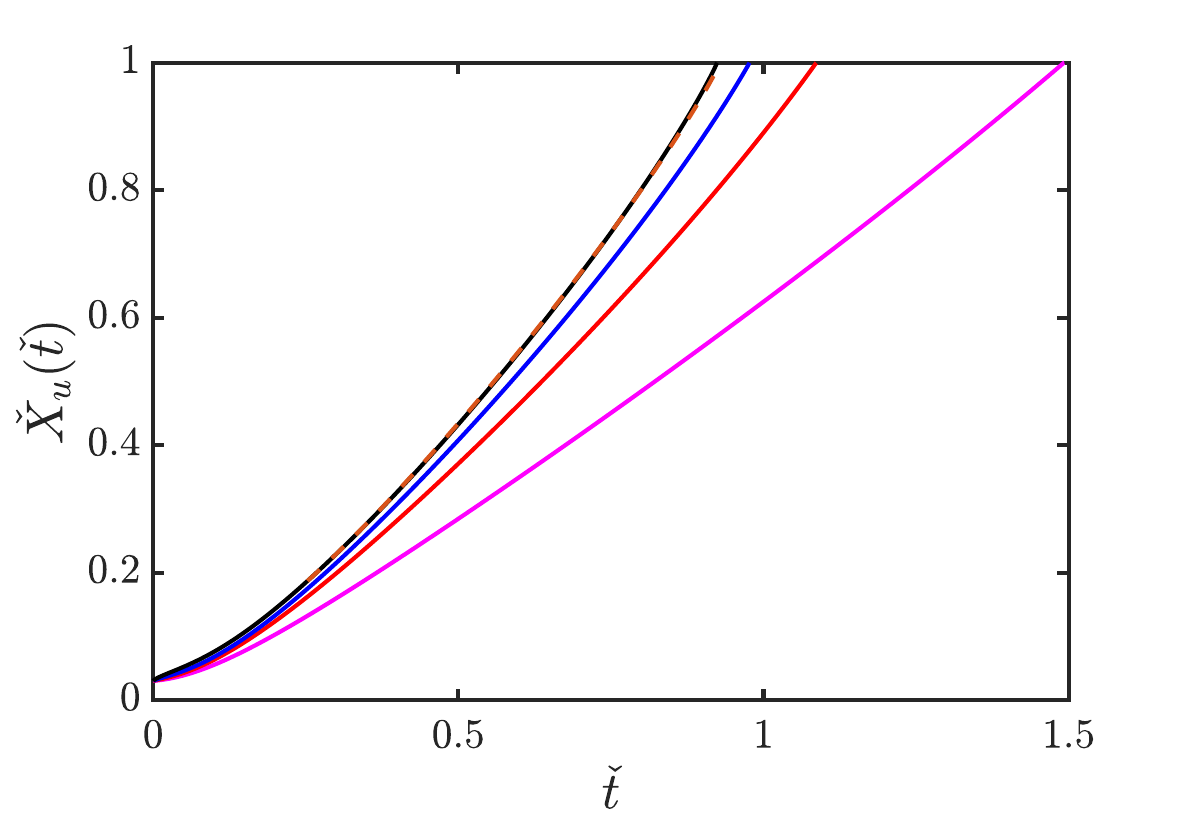}
    \end{subfigure}
    \\[-2\baselineskip]
    \begin{subfigure}{0.49\textwidth}
        \caption{}
        \includegraphics[width = \textwidth]{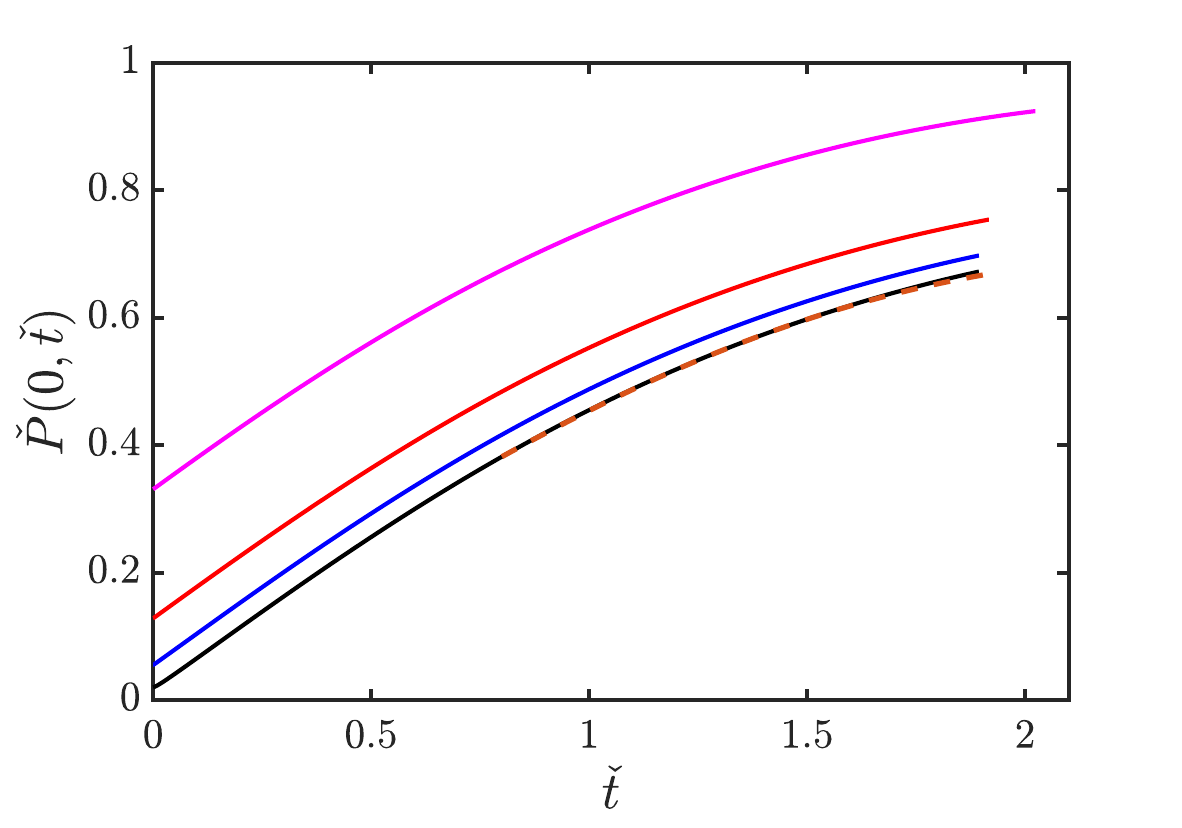}
    \end{subfigure}
    \hfill
     \begin{subfigure}{0.49\textwidth}
        \caption{}
        \includegraphics[width = \textwidth]{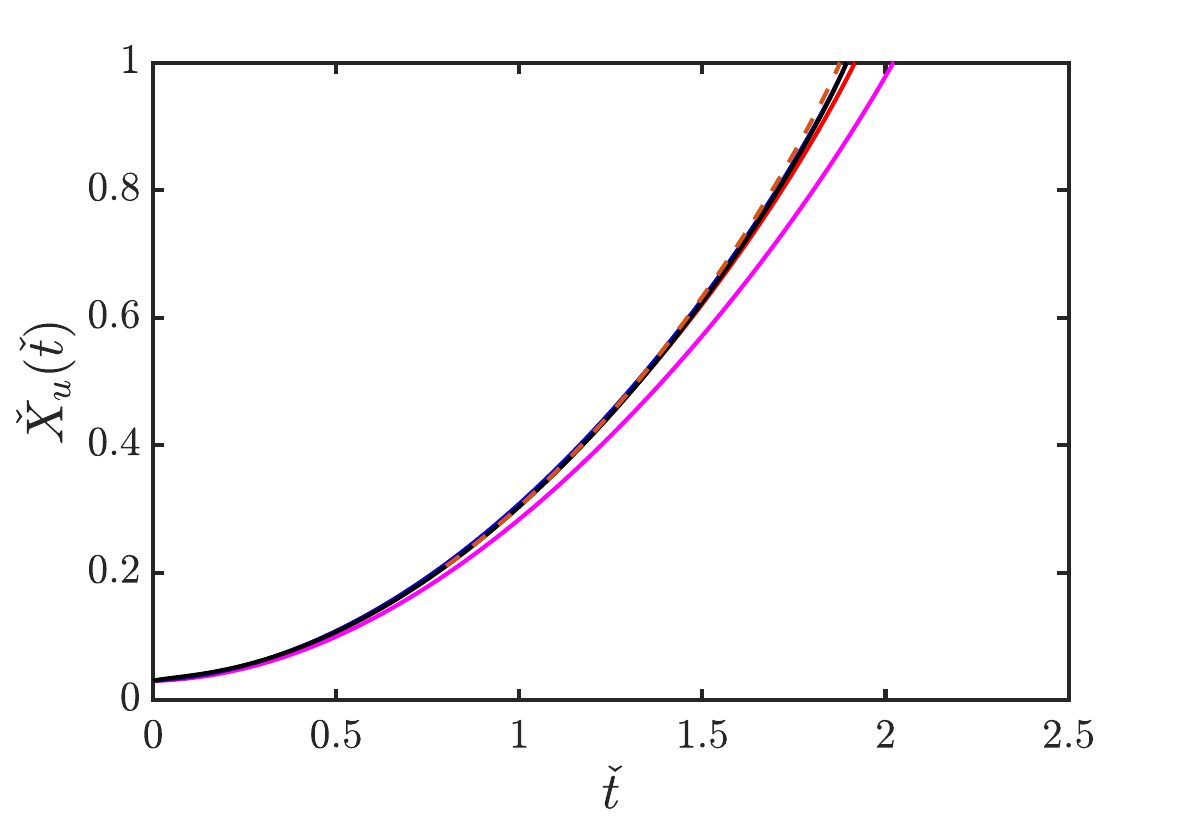}
    \end{subfigure}
    \\[-2\baselineskip]
     \begin{subfigure}{0.49\textwidth}
        \caption{}
        \includegraphics[width = \textwidth]{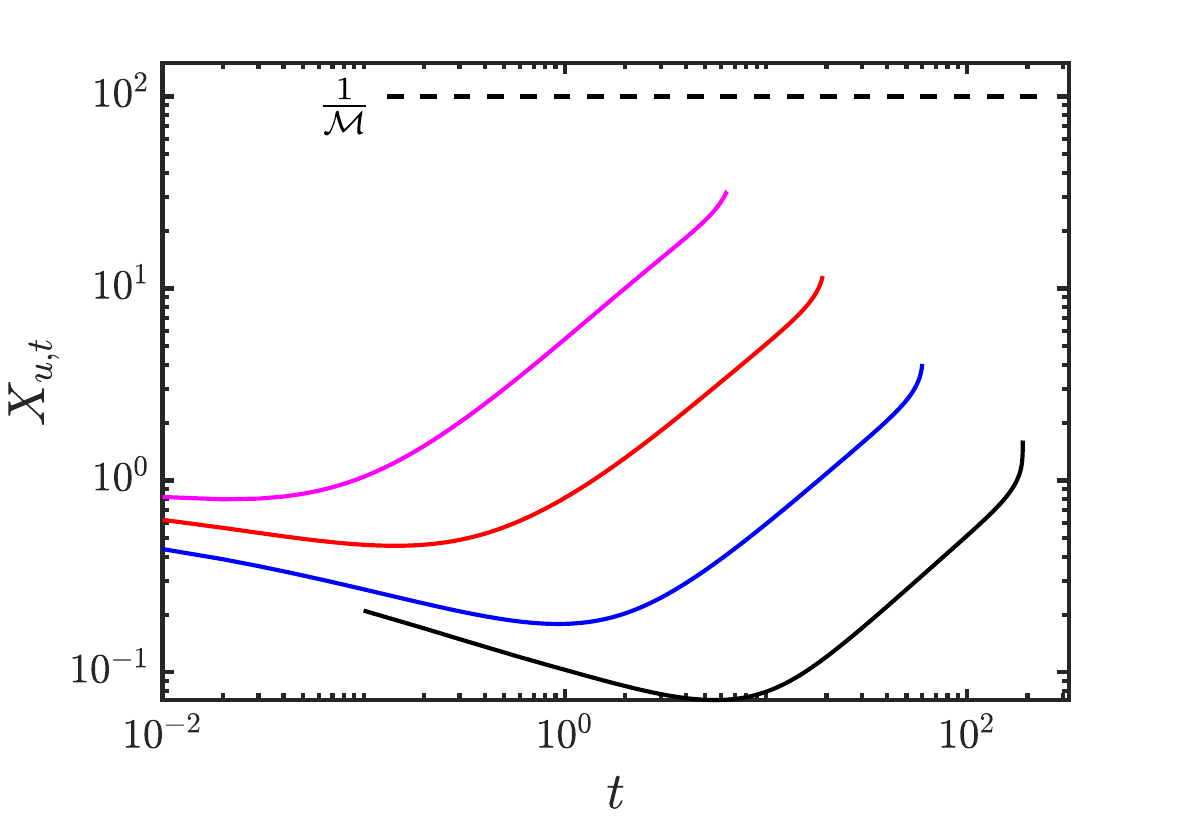}
    \end{subfigure}
    \hfill
    \begin{subfigure}{0.49\textwidth}
        \caption{}
        \includegraphics[width = \textwidth]{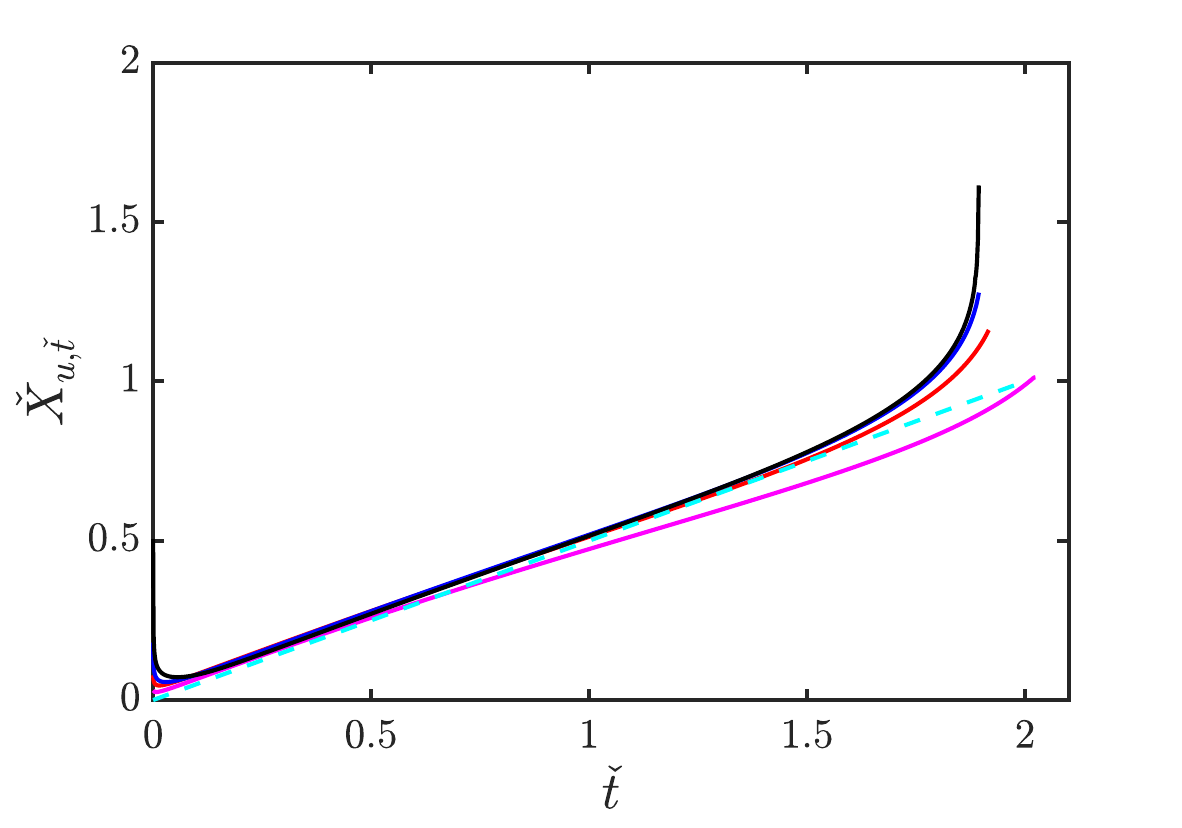}
    \end{subfigure}
     \caption{The solutions of (\ref{eq:put2}). (a) and (b) are for $\mathcal{M} = 0.1$, while (c), (d), (e), (f) corresponds to $\mathcal{M} = 0.01$. (a) and (c) show the pressure at the origin and (b) and (d) show the upper contact line, all replotted in the outer coordinates of (\ref{eq:put2toc}), given by $P = (\mathcal{L}/\mathcal{M} \zeta)^{1/2} \check{P}$,  $X_u = \mathcal{L} \check{X}_u$, and $t = (\mathcal{L}^3 \mathcal{M} \zeta)^{1/2} \check{t}$. The solid curves represent the full numerical solution of (\ref{eq:put2}), while the dashed orange curves correspond to the outer approximation obtained by numerically solving (\ref{eq:put2toc}). For both values of  $\mathcal{M}$, the outer problem was initialized using the full solution for $\zeta = 1$ at $t = 80$, by which time the late-time inner-outer structure has fully developed. (e) shows the speed of the upper contact line for $\mathcal{M} = 0.01$, which is compared to the speed of incompressible propagation (dashed black curve). In (f) we again replotted the curves presented in (e) in the outer coordinates, the dashed cyan curve $\check{t}/2$ is plotted for comparison. }
     \label{fig:CollapsedCurves}
\end{figure}

To investigate further the underlying scaling relationships, we replot simulation results using variables appropriate to region $\Pi_2$ (the blue region in figure~\ref{fig:overview}). Despite parameter variation over three orders of magnitude, the gas pressure at the source, the overall pressure distribution and the location of the upper contact line show only minor relative variation when plotted relative to $\check{t}=t/(\mathcal{L}^3\mathcal{M}\zeta)^{1/2}$ (figure~\ref{fig:CollapsedCurves}a-d); the outlying case with $\zeta=10^{-3}$ lies on the boundary with region $\Pi_1$.  The speed of the upper contact line is shown in regular coordinates in figure~\ref{fig:CollapsedCurves}(e) for multiple values of $\zeta$, with $\mathcal{M} = 0.01$, at the border between regions $\Pi_2$ and $\Pi_3$. Initially, the contact line undergoes a brief deceleration, followed by a phase of approximately constant acceleration. As it approaches the outlet, a sharp increase in speed is observed, driven by the rapid decrease in viscous resistance from the draining liquid. For reference, the constant incompressible contact line speed, given by $1/\mathcal{M}$, is indicated in figure~\ref{fig:CollapsedCurves}(e).  Across all values of $\zeta$ considered, the contact-line speed remains below this incompressible limit throughout.  When the same data are replotted in the outer coordinate of problem $\Pi_2$, the curves again collapse (figure~\ref{fig:CollapsedCurves}f) with $\check{X}_{u,\check{t}}\approx \tfrac{1}{2}\check{t}$. The quadratic dependence of contact-line location on time for parameters at the interface of regions $\Pi_2$ and $\Pi_3$ (figure~\ref{fig:CollapsedCurves}d,e,f), which (following (\ref{eq:xulv})) we attribute to dynamic pressure changes at the source, contrasts with the flux-driven linear dependence in the incompressible limit (\ref{eq:clsp}).

\subsection{Time-dependent injection}
\label{sec:unsteady}
Finally, we briefly consider two cases of time-dependent injection to examine how variations in the injection rate affect the total mass stored in the channel at the breakthrough time. Specifically, we investigate an increasing rate, $\mathcal{Q}(\Omega t) = 1 + \Omega t$, and a decreasing rate, $\mathcal{Q}(\Omega t) = 1 - \Omega t$, as defined in (\ref{eq:massg}). Since the breakthrough time is not known \textit{a priori}, we take $\Omega \ll 1$ to prevent mass withdrawal from the system.

Figure~\ref{fig:UnsteadyInj} compares the pressure evolution at the source and at the upper contact line for constant and time-dependent injection rates, with $\mathcal{L} = 100$, $\mathcal{M} = 10^{-2}$, $\zeta = 10^{-1}$ and $\Omega = 10^{-2}$ (on the region $\Pi_2/\Pi_3$ boundary). The initial pressure field (\ref{eq:PIC}) may be substituted into (\ref{eq:InMas}) to evaluate the initial mass of gas in the channel, giving $M_{g0} = 3$  for $\zeta = 10^{-1}$. For steady injection, the breakthrough time is $t_b \approx 60$, corresponding to a final mass of $M_g(t_b) \approx M_{g0} + 60$. When the injection rate increases over time, the pressure rises more rapidly, and the elevated pressure scale and steeper pressure gradient at $X_u$ lead to faster spreading, resulting in an earlier breakthrough at  $t_b \approx 55$  with  $M_g(t_b) \approx M_{g0} + 71$. Conversely, a decreasing injection rate shortens the pressure rise time, allowing the pressure to saturate near the source by the time that breakthrough occurs (figure~\ref{fig:UnsteadyInj}a).  Quicker pressure saturation is consistent with the early-time problem in (\ref{eq:Boyles}), where a decreasing injection rate leads to a decline in $P_{0,t}$. Spreading is also slowed, so that the breakthrough time is delayed to $t_b \approx 68$, at which time the mass in the channel ($M_g(t_b) \approx M_{g0} + 45$) is less than the cases with steady or increasing injection rate. These findings suggest that an increasing injection rate enhances storage efficiency, since the total gas volume remains the same at breakthrough across all cases according to (\ref{eq:dimVol}). 

\section{Discussion}
{\color{black}\subsection{Dimensional scales}}
\label{sec:discussion}

We have used a simplified model of gas injection into a confined liquid-filled porous medium in order to investigate the role of gas compressibility.  In a long-wave limit, the spreading of the gas is modelled by coupled evolution equations (\ref{eq:put2}) for the gas pressure and interface height.  Focusing on the regime $\mathcal{M}\ll 1$ (high gas mobility, leading to spreading being confined to a thin layer of gas that advances above the brine), $\mathcal{L}\gg 1$ (a long domain length) and $\zeta\ll 1$ (weak gas compressibility), we identified three dominant regions of parameter space (figure~\ref{fig:overview}) in which spreading is regulated by different dominant balances: in region $\Pi_1$, compressible effects are transient and spreading is essentially incompressible, being regulated by the source strength; in $\Pi_2$, compressible effects regulate dynamics in the thin gas film and hence the overall spreading rate; in $\Pi_3$ and $\Pi_4$, injection drives a steady increase of the pressure in the main gas bubble and spreading is confined to an ultra-thin film of gas. {\color{black} Figures~4--11 provide both qualitative and quantitative validation of the behaviour across all regimes, at parameter values indicated in figure~\ref{fig:overview}.} A table summarising these codimension-0 regions of $(\zeta,\mathcal{M})$-parameter space, their codimension-1 boundaries and the codimension-2 points where the boundaries intersect is provided in table~\ref{tab:regionsummary}.

\begin{figure}
    \centering
    \begin{subfigure}{0.49\textwidth}
        \caption{}
        \includegraphics[width = \textwidth]{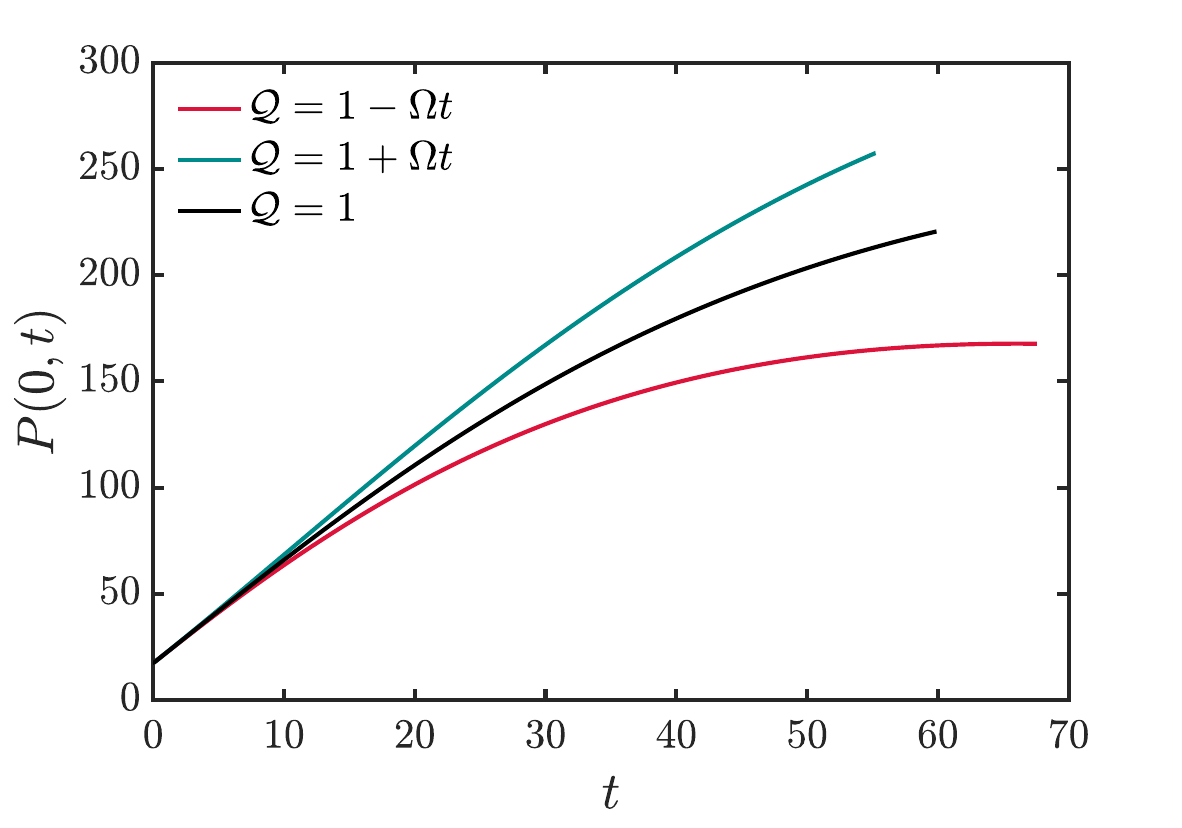}
    \end{subfigure}
    \hfill
    \begin{subfigure}{0.49\textwidth}
        \caption{}
        \includegraphics[width = \textwidth]{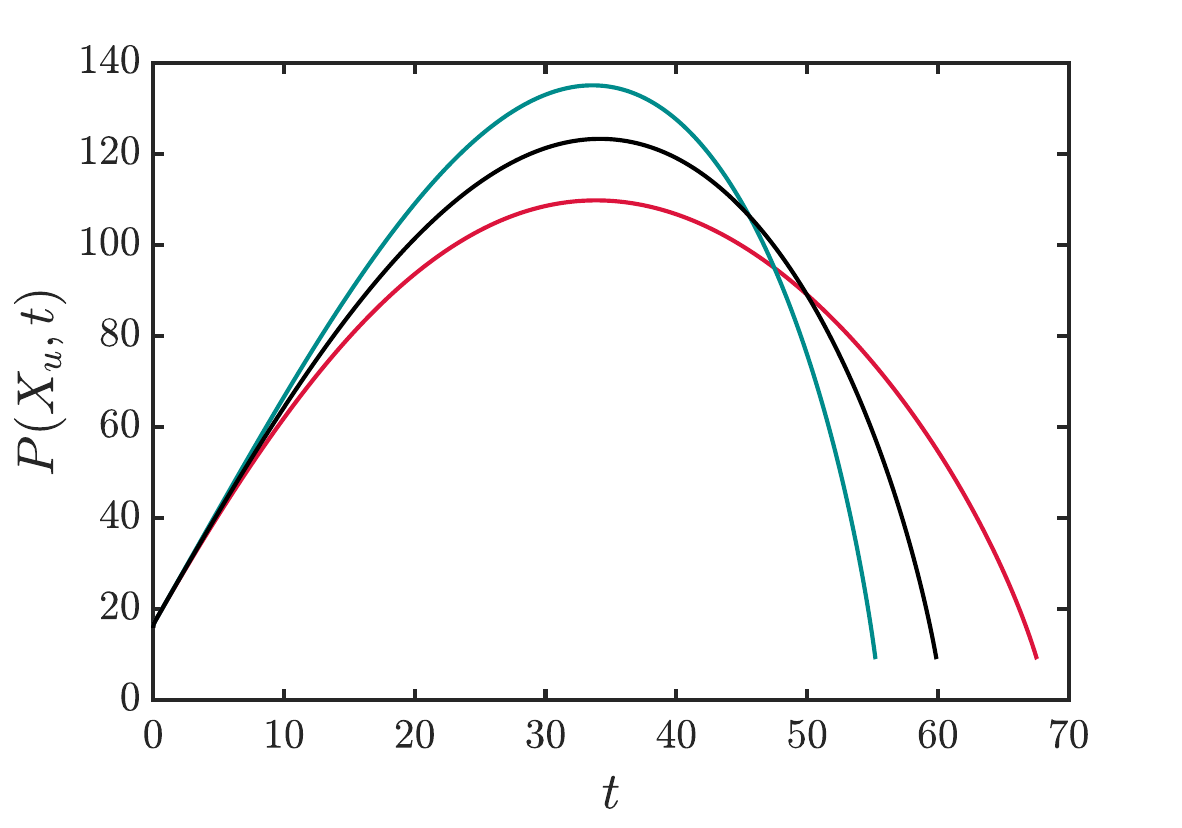}
    \end{subfigure}
     \caption{Numerical simulations of (\ref{eq:put2}) for time-dependent injection rates, with $\mathcal{M} = 10^{-2}$, $\zeta = 0.1$, $\mathcal{L} = 100$, and $\Omega = 10^{-2}$. The corresponding steady-injection case is shown for comparison. (a) Evolution of the pressure at the origin. (b) Evolution of the pressure at the upper contact line.
 }
     \label{fig:UnsteadyInj}
\end{figure}

It is helpful to revisit the predicted time and pressure scales in these regions,  expressing them using dimensional quantities.  In the incompressible region $\Pi_1$, the scales for pressure, breakthrough time and rise time are
\addtocounter{equation}{1}
\begin{equation}
\label{eq:pi1scales}
    p_g^* - p_{g0} \sim \frac{q \mu_w L}{k_0 \rho_{g0} H} , \quad t_b^* \sim \frac{  \rho_{g0} H L \mu_g}{q \mu_w}, \quad t_r^* \sim \frac{\mu_w L \rho_w g H^2}{q \mu_g c^2},
    \tag{4.1$a$,$b$,$c$}
\end{equation}
respectively. Here, the pressure scale (above the reference pressure $p_{g0}$ at the downstream end of the aquifer) is primarily set by the viscous pressure drop along the liquid-filled column. The breakthrough time (\mainref{eq:pi1scales}{4.1}$b$) is set by the source-driven transit time, $\rho_{g0} H L / q$, reduced by the mobility ratio, $\mu_g / \mu_w$, to account for the formation of a thin gas film over the liquid, which facilitates faster spreading. The much shorter scale for the pressure rise time (\mainref{eq:pi1scales}{4.1}$c$) is governed by a balance between viscous dissipation, gas compressibility, and buoyancy. The full dimensional gas pressure field for the incompressible problem $\Pi_1$ is, from (\ref{eq:poutinc}),
\begin{equation}
    p_g^* = p_{g0} + \frac{q \mu_w L \mathcal{Q}
    }{k_0 \rho_{g0} H} - \frac{q^2 \mu_w^2 \mathcal{V}^* \mathcal{Q}}{3\rho_{g0}^2 \mu_g k_0 H^2} \left[1 + 2 \left(\frac{ \mu_g \rho_{g0} H}{\mu_w q} \frac{x^*}{\mathcal{V^*}} \right)^{3/2} \right], \quad \mathcal{V}^* = \int_0^{t^*} \mathcal{Q} \, \mathrm{d} t^*,
\end{equation}
for $x^*$ between the source and the upper contact-line location $X_u^*(t^*)= q \mu_w \mathcal{V}^*/(\rho_{g0} H \mu_g)$.  The channel length $L$ dictates the large uniform pressure component needed to drive the motion; this builds in the gas through rapid compression over the timescale $t_r^*$. However, this pressure component is decoupled from the spatially non-uniform spreading pressure field, which remains independent of the domain length.

The corresponding dimensional scales in region $\Pi_2$ of figure~\ref{fig:overview}, when compressible effects become significant, are
\addtocounter{equation}{1}
\begin{equation}
\label{eq:pi2scales}
    p_g^* - p_{g0} + \rho_{g0} c^2 \sim \left[\frac{q \mu_w c^2L}{H k_0 } \right]^{1/2}, ~~ t_b^* \sim \left[\frac{H \mu_g^2 L^3}{q k_0 \mu_w c^2} \right]^{1/2}, ~~ t_r^* \sim \left[\frac{\rho_w^2 g^2 H^5 k_0 \rho_{g0}^2 \mu_w L }{\mu_g^2 q^3 c^2}\right]^{1/2}.
    {\tag{4.3$a$,$b$,$c$}}
\end{equation}
The pressure scale (\mainref{eq:pi2scales}{4.3}$a$) is given by the geometric mean of the viscous pressure drop along the liquid-filled channel and the compressible pressure scale $\rho_{g0}c^2$. The breakthrough time scale (\mainref{eq:pi2scales}{4.3}$b$) emerges from a leading-order balance between gas compressibility and viscous dissipation in both the liquid and gas. The shorter pressure rise time (\mainref{eq:pi2scales}{4.3}$c$) is determined by all physical effects in the model. The scales (\mainref{eq:pi2scales}{4.3}$a,b,c$) show how, for a more compressible gas (with smaller $c^2$), the pressure required to displace the liquid falls but the time taken to deliver the gas rises. As demonstrated in (\ref{eq:InMas}), the breakthrough time determines the mass of gas delivered when the source injection is steady, via $q t_b^*$.

\begin{table}
    \centering
    \begin{tabular}{c c c c }
    Limit  & \multicolumn{3}{c}{Attributes} \medskip\\ \cline{2-4}\noalign{\medskip}%\hline
$\Pi_{a}$ & $\mu_g\sim \mu_l$ & $q \mu_g L \sim k_0 \rho_{g0}^2 H c^2$ & $t_r\ll t_b$ \\
$\Pi_{b}$  & $q^3\mu_g^3L^3\sim k_0^3c^4\rho_{g0}^5 \rho_w g H^4$ & $q \mu_g^3 L \sim k_0 \rho_{g0} \rho_w g H^2 \mu_w^2 $ & Buoyancy near $X_l$; $t_r\ll t_b$ \\
$\Pi_{c}$  & $q\mu_g^2 L \sim k_0 \rho_{g0} \rho_w g H^2 \mu_w$ & $q^2 \mu_g^2 L^2\sim \rho_{g0}^3 c^2 k_0^2 \rho_w g H^3$ & Buoyancy near thin film base  \\
\\ 
$\Pi_{12}$   & $q \mu_w L \sim \rho_{g0}^2 c^2 k_0 H$ & Thin film; $t_r\ll t_b$ & Quasi-steady $P_b$, Flux-driven \\
     $\Pi_{23}$  & $q\mu_g^2 L \sim k_0 \rho_{g0} \rho_w g H^2 \mu_w$ & Compressible thin film & Flux-driven\\
$\Pi_{34}$  & $q^2 \mu_g^2 L^2\sim \rho_{g0}^3 c^2 k_0^2 \rho_w g H^3$ & 
Ultra-thin film, $t_r\sim t_b$ &  Dynamic $P_b$ \\
$\Pi_{14}$  & $\mu_g^2 \rho_{g0} c^2 \sim \mu_w^2 \rho_w g H$ & Incompressible thin film   & \\
\\
 $\Pi_1$ & Incompressible thin film & $t_r\ll t_b$ & Quasi-steady $P_b$,  Flux-driven\\
     $\Pi_2$ & Compressible thin film & $t_r\ll t_b$ & Quasi-steady $P_b$, Flux-driven\\
     $\Pi_3$ & Compressible ultra-thin film & $t_r\sim t_b$ & Dynamic $P_b$, Flux-driven \\
     $\Pi_4$ & Incompressible ultra-thin film & $t_r\sim t_b$ & Dynamic $P_b$, Pressure-driven \end{tabular}
    \caption{    { A summary of asymptotic limits.  
Codimension-2 problems $\Pi_a$, $\Pi_b$, $\Pi_c$ are formulated in distinguished asymptotic limits specified by two parameter balances (``$\sim$" denotes comparable magnitude), given using dimensionless parameters in figure~\ref{fig:overview} and here using dimensional parameters.  These general problems capture multiple competing physical effects; only a subset of descriptors are given in the final column.  Codimension-1 problems $\Pi_{12}$, $\Pi_{23}$, $\Pi_{34}$, $\Pi_{14}$ are specified by a single parameter balance; descriptors are given in columns 2 and 3; $P_b$ is shorthand for the near-source pressure of the gas bubble that connects to a thin gas film.  Codimension-0 problems $\Pi_1$, $\Pi_2$, $\Pi_3$, $\Pi_4$ occupy regions of $(\zeta,\mathcal{M})$-parameter space (figure~\ref{fig:overview}) and represent the most specific balances of physical effects; primary distinguishing features are summarised in columns 2-4.  }}
    \label{tab:regionsummary}
\end{table}

In region $\Pi_3$ (the green region in the parameter map, figure~\ref{fig:overview}; Appendix~\ref{app:verylowviscosity}), the dimensional scales simplify to
\addtocounter{equation}{1}
\begin{equation}
\label{eq:pi3scales}
        p_g^* - p_{g0} + \rho_{g0} c^2\sim  \frac{q \mu_g L}{k_0 \rho_{g0}H} \left( \frac{\rho_{g0} c^2}{\rho_w g H} \right)^{1/2}, \quad t_b^* \sim t_r^* \sim \frac{\rho_{g0} H L }{q} \left( \frac{\rho_w g H }{\rho_{g0} c^2} \right)^{1/2}.
        {\tag{4.4$a$,$b$}}
\end{equation}
The pressure scale is independent of the liquid viscosity but now depends on buoyancy: this is because the very thin gas layer displaces the liquid by slightly depressing it, overcoming hydrostatic pressure rather than displacing a large column of liquid along the channel against viscous resistance. The timescales for breakthrough and pressure rise are governed by the source-driven transit time, modulated by compressibility and buoyancy effects. Here we have taken the initial bubble length $L_0$ to be order unity. If instead we let $L_0 \rightarrow 0 $ the dimensionless scales become $P \sim (\mathcal{L}^2/ \zeta L_0)^{1/2}$, $t_b \sim (\zeta \mathcal{L}^2 L_0)^{1/2}$, showing how the pressure scale increases for a smaller initial bubble and the breakthrough time decreases; the scale for the breakthrough time is consistent with (\ref{eq:xulv}).  The final region, $\Pi_4$ in figure~\ref{fig:overview} (Appendix~\ref{app:verylowviscosity}), shares the pressure scale of region $\Pi_1$ (\mainref{eq:pi1scales}{4.4}$a$) and the timescale of region $\Pi_3$ (\mainref{eq:pi3scales}{4.4}$b$).

At the breakthrough time, the gas is predominantly stored in the film (the outer region) for $\mathcal{L}^{-1}\ll \mathcal{M}\ll 1$, and in the bubble (the inner region) for $\mathcal{ML}\ll 1$.  Scaled relative to $\rho_{g0}HL$, the mass of gas is $O(\mathcal{M})$ in region $\Pi_1$, rising to $(\mathcal{L} \mathcal{M} \zeta)^{1/2}$ in region $\Pi_2$.  In the incompressible case ($\Pi_1$), the viscosity ratio determines the film thickness, however the mass of gas is independent of other flow-related parameters. The mass of stored gas when $\mathcal{ML}\ll 1$, is $O(L_0\zeta^{1/2})$ and the dimensional mass is of order $LL_{g0}q\mu_g /[k_0 (\rho_{g0} \rho_w g H)^{1/2} c]$.  Here the pressure (and therefore gas density) is elevated by the flow of gas in the film. 

Having reviewed the primary regions, we comment briefly on the boundaries between them.  These are specified by the thresholds $\Pi_{12}$, $\Pi_{23}$, $\Pi_{34}$, $\Pi_{14}$ summarised in table~\ref{tab:regionsummary}.   A primary condition defining when compressibility strongly influences spreading is the requirement $\mathcal{M}\ll\mathcal{L}\zeta $ when $\mathcal{L}^{-2}\ll \zeta \ll \mathcal{L}^{-1}$, a threshold defined by the boundary between regions $\Pi_1$ and $\Pi_2$ in figure~\ref{fig:overview}.  Expressed in terms of input parameters, this requires that the pressure generated by viscous pressure drop along the channel is sufficiently large to cause appreciable density changes in the gas, \hbox{i.e.}
\begin{equation}\label{eq:Bound}
    \frac{q \mu_w L}{k_0\rho_{g0}H} \gg \rho_{g0} c^2.
\end{equation}
Thus compressibility effects are promoted by rapid injection into long, narrow, low-permeability channels.  An alternative pathway arises at much lower viscosity ratios, involving a transition from flux-driven to pressure-driven spreading on crossing boundary $\Pi_{14}$, followed by a transition to flux-driven spreading on crossing boundary $\Pi_{34}$ with   
\begin{equation}
\left(\frac{q \mu_g L}{k_0 \rho_{g0} H}\right)^2
    \gg (\rho_{g0} c^2) (\rho_w g H).
\end{equation}
Here viscous losses in the gas must exceed a threshold defined by compressive effects and buoyancy.

{\color{black}\subsection{Practical implications}}
The results for time-dependent injection rates (figure~\ref{fig:UnsteadyInj}) suggest that variations in injection rate may provide an effective means of controlling key aspects of flow and storage. Depending on the primary concerns at a given storage site, either an increasing or decreasing injection rate may offer advantages. An increasing injection rate can lead to elevated pressure buildup and more rapid spreading, potentially compromising reservoir integrity; however, if the reservoir is sufficiently robust, this strategy enables greater mass storage by the time of breakthrough. Conversely, a decreasing injection rate slows the spreading and results in less total mass delivered at breakthrough. Nevertheless, the associated reduction in pressure scale may reduce mechanical stress on the reservoir rock, making this approach preferable in settings where the risk of fracturing is a primary concern.

To illustrate the practical implications of compressibility, table~\ref{tab:Sleipner} reports ranges of the dimensionless parameters in (\ref{eq:put2}) for hydrogen storage, together with values calculated from the data provided in table~4 of \citet{zheng_self-similar_2019} for two CO$_2$ storage sites: the Sleipner aquifer in the North Sea \citep{boait_spatial_2012} and the In Salah aquifer in Algeria \citep{vasco_satellitebased_2010}. Estimates of the speed of sound for CO$_2$ were obtained from \citet{NIST}. For hydrogen, the ranges were computed from the typical dimensional parameters presented in table~\ref{tab:kd}, yielding dimensionless parameters spanning several orders of magnitude. This reflects the diversity of potential storage sites in terms of geometry, thermodynamic conditions, and pore-scale permeability. The lower end of $\mathcal{L}$, corresponding to extreme cases of very slow injection into a short, highly permeable, and deep channel, lies outside the validity of the present model. Therefore, restricting attention to $\mathcal{L} \gtrsim 100$ places hydrogen in the compressible regime $\Pi_2$ (based on the threshold $\Pi_{12}$; $\mathcal{M} \ll \theta$), and potentially close to the boundary $\Pi_{23}$ where ultra-low viscosity effects may influence spreading. The parameter ranges in table~\ref{tab:Sleipner} for Sleipner and In Salah data places the CO$_2$ spreading in regions $\Pi_1$ and $\Pi_2$, respectively. This suggests that spreading at Sleipner is effectively incompressible, whereas at In Salah compressibility may be significant, thereby slowing the rate of plume evolution.
\begin{table}
  \begin{center}
  \begin{tabular}{l@{\hspace{0.5cm}}l@{\hspace{0.5cm}}l@{\hspace{0.5cm}}l}
      Parameters  & Hydrogen &CO$_2$ at Sleipner &CO$_2$ at In Salah\\[4pt]
       $\mathcal{\zeta} = \Delta \rho g H/(\rho_{g0} c^2)$ & $10^{-2}$ -- 1 & $10^{-4}$  & $10^{-3}$\\[2pt]
      $\mathcal{M} = \mu_g/\mu_w$ & 10$^{-2}$  & 0.1 & 0.2 \\[2pt]
       $\mathcal{L} = q \mu_g L/(k_0 \rho_{g0} \Delta \rho g H^2)$ & $1$ -- $10^7$ & 350 & 10$^3$ \\[2pt]  
      \end{tabular}
  \caption{{ Estimated dimensionless parameter ranges for hydrogen and CO$_2$. The hydrogen ranges (column 2) are derived from table~\ref{tab:kd}, reflecting the wide variability in geometry, thermodynamic conditions and permeability of potential underground storage sites. Column 3 lists estimates for CO$_2$ storage at the Sleipner aquifer in the North Sea \citep{boait_spatial_2012}, based on the data in table 4 of \citet{zheng_self-similar_2019}. Column 4 gives corresponding estimates for CO$_2$ storage at the In Salah aquifer in Algeria, also from \citet{zheng_self-similar_2019}.}
  }
  \label{tab:Sleipner}
  \end{center}
\end{table}
{\color{black}\subsection{Model limitations and final remarks}}
The present model rests on numerous assumptions, which may limit its applicability to practical applications. In particular, inertial effects may become significant in coarse-grained porous media, especially for gases, which typically exhibit low viscosity. An extension of Darcy's law (the Darcy--Forcheimer equation) accounts for the additional inertial drag in high-Reynolds-number flow \citep{majdabadi_farahani_darcyforchheimer_2024}. At the opposite end of the pore-size spectrum, when pore diameters approach the mean free path of gas molecules, the continuum assumption can break down. At these small scales, molecular interactions with solid boundaries become more frequent than intermolecular collisions, giving rise to non-continuum transport phenomena. This includes pressure-dependent permeability through the emergence of slip flow \citep{wu_no_1998} and Knudsen diffusion, a mechanism driven by molecule–wall interactions that dominate over bulk-phase collisions \citep{darabi_gas_2012}. A simple model to account for slip flow, originally proposed by \citet{Klink_1941}, introduces a nonlinear correction to the permeability that is inversely proportional to pressure. In the context of this study, lower pressures were observed for more compressible gases, which may enhance slip flow relative to weakly compressible gases.

We presented a model for a regime where capillary forces are subdominant to gravitational forces, in which case the interface will be sharp and the pressure across the interface continuous. {\color{black} This modelling assumption is consistent with recent experiments by \citet{mortimer_dynamic_2024}, which showed that the interface remained relatively sharp under slow air injection into a water-saturated tank of glass beads.} However, when capillary forces become significant, the sharp-interface approximation may no longer be valid, and it may become necessary to incorporate a macroscale capillary pressure jump into the model. Increased capillary forces also promote residual trapping of isolated gas bubbles through capillary snap-off \citep{lysyy_pore-scale_2022}. The extent of this trapping is governed by the system’s wettability and the heterogeneity of the pore structure. In CO$_2$ storage, residual trapping is beneficial since storage is intended to be permanent; in contrast, this may complicate recovery in hydrogen storage. However, once isolated, trapped bubbles may undergo Ostwald ripening, whereby spatial variations in capillary pressure drive diffusive mass transfer from smaller to larger bubbles through the surrounding liquid \citep{zhang_porescale_2023}. This redistribution could promote the formation of larger, more mobile gas clusters, potentially enhancing recoverability.  

Given the potential importance of capillary effects, it is useful to identify conditions under which the sharp-interface approximation is valid. \citet{zheng_self-similar_2019} extended the incompressible injection models of \citet{pegler_fluid_2014} and \citet{zheng_flow_2015} by incorporating partial saturation and a saturation-dependent capillary pressure. Their formulation employed a Brooks–Corey capillary–pressure saturation relation together with the multiphase extension of Darcy’s law \citep{bear1972dynamics}, which captures reductions in permeability due to partial saturation. The resulting dimensionless formulation depends on three key parameters: the {viscosity} ratio scaled by the endpoint relative permeability of the non-wetting phase, $k_{rn0}/\mathcal{M}$; the Bond number, $Bo = \Delta \rho g H / p_e$, where $p_e$ is the capillary entry pressure; and an empirical pore-size distribution parameter, $\Lambda$. In this framework, a sharp interface is recovered in two limits: $Bo \to \infty$ (see the estimate in \S~\ref{sec:Model}), where buoyancy forces dominate over all capillary entry pressures, and $\Lambda \to \infty$, corresponding to monodisperse pores with constant identical entry pressures, so that displacement occurs as a single advancing front rather than a diffuse transition zone.

In addition to these capillary-driven interfacial phenomena, fluid displacement may also be destabilised dynamically through the onset of viscous fingering, which arises when a low-viscosity fluid displaces a more viscous one. A detailed discussion of how this instability impacts underground hydrogen storage can be found in \citet{paterson_implications_1983}. It is noteworthy, however, that certain features of liquid displacement by a compressible gas can act to delay the onset and mitigate the severity of such instabilities. In particular, the large density contrast between hydrogen and brine has a stabilising effect on the interface, as demonstrated by \citet{Taylor_1958}. More recently, \citet{cuttle_compression-driven_2023} showed how higher gas compressibility can have an effect comparable to that of an elevated capillary number in delaying the onset of fingering.

While interfacial instabilities such as viscous fingering shape the large-scale displacement pattern, the rheological properties of the displacing fluid itself can also influence the flow dynamics. In addition to the shear viscosity of the fluids, the compressible Navier–Stokes equations reveal an additional viscous contribution associated with volumetric deformation, quantified by the bulk viscosity, $\mu_b = \lambda + \frac{2}{3} \mu$, where $\lambda$ is the second (or volume) viscosity \citep{li_continuum_2017} which is not accounted for in Darcy's law. As a fundamentally pore-scale effect, the role of bulk viscosity, and its potential to be systematically upscaled to reservoir-scale flow, remains unclear. However, when energy dissipation due to volumetric deformation is negligible, the Stokes hypothesis ($\mu_b = 0$) is often invoked \citep{graves_bulk_1999}. This approximation has been shown to hold exactly for monatomic gases such as helium, and approximately for weakly compressible polyatomic gases like hydrogen \citep{graves_bulk_1999}. However, for highly compressible gases such as CO$_2$, the bulk viscosity can exceed the shear viscosity by several orders of magnitude \citep{pan_role_2017} and may therefore have a significant influence on flow.

Gas compression leads to localized temperature increases, resulting in both spatial and temporal temperature fluctuations within the injected gas. Capturing these effects would require a more sophisticated equation of state, coupled with conservation of energy. However, in porous media, the key question is whether heat exchange with the solid matrix is sufficient to suppress such variations. In this study we considered isothermal compression. When analysing temperature and pressure variations during cyclical injection and withdrawal in underground compressed-air storage, \citet{kushnir_thermodynamic_rev_2012} identified the dominant factor distinguishing adiabatic from isothermal behaviour in a porous medium as the rate of heat transfer between the gas and the solid matrix. This rate can be quantified by the thermal effusivity, $e = \sqrt{\kappa \rho c_p}$,
where $\kappa$ is the thermal conductivity, $\rho$ the density, and $c_p$ the specific heat capacity \citep{Dante2015_FrictionMaterials}. In underground reservoirs the matrix is typically sandstone. Measurements by \citet{labus_thermal_2018} give sandstone effusivity $e_s \approx 1400$ -- $3700~\mathrm{J\,m^{-2}\,K^{-1}\,s^{-1/2}}$. For hydrogen, typical reservoir conditions yield order-of-magnitude estimates $c_p$ $\sim 10~\mathrm{kJ\,kg^{-1}\,K^{-1}}$ and $\kappa \sim 0.1~\mathrm{W\,m^{-1}\,K^{-1}}$ \citep{NIST}, giving hydrogen effusivity $e_g \sim 10$ -- $10^2~\mathrm{J\,m^{-2}\,K^{-1}\,s^{-1/2}}$. Since $e_g \ll e_s$, the solid acts as an efficient heat sink, absorbing heat rapidly and keeping the gas temperature nearly constant over the long compression timescale $t_r$. However, this may not hold for CO$_2$ due to its relatively high density. \citet{kushnir_thermodynamic_rev_2012} further noted that in high-effusivity rocks, temperature variations can generate thermal stresses that modify pore structure and permeability, while conductive heat loss into the matrix increases gas density and may thereby enhance storage efficiency.

Since shear viscosity and bulk viscosity are pressure- and temperature-dependent, the spreading dynamics may be influenced by thermodynamic effects arising from gas compression. However, the significance of these effects depends upon the gas under consideration. For the range of typical pressures and temperatures of underground reservoirs ($T < 150\,^{\circ}C$, $P < 50\,\si{MPa}$), hydrogen exhibits minimal variation in viscosity. {\color{black} We therefore expect that this neglected variation may affect the quantitative agreement with the model, but it is unlikely to substantially alter the overall dynamics or the specific flow regime.} In contrast, CO$_2$ can experience an order-of-magnitude increase in viscosity at elevated pressures \citep{heinemann_enabling_2021}. For CO$_2$, this would generate a viscosity gradient along the channel, modifying both flow velocities and the pressure required to drive the fluid.

A distinguishing feature of hydrogen storage, relative to other gases, is the approximately seasonal cycle of injection and withdrawal. Such periodic injection behaviour has been explored in the context of {underground} thermal energy storage, where hot, dense fluids are cyclically injected and extracted \citep{dudfield_periodic_2014-1}, and more recently adapted to hydrogen storage scenarios \citep{whelan_periodic_2025}. Notably, \citet{whelan_periodic_2025} showed that when the injection velocity exceeds the characteristic buoyancy speed, the system approaches a state in which oscillations remain confined near the source. Conversely, when buoyancy dominates, oscillations propagate throughout the domain. Although the present study focused mainly on a steady injection rate, the model formulation readily accommodates periodic forcing. As injection cycles drive pressure fluctuations, the associated compression and decompression of the gas will also oscillate. Understanding how the phase relationship between these oscillatory components influences the large-scale flow may be a fruitful direction for future investigation.

In practical settings, stratification and curvature may also play important roles. Curvature of the confining boundaries due to tectonic deformation may be more favourable for trapping buoyant gas \citep{hagemann_mathematical_2015}. While we have assumed permeability to be isotropic and uniform, natural reservoirs typically exhibit stratification, with porous channels forming layered structures that result in higher horizontal than vertical permeability. These directional variations are further compounded by spatial variation in the permeability arising from the varying pore structure, which can significantly influence large-scale flow behaviour.  Additionally, the present model is readily generalisable to an axisymmetric geometry with a point source. A key qualitative difference from the planar case is that, depending on a viscous/buoyancy ratio \citep{guo_axisymmetric_2016,hutchinson_evolution_2023}, the injected gas plume may not extend to the lower domain boundary, a consequence of the point source {\citep{guo_axisymmetric_2016}}.

As noted in the introduction, \citet{cuttle_compression-driven_2023} studied a mathematically equivalent set-up involving liquid displacement in a radial Hele-Shaw cell, focusing on how gas compressibility influences the onset of interfacial instabilities. Our findings are in qualitative agreement with their observations of bulk flow dynamics in the absence of such instabilities. Specifically, they reported a non-monotonic pressure evolution, \hbox{i.e.} an initial rise due to gas compression, followed by a decline as viscous resistance diminished due to liquid drainage. They also found that the pressure scale decreased with increasing compressibility and that breakthrough times were correspondingly delayed. Notably, these bulk features were relatively insensitive to the presence of instabilities and exhibited only weak dependence on the capillary number, suggesting that the overall dynamics are primarily governed by a balance between gas compressibility and viscous dissipation in the displaced liquid. Building on these insights, we incorporated buoyancy effects and viscous dissipation in the gas. In contrast to their model, where the gas pressure is assumed spatially uniform and the dynamics are entirely coupled to viscous dissipation in the liquid, we revealed that, in the high-gas-mobility limit $\mathcal{M}\ll 1$, the flow develops an inner-outer structure. Near the source, for $\mathcal{L}^{-1}\ll \mathcal{M}\ll 1$, the gas pressure equilibrates and the flow is incompressible at late-times, while in the outer region, compressibility and viscous dissipation govern the spreading; for smaller $\mathcal{M}$, the bubble pressure can continue to rise throughout spreading.

In summary, we have investigated the influence of compressibility on flow dynamics using a long-wave model, exploring a broad range of parameter space. We presented reduced models and scalings for the dominant balances across distinct regions of parameter space, testing reduced-order models against numerical results. Our findings show that gas compression slows the spreading rate and increases the mass-to-volume ratio of the gas plume by the end of the injection cycle, compared to the incompressible case. These effects are particularly relevant for underground gas-storage applications, such as hydrogen storage, where accurate plume evolution predictions are crucial. Practically, our results suggest that elongated aquifers ($H/L \ll 1$), with relatively low permeability, amplify compressibility effects, leading to more efficient storage. We have demonstrated that, even for weakly compressible gases, compressibility effects may not be purely transient, but can persist throughout the entire spreading process when the channel is sufficiently long.

\section*{Acknowledgements}
LM would like to acknowledge the funding support from the EPSRC Industrial Decarbonisation Research and Innovation Centre (IDRIC MIPs 7.4 \& 7.8; EP/V027050/1).
For the purpose of open access, the authors have applied a Creative Commons Attribution (CC-BY) licence to any author accepted manuscript version arising.  {The authors acknowledge numerous helpful comments from reviewers.}

\section*{Declaration of interests} 
The authors report no conflict of interest.

\section*{Data availability statement}
The \texttt{MATLAB} code required to reproduce the numerical simulations of this study is available at 
\href{https://github.com/PeterCastellucci/Compressible_Evolution_Equations}{https://github.com/PeterCastellucci/Compressible\_Evolution\_Equations}.

\section*{Author ORCiDs}

PC: 0009-0000-8798-0213; RB: 0000-0003-1345-7029; LM: 0000-0003-2072-8480; ILC: 0000-0003-0284-9318; OEJ: 0000-0003-0172-6578

\appendix

\section{Numerical methods}
\label{app:nummeth}

To solve \eqref{eq:put2} we employ a coordinate transformation, mapping $(0,X_l(t))$ and $(X_l(t), X_u(t))$  onto fixed domains. We use the method of lines, discretising space using a finite-volume method based on central differences, and integrating the resulting system of ordinary differential equations (ODEs) in time using MathWorks \texttt{MATLAB}~R2024b’s stiff solver \texttt{ode15s}. This solver employs variable-step, variable-order integration, resulting in a scheme that achieves second-order accuracy in both space and time. For each simulation, the error in global mass conservation is maintained below 1\%. To enhance computational efficiency, we numerically compute the Jacobian matrix and supply it explicitly to the solver. By avoiding the overhead of the solver recalculating the Jacobian at each iteration, this approach significantly reduces computation time.

In the early-time formulation, where the compressed pressure and dynamic pressure are decoupled, we time-step $P_0(t)$ using \eqref{eq:Boyles}, while the interface height  $F(x,t)$ is evolved independently using (\ref{eq:put3}$b$). At each time step, the solution of \eqref{eq:Boyles} is substituted into \eqref{eq:put3}, allowing us to solve for $P_{1x}$ via the resulting ODE.

When solving the outer problem $\Pi_{2}$ we solve on the domain $\bar{x} = (\delta, \check{X}_u(\check{t})) $, where $\delta \ll 1$ to capture the singular behaviour from the boundary condition (\ref{eq:put2toc}$e$). In order to improve the resolution of the singularity at the origin, we use the transformation $\bar{x} = \exp(\xi) \check{X}_u$. A uniform grid in $\xi$ then clusters grid points near the origin. We discretize in space using a second-order-accurate finite-volume method and use \texttt{ode15s} to time-step these equations.

\section{Problem $\Pi_b$: Buoyancy}
\label{app:buoyancy}

A distinguished limit appears when $\phi\equiv \mathcal{L} \mathcal{M}^2\sim 1$ and $\theta\equiv \zeta\mathcal{L} \sim \mathcal{M} \ll 1$ (problem $\Pi_b$, figure~\ref{fig:overview}).  Here, buoyancy effects become dominant in the inner region of problem $\Pi_{12}$, with (\ref{eq:put2ti}) replaced by
\begin{subequations}
\begin{align}
0 &= \left[ (1-\tilde{F}) \tilde{P}_0 \tilde{P}_{1\tilde{x}}\right]_{\tilde{x}}, & \\
\tilde{F}_{\tilde{t}} &=\left[ \tilde{F}(\tilde{P}_{1\tilde{x}}+\phi^{-1}\tilde{F}_{\tilde{x}})\right]_{\tilde{x}}, & \tilde{X}_l<\tilde{x},\\
 - \tilde{P}_0 \tilde{P}_{1\tilde{x}}&=\mathcal{Q}(\mathcal{M}/\theta), & (\tilde{x}=0),\\
     \tilde{X}_{l,\tilde{t}}&=- (\tilde{P}_{\tilde{x}}+\phi^{-1}\tilde{F}_{\tilde{x}}), & (\tilde{x}={\tilde{X}_l+}).
\end{align}
The interface evolution is governed by the mixed hyperbolic-parabolic evolution equation
\begin{equation}
    \tilde{F}_{\tilde{t}}+\frac{\mathcal{Q}}{\tilde{P}_0}\frac{\mathcal{M}}{\theta}\frac{\tilde{F}_{\tilde{x}}}{(1-\tilde{F})^2}=\frac{1}{\phi}\left[\tilde{F}\tilde{F}_{\tilde{x}}\right]_{\tilde{x}}.
\end{equation}
\label{eq:put2b}
\end{subequations}
This corresponds to (\ref{eq:put2ti}$a$-$e$) with buoyancy effects restored.  For $\mathcal{L}^{-1/2}\ll \mathcal{M}\ll 1$, buoyancy effects are confined to a boundary layer near the lower contact line, perturbing its location. For $\phi \ll 1$, the inner region splits into two zones: a short region in which $\tilde{F}_{\tilde{t}}=\phi^{-1}(\tilde{F}\tilde{F}_{\tilde{x}})_{\tilde{x}}$ to leading order, with the lower contact line receding towards the origin; and a longer unsteady region of mixed parabolic-hyperbolic form with $F$ close to 1 that allows matching to the outer region.  We write the latter region using $\tilde{F}=1-\tilde{f}(\tilde{x},\tilde{t})$ with $\vert\tilde{f}\vert\ll 1$, so that to leading order
\begin{equation}
    \tilde{f}_{\tilde{t}}+\frac{\mathcal{Q}}{\tilde{P}_0}\frac{\mathcal{M}}{\theta}\frac{\tilde{f}_{\tilde{x}}}{\tilde{f}^2}=\frac{1}{\phi}\tilde{f}_{\tilde{x}\tilde{x}}.
\label{eq:put2c}
\end{equation}
The outer region matches via the first two terms, where $\tilde{f}$ becomes very small over long lengthscales, while the nonlinear diffusion region matches to (\ref{eq:put2c}) via the first and third terms where $\tilde{f}$ is larger. Eq.~(\ref{eq:put2c}) admits a similarity solution involving a balance of all three terms with $\tilde{f}=(\mathcal{Q}\mathcal{M}/\tilde{P}_0\theta)^{1/2} (\phi t^{1/4})g(\xi)$, $\xi=(\phi/t)^{1/2} x$ provided the time-dependence of $\mathcal{Q}$ and $\tilde{P}_0$ is neglected.  However, we use (\ref{eq:put2b}$a$) to argue that the gas flux through this region is uniform, allowing the source (\ref{eq:put2b}$c$) to communicate directly to the outer region, so that recession of the lower contact line does not have a prominent effect on spreading further downstream.

\section{Very low viscosity ratio}
\label{app:verylowviscosity}

Here, we discuss features that emerge at the base of regions $\Pi_1$ and $\Pi_2$ in figure~\ref{fig:overview}, where the viscosity ratio is very low.  When $\mathcal{M}$ is very small in the early-time problem (\ref{eq:put3}), $F_0$ and $X_l$ are effectively frozen at their initial state (\ref{eq:ini}) and (\ref{eq:put3}$a$,$c$) can be integrated to give 
\begin{equation}
    P_{1x}=\frac{\Delta_0(P_{0t}L_0-\mathcal{Q})}{P_0(X_u(0)-x)}-\frac{X_u(0)-x}{2\Delta_0}\frac{P_{0t}}{P_0}.
\end{equation}
This solution accommodates a mass flux $P_0 P_{1x}(1-F)=P_{0t}L_0-\mathcal{Q}$ as $x\rightarrow X_u(0)-$, consistent with (\ref{eq:Boyles}) for $\mathcal{M}\ll 1$.  {Here,} $P_1$ diverges logarithmically as $x\rightarrow X_u(0)-$ and is regularised by viscous effects over short lengthscales as follows.  

Set $x=X_u(0)+\mathcal{M}^{1/3}\breve{x}$, $F=1-\mathcal{M}^{1/3}\breve{F}(\breve{x},t)$.  Then (\ref{eq:put3}) gives, to leading order in $\mathcal{M}\ll 1$ 
\begin{equation}
    0=[\breve{F}P_{1\breve{x}}]_{\breve{x}}, \quad \breve{F}_t=-P_{1\breve{x}\breve{x}}.
    \label{eq:short}
\end{equation}
This matches to the frozen region via $\breve{F}\approx -\breve{x}/\Delta_0$ for $-\breve{x}\gg 1$.  The mass flux of gas is uniform through this short region, with value $\breve{F}P_0P_{1\breve{x}}\approx P_{0t}L_0-\mathcal{Q}$; (\ref{eq:short}) gives the local evolution equation
\begin{equation}
    \breve{F}_t+\frac{\mathcal{Q}-P_{0t}L_0}{P_0}\frac{\breve{F}_{\breve{x}}}{\breve{F}^2}=0.
    \label{eq:shortevo}
\end{equation}
This variation of the hyperbolic transport problem for the interface shape in the inner region (\ref{eq:innerhyp}) accounts for compressible effects through transient variations in $P_0$, arising over short timescales.  The solution of (\ref{eq:shortevo}) satisfying $\breve{F}(\breve{x},0)=-\breve{x}/\Delta_0$ in $\breve{x}<0$ can be evaluated (via characteristics) as
\begin{equation}
    \breve{F}^2(\breve{x}-\breve{F}\Delta_0)=\int_0^t \frac{\mathcal{Q}}{P_0}\,\mathrm{d}t-L_0\log\left(\frac{P_0(t)}{P_0(0)}\right).
\end{equation}
This links the upstream limit $\breve{F}\approx -\breve{x}/\Delta_0\rightarrow \infty$ with the downstream limit 
\begin{equation}
    {F}\approx 1- \left(\frac{\mathcal{M}}{x-X_u(0)}\right)^{1/2} \left[\int_0^t \frac{\mathcal{Q}}{P_0}\,\mathrm{d}t-L_0\log\left(\frac{P_0(t)}{P_0(0)}\right) \right]^{1/2}
    \label{eq:newsource}
\end{equation}
(expressed in the original variables; recall $P=P_0/\zeta+P_1+\dots$).  Thus, over short timescales, the outer region is supplied directly by a flux that accommodates dynamic variations in $P_0$.  This adjustment can be expected to have an influence along the line $\mathcal{M}\zeta \mathcal{L}^3\sim 1$, when the timescale of problem $\Pi_2$ is no longer long in comparison to the timescale of the early evolution.  This line radiates from the distinguished limit $\zeta\sim 1/\mathcal{L}^2$, $\mathcal{M}\sim 1/\mathcal{L}$.

{\textbf{Problem $\Pi_c$.}}  We therefore reformulate (\ref{eq:put2}) setting $P=\mathcal{L}^2\ddot{P}(\bar{x},t)$, $F=1-\mathcal{M}\ddot{F}(\bar{x},t)$, $x = \mathcal{L} \bar{x}$, to give (for $\mathcal{L}\gg 1$),
\begin{subequations}
\label{eq:put21}
\begin{align}
\left[\ddot{F}\ddot{P}\right]_t &= \left[\ddot{F} \ddot{P} \ddot{P}_{\bar{x}}\right]_{\bar{x}}, & (0<\bar{x}<\bar{X}_u),\\
-\ddot{F}_t &= \ddot{P}_{\bar{x}\bar{x}}, & (0<\bar{x}<\bar{X}_u),\\
\ddot{P}+\ddot{P}_{\bar{x}} (1 - \bar{X}_u) &=\frac{1}{\zeta \mathcal{L}^2},  &(\bar{x}=\bar{X}_u),\\
    \bar{X}_{u,t}&=- \ddot{P}_{\bar{x}},& (\bar{x}=\bar{X}_u-),
\end{align}
with
\begin{equation}
    \ddot{F}\approx \left[\frac{1}{\bar{x}} \left( \frac{1}{\mathcal{M}\zeta \mathcal{L}^3}\int_0^t \frac{\mathcal{Q}}{\ddot{P}(0,t)}\,\mathrm{d}t-\frac{L_0}{\mathcal{M}\mathcal{L}}\log\left(\frac{\zeta \mathcal{L}^2 \ddot{P}(0,t)}{P_0(0)}\right) \right)\right]^{1/2}, \quad (\bar{x}\rightarrow 0).
    \label{eq:DLBC}
\end{equation}
\end{subequations}
This problem ($\Pi_c$ {in figure~\ref{fig:overview}}) has two independent parameters, $\zeta\mathcal{L}^2$ and $\mathcal{M}\mathcal{L}$, which are both $O(1)$ in the distinguished limit.  We can state (\ref{eq:put21}$e$) alternatively as
\begin{equation}
\label{eq:FBC}
\mathcal{ML}(\ddot{F}^2\bar{x})_t  \approx \frac{\mathcal{Q}/(\zeta \mathcal{L}^2)-{L_0} \ddot{P}_t}{\ddot{P}}, \quad (\bar{x}\rightarrow 0).
\end{equation}
{Here the flux entering the thin film (on the left) balances the source flux and dynamic (compressive) changes to the pressure in the main gas bubble.  The parameter $\mathcal{M}\mathcal{L}$ regulates the flux into the thin film while the parameter $\zeta \mathcal{L}^2$ regulates whether spreading is driven by the mass flux at the source or by the evolving bubble pressure.}

When $1\ll \mathcal{ML}\sim \zeta \mathcal{L}^2\ll \mathcal{L}$, we recover from (\ref{eq:put21}) the outer limit (\ref{eq:put2to}) of problem $\Pi_{12}$ by setting $\ddot{P}=\hat{P}/(\mathcal{LM})$ and $t=\mathcal{LM}\tilde{t}$.  The log term in (\ref{eq:put21}$e$) provides an $O(1/(\mathcal{ML}))$ correction.

When $\max((\zeta \mathcal{L}^2)^{1/2},\zeta\mathcal{L}^2)\ll \mathcal{ML}\ll \mathcal{L}$, we set $\ddot{P}=(\zeta \mathcal{L}^2)^{-1}+(\mathcal{ML})^{-1}\hat{P}_1$ and $t=\mathcal{ML}\tilde{t}$ recovers (\ref{eq:put2toa}) from (\ref{eq:put21}), {\hbox{i.e.}} the outer limit of problem $\Pi_1$, with error $O(L_0\zeta/\mathcal{M}^2)$.  

When $1\ll \mathcal{ML}\ll \zeta \mathcal{L}^2$, we recover from (\ref{eq:put21}) the outer problem (\ref{eq:put2toc}) (with error $1/(\zeta\mathcal{L}^2)$) by setting $\ddot{P}=\check{P}/(\mathcal{ML}^3\zeta)^{1/2}$, $t=(\mathcal{ML}^3\zeta)^{1/2}\check{t}$, but with the revised source condition
\begin{equation}
    \check{F}\approx \left[\frac{1}{\bar{x}} \left(\int_0^{\check{t}} \frac{\mathcal{Q}}{\check{P}(0,\check{t}^{\prime})}\,\mathrm{d}\check{t}^{\prime}-\frac{L_0}{\mathcal{M}\mathcal{L}}\log\left(\frac{(\mathcal{ML})^{-1}\check{P}(0,\check{t})}{P_0(0)}\right) \right)\right]^{1/2}.
    \label{eq:twoterms}
\end{equation}
The log term identifies the lower boundary of problem $\Pi_2$ as the line $\mathcal{ML}\sim 1$.  Along this boundary, the two terms in (\ref{eq:twoterms}) can be expected to balance.  

Below this boundary, we set $t=(\zeta \mathcal{L}^2)^{1/2}\dot{t}$, $\ddot{P}=(\zeta \mathcal{L}^2)^{-1/2}\dot{P}$, $\bar{X}_u(t) = \dot{X}_u(\dot{t})$.  This recovers from (\ref{eq:put21}) the problem
\begin{subequations}
\label{eq:put22}
\begin{align}
\left[\dot{F}\dot{P}\right]_{\dot{t}} &= \left[\dot{F} \dot{P} \dot{P}_{\bar{x}}\right]_{\bar{x}}, & (0<\bar{x}<\dot{X}_u),\\
-\dot{F}_{\dot{t}} &= \dot{P}_{\bar{x}\bar{x}}, & (0<\bar{x}<\dot{X}_u),\\
\dot{P}+\dot{P}_{\bar{x}} (1 - { \dot{X}_u}) &=0,  &(\bar{x}=\dot{X}_u),\\
{\dot{X}_{u,\dot{t}}}&=- \dot{P}_{\bar{x}},& (\bar{x}=\dot{X}_u-),\\
\dot{P}_{\dot{t}}&=\mathcal{Q}/{L_0}, & (\bar{x}=0),
\end{align}
\end{subequations}
which holds for $\mathcal{ML}\ll 1$ and $1\ll \zeta \mathcal{L}^2$ (region $\Pi_3$ in figure~\ref{fig:overview}).  The pressure at the origin rises independently of the degree of spreading via (\ref{eq:put22}$e$), because there is very small leakage of gas from the bubble into the thin film.  However we have now lost a boundary condition on $\dot{F}$.  This is resolved via retention of buoyancy, for example in (\ref{eq:shortevo}), which becomes
\begin{equation}
    \breve{F}_t+\frac{\mathcal{Q}-P_{0t}L_0}{P_0}\frac{\breve{F}_{\breve{x}}}{\breve{F}^2}=\mathcal{M}^{1/3}\breve{F}_{\breve{x}\breve{x}}.
    \label{eq:shortevob}
\end{equation}
If the film becomes sufficiently thin, the flux it carries is too small to allow gas to escape the bubble, and the supply is balanced by a continuous increase in pressure. In this state, the advection term in (\ref{eq:shortevob}) is then very weak, allowing buoyancy to regulate the transition between the bubble and the thin gas film further downstream.  Equation (\ref{eq:shortevob}) suggests that, under (\ref{eq:put22}$e$), buoyancy becomes dominant over a short lengthscale $x\sim \mathcal{M}^{1/2}$ near $X_u(0)$ and can be expected to modify the matching condition on $\dot{F}$.

On the boundary $\zeta \mathcal{L}^2\sim 1$ with $\mathcal{ML}\ll 1$, (\ref{eq:put21}) becomes
\begin{subequations}
\label{eq:put23}
\begin{align}
\left[\ddot{F}\ddot{P}\right]_t &= \left[\ddot{F} \ddot{P} \ddot{P}_{\bar{x}}\right]_{\bar{x}}, & (0<\bar{x}<\bar{X}_u),\\
-\ddot{F}_t &= \ddot{P}_{\bar{x}\bar{x}}, & (0<\bar{x}<\bar{X}_u),\\
\ddot{P}+\ddot{P}_{\bar{x}} (1 - \bar{X}_u) &={1}/({\zeta \mathcal{L}^2}),  &(\bar{x}=\bar{X}_u),\\
    \bar{X}_{u,t}&=- \ddot{P}_{\bar{x}},& (\bar{x}=\bar{X}_u-), \\
\ddot{P}_t&={\mathcal{Q}}/(L_0 \zeta \mathcal{L}^2), & (x\rightarrow 0).
\end{align}
\end{subequations}
This is distinguished from (\ref{eq:put22}) by the additional term in (\ref{eq:put23}$c$), elevating the pressure. 

The final major sublimit of (\ref{eq:put21}) arises for $\mathcal{M}\ll \zeta^{1/2}$ and $\zeta \mathcal{L}^2\ll 1$ (region $\Pi_4$ in figure~\ref{fig:overview}).  Here, set $\ddot{P}=(\zeta\mathcal{L}^2)^{-1}+\mathring{P}_1/(\zeta \mathcal{L}^2)^{1/2}$, $\ddot{F}=\mathring{F}(\bar{x},\mathring{t})$, $t=(\zeta \mathcal{L}^2)^{1/2}\mathring{t}$ and $\bar{X}_u (t)= \mathring{X}_{u}(\mathring{t})$, giving with error $O(\mathcal{M}/\zeta^{1/2})$
\begin{subequations}
\label{eq:put24}
\begin{align}
\mathring{F}_{\mathring{t}} &= \left[\mathring{F} \mathring{P}_{1\bar{x}}\right]_{\mathring{x}}, & (0<\bar{x}<\mathring{X}_u),\\
-\mathring{F}_{\mathring{t}} &= \mathring{P}_{1\bar{x}\bar{x}}, & (0<\bar{x}<\mathring{X}_u),\\
\mathring{P}_1 + \mathring{P}_{1x} (1 - \mathring{X}_u) &= 0,& (\bar{x}=\mathring{X}_u-), \\
\mathring{X}_{u,\mathring{t}}&=- \mathring{P}_{1\bar{x}},& (\bar{x}=\mathring{X}_u-),\\
\mathring{P}_{1\mathring{t}}&=\mathcal{Q}/{L_0}, & (\bar{x}=0).
\end{align}
\end{subequations}
This resembles the incompressible problem $\Pi_1$, but is driven by a pressure condition instead of an interfacial condition.  Finally, along the line $\mathcal{M}\sim\zeta^2\ll 1/\mathcal{L}$, the $(F^2 \bar{x})_{\mathring{t}}$ term is restored to the boundary condition (\ref{eq:put24}); this term balances $\mathcal{Q}$ on moving back into the incompressible region $\Pi_1$. Combining equations (\ref{eq:put24}$a$,$b$), integrating up to $\mathring{X}_u$, and applying the boundary conditions (\ref{eq:put24}$d$,$e$) relates the pressure gradient to the velocity of the upper contact line
\begin{equation}
\label{eq:solPx4}
    \mathring{P}_{1x} = - \frac{\mathring{X}_{u,\mathring{t}}}{(1+F)}.
\end{equation}
Substituting this expression into (\ref{eq:put24}$b$) leads to the hyperbolic problem (\ref{eq:hyp}), which can be integrated along characteristics to give
\begin{equation}
\label{eq:sol4}
    \mathring{F} = \left(\frac{\mathring{X}_u(\mathring{t})}{\bar{x}} \right)^{1/2} - 1.
\end{equation}
This solution has the equivalent form as the incompressible solution (\ref{eq:Peg}), but here $\mathring{X}_u$ is determined by the buildup of pressure in the gas bubble via (\ref{eq:put24}$e$), rather than being driven directly by the source. Integrating (\ref{eq:solPx4}), applying (\ref{eq:put24}$e$) and using (\ref{eq:sol4}) to evaluate the integral of $1/(1+\mathring{F})$, yields
\begin{equation}
\label{eq:Psol4}
    \mathring{P} = \frac{\mathring{\mathcal{V}}(\mathring{t})}{L_0} -\frac{2 \bar{x}^{3/2}}{3\mathring{X}_u^{1/2}} \mathring{X}_{u,\mathring{t}}, \quad \mathring{\mathcal{V}}(\mathring{t}) = \int_0^{\mathring{t}} \mathcal{Q} \,\mathrm{d}\mathring{t}^\prime.
\end{equation}
Combining the boundary condition (\ref{eq:put24}$c$) with (\ref{eq:Psol4}) gives $\mathring{X}_{u,\mathring{t}}(1-\tfrac{1}{3}\mathring{X}_u)=\mathring{\mathcal{V}}/L_0$, with solution
\begin{equation}
    \label{eq:Xu4}
    \mathring{X}_u(\mathring{t}) = 3\left[ 1- \left(1 - \frac{2}{3 L_0} \int_0^{\mathring{t}} \mathring{\mathcal{V}}(\mathring{t}^\prime) \,\mathrm{d} \mathring{t}^{\prime} \right)^{1/2} \right].
\end{equation}
Taking $\mathcal{Q} = 1$, (\ref{eq:Xu4}) reveals that for small $\mathring{t}$, $\mathring{X}_u \approx \mathring{t}^2/(2 L_0)$, or equivalently $X_u\approx t^2/(2\zeta\mathcal{L}L_0)$. Eq (\ref{eq:Xu4}) predicts a breakthrough time of $t_b = ((5/3) L_0 \zeta \mathcal{L}^2)^{1/2}$. 

\bibliographystyle{jfm}
\bibliography{jfm}

\end{document}